\documentclass{cernrep} 
\usepackage{texnames}
\usepackage[T1]{fontenc}
\usepackage[bookmarks, colorlinks=true, linktoc=page, pdftex, linkcolor=red, citecolor=red, urlcolor=red]{hyperref}
\usepackage{fixltx2e}

\graphicspath{{pics/}}
\usepackage{paralist}	
\usepackage{siunitx}
\usepackage{booktabs}
\usepackage{subfig}

\newcommand{\fig}[1]{Fig.~\ref{#1}}
\newcommand{\tab}[1]{Tab.~\ref{#1}}

\usepackage{tikz}
\usepackage{tikz-3dplot}
\usetikzlibrary{positioning,shapes,arrows,calc}
\usetikzlibrary{decorations.pathmorphing}
\usetikzlibrary{decorations.markings}
\usetikzlibrary{decorations.text}
\usetikzlibrary{patterns,fadings}
\usetikzlibrary{shapes.multipart}
\usetikzlibrary{plotmarks}
\title{Lattice Design and Dynamic Aperture Studies for the FCC-ee Top-Up Booster Synchrotron}
\author{B. Haerer, T. Tydecks}
\institute{CERN, Geneva, Switzerland}

\begin{document}

\sisetup{per-mode = reciprocal,separate-uncertainty}

\begin{abstract}
The Future Circular Collider (FCC) study investigates the feasibility of circular colliders in the post-LHC era. 
The sub-study FCC-ee is a \SI{100}{\kilo\metre} electron positron collider in the energy range of 90-\SI{365}{\giga\electronvolt}.
In order to achieve a design luminosity in the order of 
\SI{e36}{\per\centi\meter\squared\per\second} continuous top-up injection is required. The injector chain therefore includes a \SI{100}{\kilo\meter} booster synchrotron in the same tunnel as the collider rings.
This paper presents the lattice design of this booster synchrotron and the first dynamic aperture studies based on the chromaticity correction sextupole scheme.
\end{abstract}

\maketitle

\section{Introduction}
This report summarises status of the lattice design of the FCC-ee top-up booster synchrotron as presented in the Conceptual Design Report \cite{fcceeDesignReport}.
FCC-ee is an electron-positron collider with a centre-of-mass energy up to \SI{365}{GeV} designed in the context of the Future Circular Collider study (FCC), which investigates the feasibility of \SI{100}{km} circular colliders for future high energy physics research \cite{fcceeDesignReport}-\cite{fccDesignReportVol4}. 
FCC-ee will be a precision measurement tool for the $Z$, $W$, and $H$ bosons as well as the top quark at the $t\bar{t}$ threshold with so far unprecedented luminosities up to the order of \SI{e36}{\per\centi\meter\squared\per\second}. 
To achieve this goal, it is mandatory to continuously top-up the stored beam intensities. 
Therefore, full-energy injection into a fully squeezed machine optics with as little as possible disturbance of the physics experiments is required. 
As a consequence, the injected beam emittance must be comparable to the collider emittance. To guarantee high efficiency and stable operation, dynamic and momentum aperture of the booster must be sufficiently large.

 \section{The FCC-ee injector complex and performance requirements}
\begin{figure}[tbp]
   \begin{center}
   \includegraphics[width=0.8\textwidth]{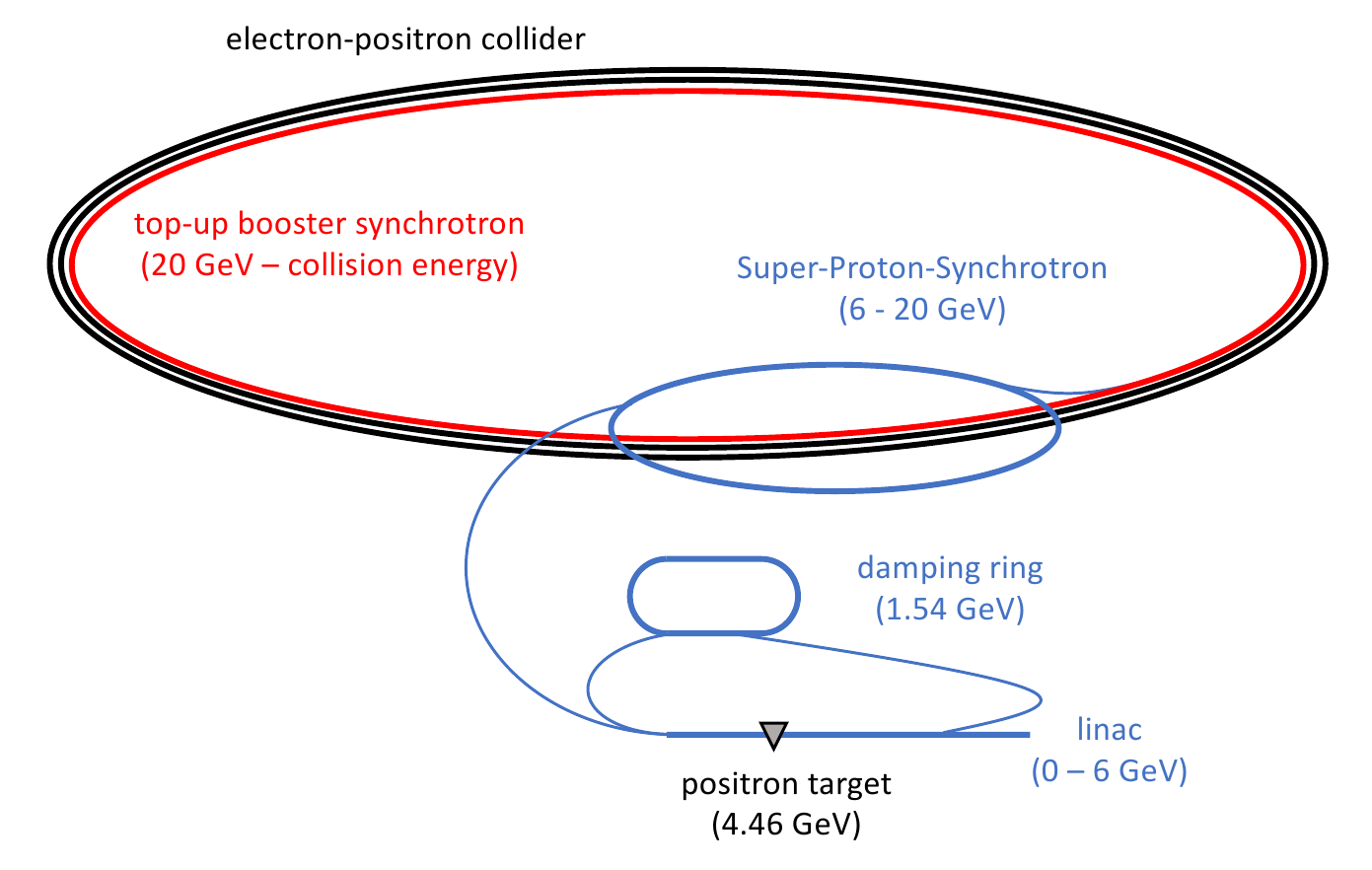}
   \end{center}
   \caption{Schematic layout of the FCC-ee injectors according to  \cite{fcceeDesignReport}. The SPS is used as  pre-booster ring.}
   \label{fig:fcceeinjectorcomplex}
 \end{figure}

 As described in the Conceptual Design Report, the FCC-ee injector chain foresees following components, which are illustrated in the schematic in Fig.~\ref{fig:fcceeinjectorcomplex} \cite{fcceeDesignReport,yannisamsterdam}:
 \begin{compactenum}
 \item A 328\,m S-Band linac with 6\,GeV extraction energy will serve as the first accelerator stage \cite{salimIPAC19,thesisSalim}. Electrons of this linac will also be used for the positron production at an energy of 4.46\,GeV.
 \item A damping ring with 242\,m circumference is used to cool the 1.54\,GeV electron and positron beams  before re-injection into the linac via a bunch compressor \cite{salimIPAC19,thesisSalim}.
 \item The next stage is the pre-booster synchrotron to increase the beam energy from 6\,GeV to 20\,GeV. The baseline foresees to use the SPS with a lattice modified for leptons and superconducting damping wigglers. As an alternative a new synchrotron is discussed as well \cite{Etisken:2694229}.
 \item The last stage will be the 100\,km top-up booster synchrotron for full-energy injection. This rapid-cycling synchrotron boosts the beam energy from \SI{20}{GeV} to the beam energy of the stored beam in the collider which ranges from \SI{45.6}{GeV} to \SI{182.5}{GeV} \cite{Haerer:IPAC2017-WEPIK031,Haerer:IPAC2018-MOPMF059}.
 \end{compactenum}
 Tab.~\ref{tab:parameterlisteAmsterdam} summarises the parameters of the injector complex. 
 The parameters for all injectors are given, but as this document focuses on the booster synchrotron, only the respective parameters will be discussed in this context.
 The injector complex has to fulfill two requirements: The overall filling time should not exceed 20 minutes and the target for the top-up injection is a current drop of less than 5\% \cite{yannisamsterdam} .
 These conditions are most difficult for the $Z$ operation mode, because as shown in Tab.~\ref{tab:parameterlisteAmsterdam} the highest number of bunches is required and also the highest bunch population.
 While for the $t\bar{t}$ mode only 50 bunches per beam are foreseen, for the $Z$ operation mode 16640 bunches are required.
 In addition, the maximum bunch population of $2.13\times10^{10}$ particles per bunch is foreseen for this operation mode.

 Details to the cycles of the FCC-ee injectors can be found in \cite{fcceeDesignReport} and \cite{yannisamsterdam}.
 1 - 8 fills of the pre-booster ring are transferred to the main booster in the bunch structure required by the collider.
 Depending on the operation mode, this corresponds to 50 - 16640 bunches. 
 As the ramp rate of the booster stays constant. The SPS was taken for reference: $\approx \SI{80}{GeV/s}$). The ramping time consequently depends on the extraction energy and is $t_{\text{r}}=\SI{0.32}{s}$ for 45.6\,GeV extraction energy and $t_{\text{r}}=\SI{2.0}{s}$ for 182.5\,GeV.
 The maximum cycle length is 51.7\,s for 45.6\,GeV operation, where the number of bunches and thus the filling time of the booster are highest.
With this scheme the current drop can be limited to $\pm3\%$ (5\% for $Z$) \cite{fcceeDesignReport}.

 \begin{table}
 \small
 \begin{center}
 \caption{Parameters of the injector chain \cite{fcceeDesignReport,yannisamsterdam}.}
 \begin{tabular}{lcccccccc}
 \toprule
   & \multicolumn{2}{c}{FCC-ee Z} & \multicolumn{2}{c}{FCC-ee W} & \multicolumn{2}{c}{FCC-ee H} & \multicolumn{2}{c}{FCC-ee tt}\\
 \midrule
 Energy (GeV) & \multicolumn{2}{c}{45.6} & \multicolumn{2}{c}{80} & \multicolumn{2}{c}{120} & \multicolumn{2}{c}{182.5}\\
 Type of filling & Full & Top-up & Full & Top-up & Full & Top-up & Full & Top-up \\
 LINAC \# bunches, \SI{2.8}{GHz} RF & \multicolumn{2}{c}{2} & \multicolumn{2}{c}{2} & \multicolumn{2}{c}{1} & \multicolumn{2}{c}{1}\\
 LINAC repetition rate (Hz) & \multicolumn{2}{c}{200} & \multicolumn{2}{c}{100} & \multicolumn{2}{c}{100} & \multicolumn{2}{c}{100}\\
 LINAC/PBR bunch popul. $\left(10^{10}\right)$ & 2.13 & 1.06 & 1.88& 0.56 & 1.88 & 0.56 & 1.38 & 0.83 \\
 \# of LINAC injections & \multicolumn{2}{c}{1040} & \multicolumn{2}{c}{1000} & \multicolumn{2}{c}{393} & \multicolumn{2}{c}{50}\\
 PBR bunch spacing (ns) & \multicolumn{2}{c}{2.5} & \multicolumn{2}{c}{22.5} & \multicolumn{2}{c}{57.5} & \multicolumn{2}{c}{450}\\
 \# PBR cycles & \multicolumn{2}{c}{8} & \multicolumn{2}{c}{1} & \multicolumn{2}{c}{1} & \multicolumn{2}{c}{1}\\
 PBR \# of bunches & \multicolumn{2}{c}{2080} & \multicolumn{2}{c}{2000} & \multicolumn{2}{c}{393} & \multicolumn{2}{c}{50}\\
 PBR cycle time (s) & \multicolumn{2}{c}{6.3} & \multicolumn{2}{c}{11.1} & \multicolumn{2}{c}{4.33} & \multicolumn{2}{c}{0.9}\\
 PBR duty factor & \multicolumn{2}{c}{0.84} & \multicolumn{2}{c}{0.56} & \multicolumn{2}{c}{0.35} & \multicolumn{2}{c}{0.08}\\
 BR \# of bunches & \multicolumn{2}{c}{16640} & \multicolumn{2}{c}{2000} & \multicolumn{2}{c}{393} & \multicolumn{2}{c}{50}\\
 BR cycle time (s) & \multicolumn{2}{c}{51.74} & \multicolumn{2}{c}{13.3} & \multicolumn{2}{c}{7.53} & \multicolumn{2}{c}{5.6}\\
 \# BR cycles & 10 & 1 & 10 & 1 & 10 & 1 & 20 & 1 \\
 \# injections/collider bucket & 10 & 1 & 10 & 1 & 10 & 1 & 20 & 1 \\
 Total number of bunches  & \multicolumn{2}{c}{16640} & \multicolumn{2}{c}{2000} & \multicolumn{2}{c}{393} & \multicolumn{2}{c}{50}\\
 Filling time (both species) (s) & 1034.8 & 103.5 & 288 & 28.8 & 150.6 & 15.6 & 224 & 11.2 \\
 Injected bunch population $\left(10^{10}\right)$ & 2.13 & 1.06 & 1.44 & 1.44 & 1.13 & 1.13 & 2.00 & 2.00\\
 \bottomrule    
 \end{tabular}
 \label{tab:parameterlisteAmsterdam}
 \end{center}
 \end{table}

\section{Layout constraints}
\begin{figure}[tbp]
    \centering
    \subfloat[FCC hadron collider]{\includegraphics[width=0.45\textwidth]{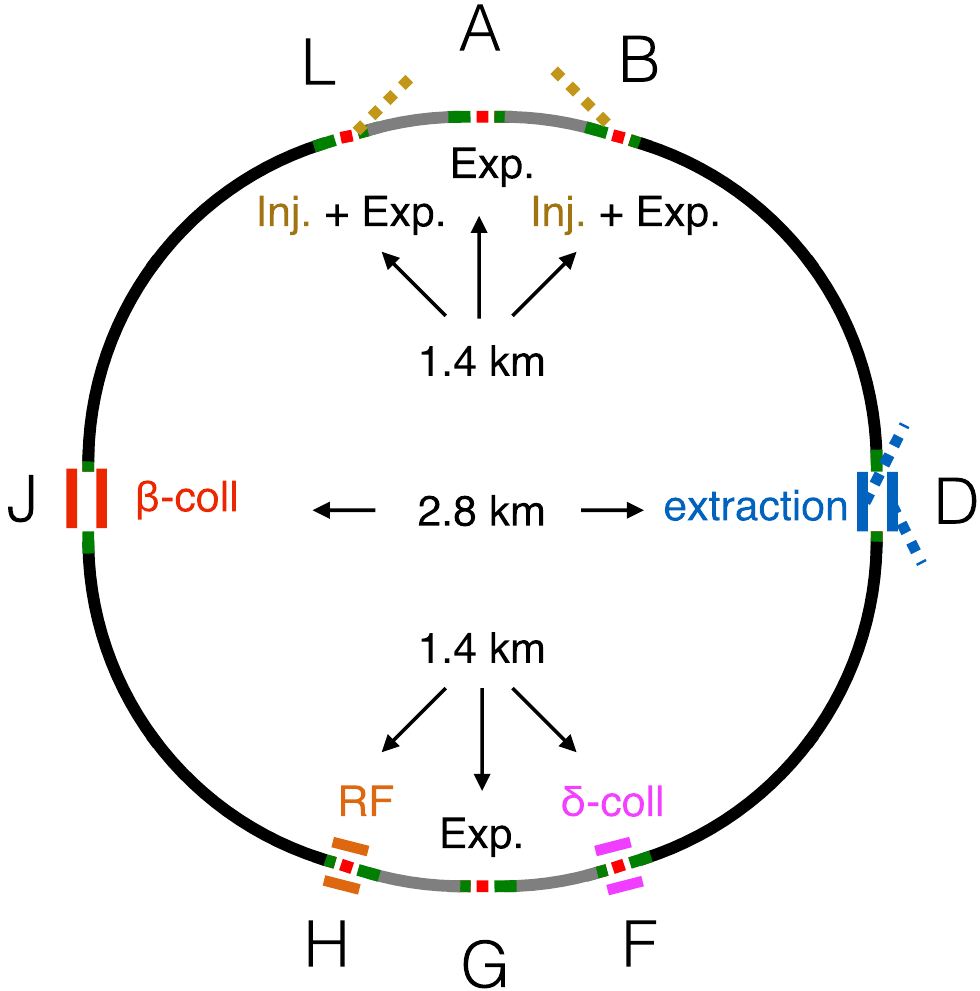}}\hfill
    \subfloat[FCC lepton collider]{\includegraphics[width=0.55\textwidth]{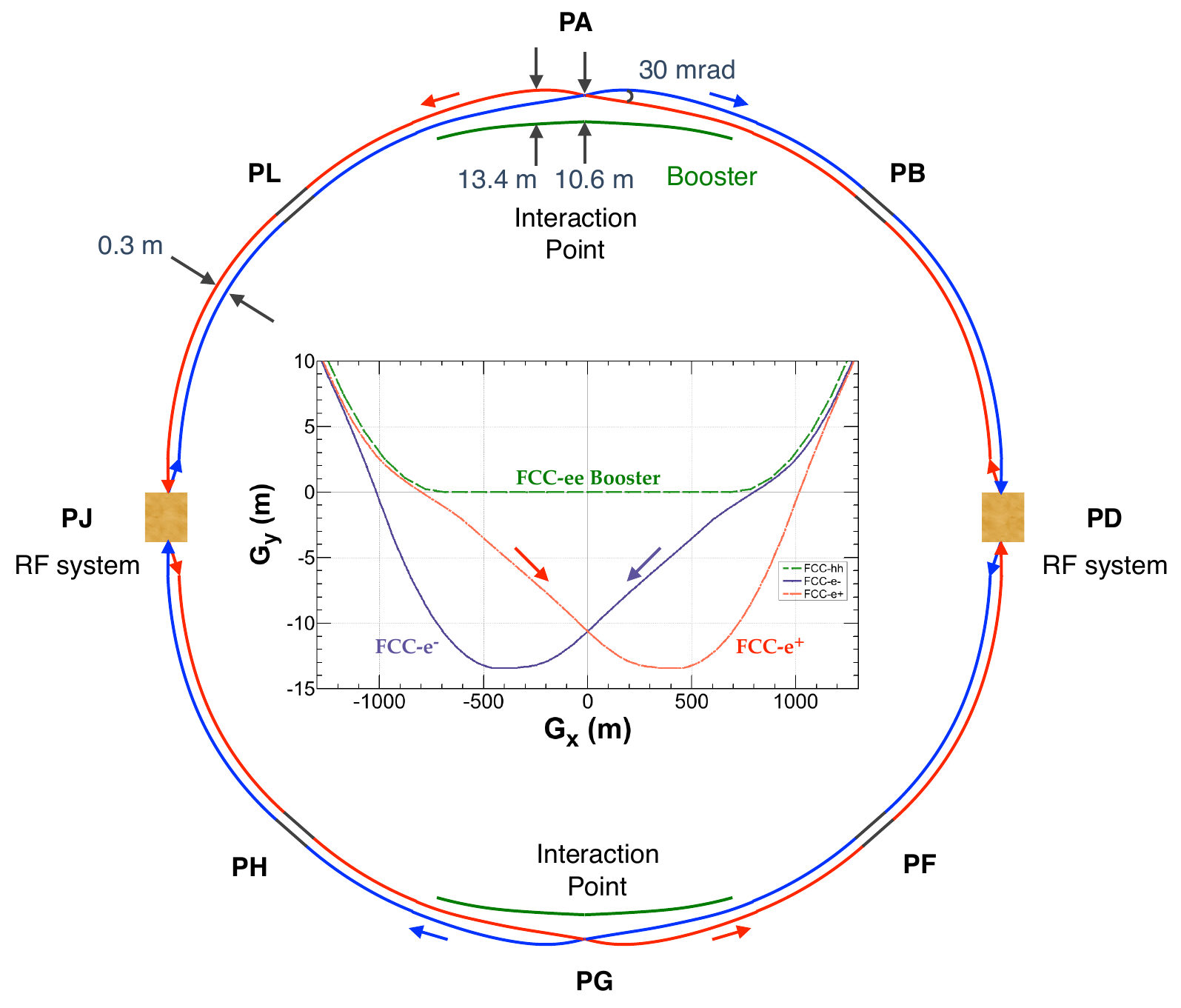}}
   \caption{Layout of the FCC colliders. (a) shows the layout of FCC-hh \cite{schulteamsterdam} and (b) shows the layout of FCC-ee as presented in \cite{fcceeDesignReport}. The booster synchrotron follows the footprint of the FCC hadron collider. The interaction regions of the FCC lepton collider are located in the straight sections around the points A and G. The interaction points will have a transverse offset of 10.6\,m to allow smooth bending of the incoming beam to mitigate synchrotron radiation background. The massive RF installations will be housed in the long straight sections around the points D and J.}
   \label{fig:FCCee_layout_oide}
 \end{figure}

The FCC-ee top-up booster shares the tunnel with the FCC-ee collider rings. 
The tunnel layout is driven by the requirements of the hadron collider shown in Fig.~\ref{fig:FCCee_layout_oide} (a):
The circumference is 97.75\,km defined by the FCC-hh design beam energy $E=\SI{50}{\giga\electronvolt}$ and the maximum available magnetic field of $B=\SI{16}{\tesla}$ based on Nb$_3$Sn technology leading to a local bending radius of \SI{13.1}{\kilo\metre} in the arc sections.

The layout of FCC-ee is shown in Fig.~\ref{fig:FCCee_layout_oide} (b). 
It includes six 
straight sections with a length of \SI{1.4}{\kilo\metre} clustered into groups of three around the points L, A, B and F, G, H. 
These straight sections are separated by short arc sections with the length of about \SI{4.3}{\kilo\metre} including dispersion suppressors.
In the centre straight sections A and G the FCC-ee interaction regions are foreseen for the two experiments.
The additional experimental caverns of FCC-hh around the points B and L will only be used for injection.
In addition, two long straight sections 
with a length of \SI{2.8}{\kilo\metre} are located around the points D and J, where the massive superconducting RF installations will be housed.
The arc sections adjacent to this straight sections have a length of 16.6 km.

\begin{figure}[tbp]
   \begin{center}
   \includegraphics[width=0.5\textwidth]{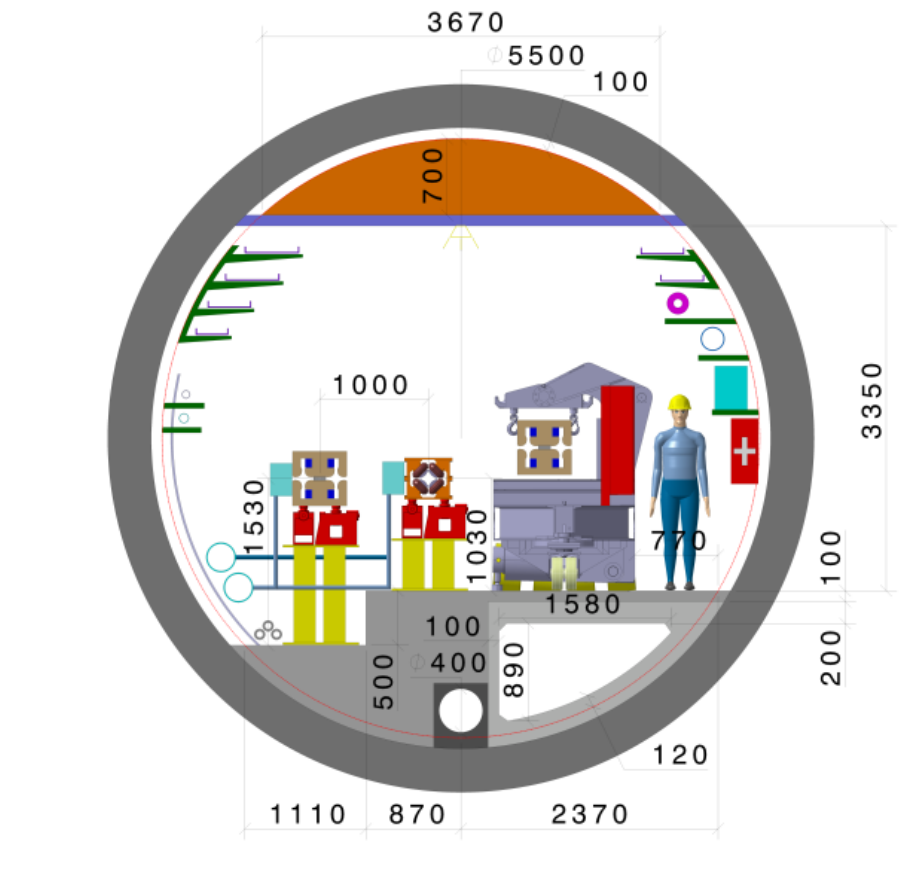}
   \end{center}
   \caption{Cross section of the FCC tunnel with FCC-ee and the top-up booster. The booster (orange yoke on the right) follows the footprint of FCC-hh, while the FCC-ee collider rings are located one metre on the outside (brown yoke on the left) \cite{fcceeDesignReport}.}
   \label{fig:FCCee_tunnel}
 \end{figure}

The booster synchrotron follows the footprint of the FCC hadron collider as shown in the cross section of the tunnel in Fig.~\ref{fig:FCCee_tunnel}.
The FCC-ee collider rings will be located further outside with a transverse offset of one metre. 
The interaction points will have an even larger offset of \SI{10.6}{m}, as a result of the requirements resulting from crossing angle and synchrotron radiation mitigation in front of the experiments.
Therefore, the booster synchrotron will bypass the FCC-ee detectors on the inside of the cavern as shown in the centre of \fig{fig:FCCee_layout_oide} (b).

The value of the injection energy is determined by the field quality and reproducibility of the magnetic field in the dipole magnets of the arc sections.
In the current design an energy of 20\,GeV is considered resulting in a magnetic field of $B=6$\,mT.


 \section{Lattice design of the FCC-ee booster and investigated optics} 
The FCC-ee booster synchrotron operates in top-up mode so the horizontal equilibrium beam emittance should not be larger than the collider's in order to not decrease luminosity.
The layout of the basic cell is therefore similar, it follows the FODO design as this lattice structure provides the largest dipole filling factor.
The cell length in the arc sections was chosen to be approximately 54\,m leading to a horizontal equilibrium emittance of \SI{1.3}{\nano\metre\radian} at the maximum operation energy of \SI{182.5}{\giga\electronvolt}.
This is about 12\% smaller than in the collider (56\,m cell length and an emittance of \SI{1.48}{\nano\metre\radian}) to allow some margin for emittance growth during beam transfer.
One cell includes four dipole magnets with a length of 11\,m each, two quadrupoles of 1.5\,m length and two sextupoles with 0.5\,m length giving a dipole filling factor of 81.5\,\%. 
Additional drift spaces for correctors, interconnections and flanges are already foreseen in the design.

Three different optics were investigated for the booster synchrotron: 
An optics with $\varphi_x=90^\degree$ and $\varphi_y=60^\degree$ phase advance per FODO cell was chosen as a starting point, as it was one of the optics used for LEP.
The baseline for the main collider rings foresees an optics with 90\textsuperscript{\degree}/90\textsuperscript{\degree}~phase advance per cell. 
An optics with same phase advance in both planes is better for chromaticity correction, because it is easier to establish the $-I$ transformations between sextupoles in both planes at the same time.
The advantages for the dynamic aperture will be discussed in detail in Sec.~\ref{sec:da}.
In addition, an optics with 60\textsuperscript{\degree}/60\textsuperscript{\degree}~phase advance per cell was investigated.
At the lowest operation energy of \SI{45}{GeV} the low synchrotron radiation power results in a small relative energy spread in the order of \num{1.e-4} and consequently a bunch length in the order of only \SI{1}{mm}. 
Investigations of  microwave instability and resistive wall effect have shown that such a short bunch length will establish unstable conditions for the beam \cite{eleonorathesis}.
The bunch length therefore had to be increased by a larger value of the momentum acceptance.
Since the geometry of the accelerator is fixed, the momentum acceptance must be changed by modifying the momentum compaction factor via the optics.
Therefore, this optics with 60\textsuperscript{\degree}/60\textsuperscript{\degree}~phase advance per cell is foreseen for the operation modes at 45.6\,GeV and 80\,GeV beam energy.
Since the change of optics also changes the equilibrium beam emittance, the optics of the booster will be adapted to obtain similar emittances as in the collider.

\begin{figure}[tbp]
 \centering
 \includegraphics[width=0.8\textwidth]{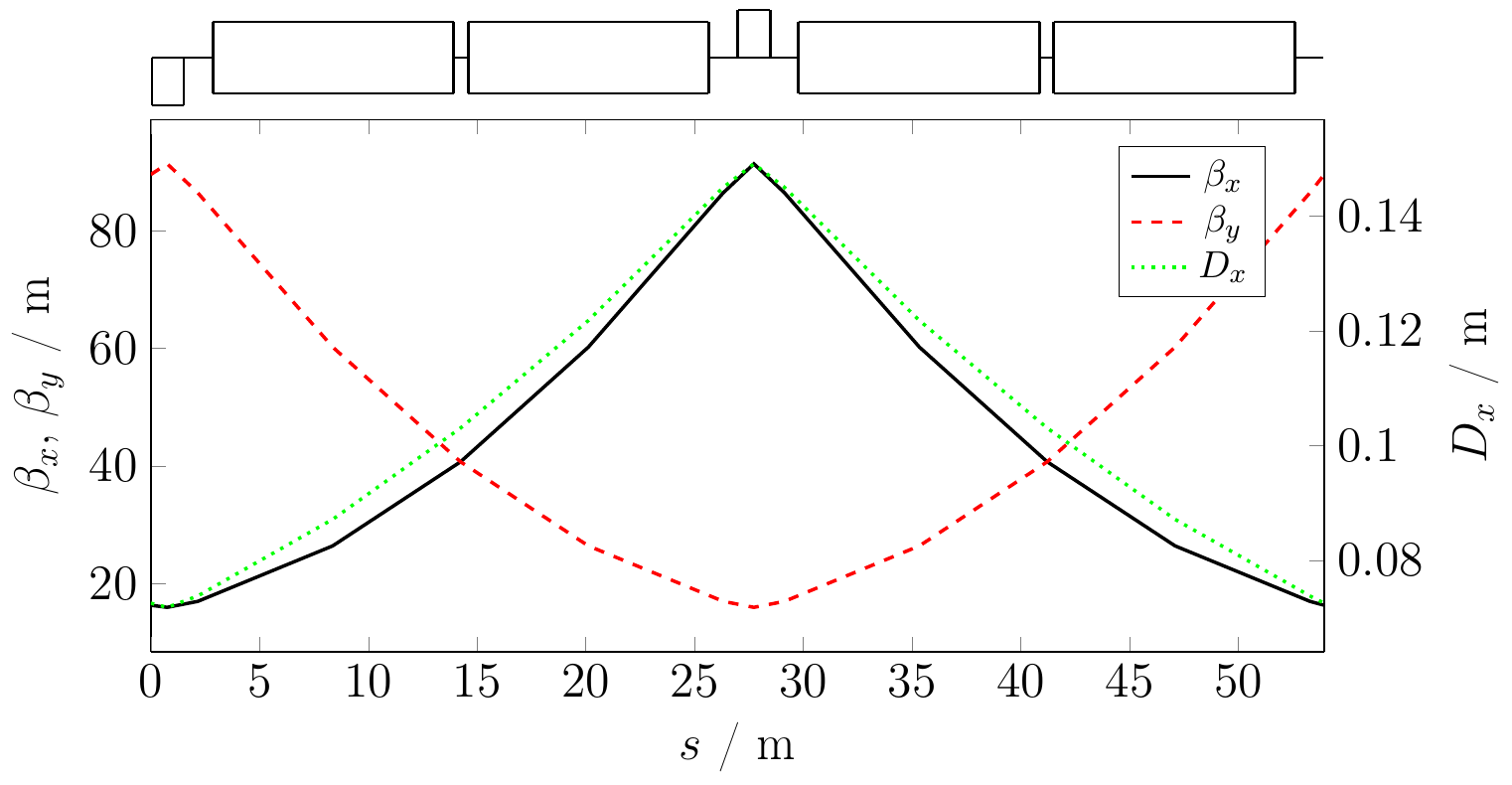}
 \caption{Beta functions and horizontal dispersion function of the FODO cell of the  FCC-ee booster synchrotron in the arc sections with a 90\textsuperscript{\degree}/90\textsuperscript{\degree}~optics. }
 \label{fig:fccfodo9090}
 \end{figure}

To summarise: As for the collider the lattice of the booster is optimised for two optics: 
an optics with 60{\textdegree} phase advance per cell is used for operation at the $Z$ peak and the $W$ pair production threshold (45.6\,GeV and 80\,GeV) and an optics with 90{\textdegree} phase advance per cell will be used for 
$H$ production and the $t\bar{t}$ production threshold (120\,GeV and 182.5\,GeV).
The resulting horizontal equilibrium emittances are compared in table \ref{tab:booster-emittances}.
The beta functions and the horizontal dispersion function of an arc FODO cell with 90\textsuperscript{\degree}/90\textsuperscript{\degree}~optics are presented in Fig.~\ref{fig:fccfodo9090}. 

 \begin{table}
 \caption{Horizontal equilibrium emittances of the booster compared to the values of the collider for all four beam energies. For 45.6\,GeV and 80\,GeV the 60{\textdegree} optics is used and for 120.0\,GeV and 182.5\,GeV optics the 90{\textdegree} optics.}
 \begin{center}
 \begin{tabular}{lcc}
 \toprule
 beam energy    & emittance booster    & emittance collider \\
 (in GeV)        & (in nm\,rad)        & (in nm\,rad)  \\
 \midrule
 45.6    &   0.24  &    0.24 \\
 80.0    &   0.73  &     0.84 \\
 120.0    &  0.55   &     0.63 \\
 182.5    &  1.30   &     1.48 \\
 \bottomrule
 \end{tabular}
 \end{center}
 \label{tab:booster-emittances}
 \end{table}

In the straight sections the horizontal dispersion is zero. 
The cell length has no effect on the equilibrium emittance and can therefore be chosen individually.
The cells in the \SI{1.4}{km} straight sections have a shorter length of exactly \SI{50}{m}, which is easier to accommodate in the relatively short straight sections.
The cell length in the long straight sections is \SI{100}{m} to maximise the room for the RF cavities and reduce the number of quadrupoles.

\begin{figure}[tbp]
 \centering
 \includegraphics[width=0.75\textwidth]{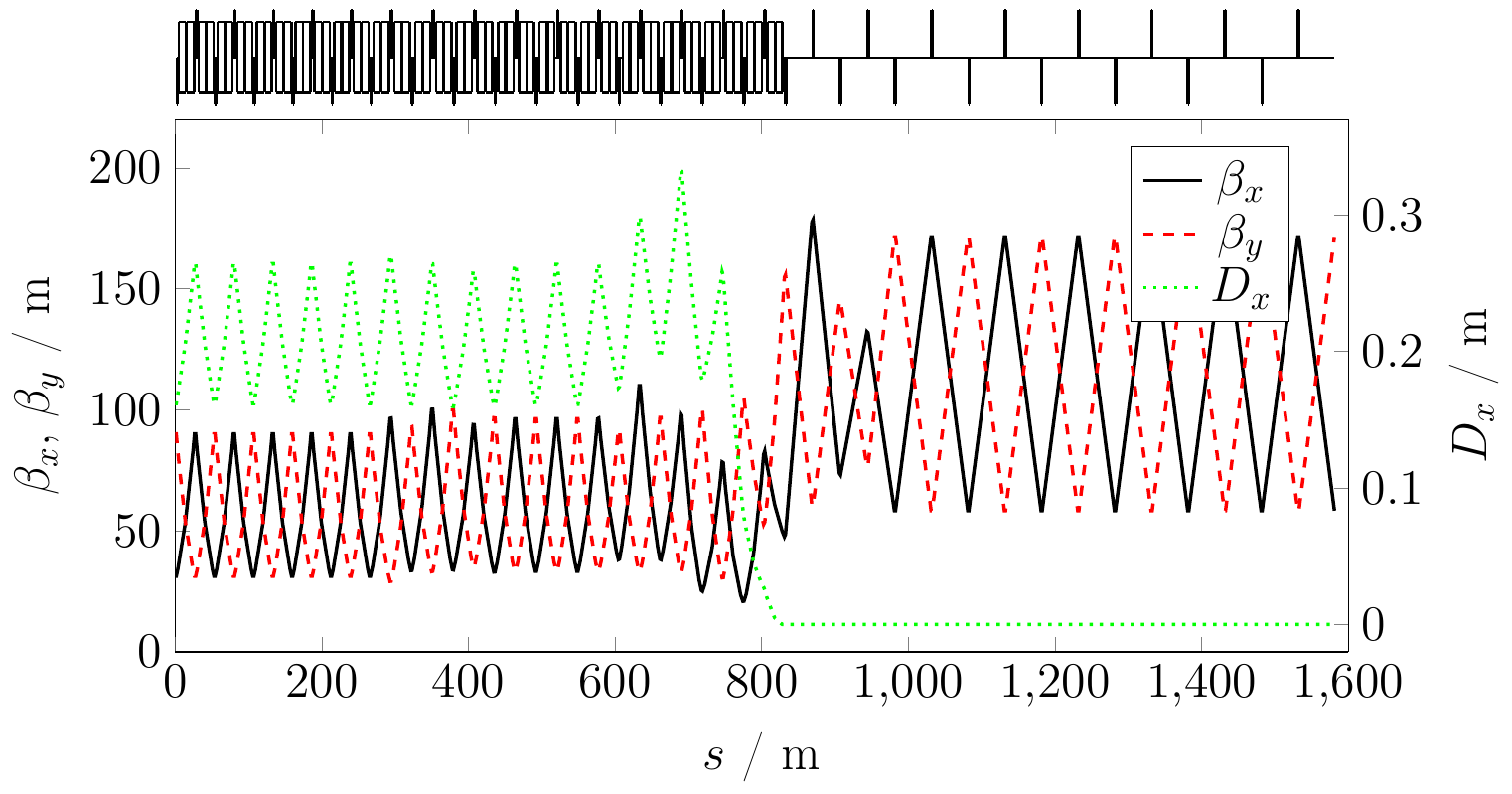}
 \caption{Transition of beta functions and horizontal dispersion function from the arc to the RF straight section for the 60\textsuperscript{\degree}/60\textsuperscript{\degree}~optics. The cell length in the arc sections is about \SI{54}{m}, in the last ten cells the cell length is \SI{56}{m} with a smaller curvature to fit the geometry of the dispersion suppressors of the hadron collider. The cell length in the straight section is \SI{100}{m} to maximise the room for the RF cavities. The quadrupoles of the \SI{56}{m} cells are used to match dispersion and beta functions.}
 \label{fig:fccboosterds}
 \end{figure}

At the beginning and at the end of each arc section dispersion suppressors are included to the lattice to allow a smooth transition of beta functions and horizontal dispersion function from arc to straight section.
The length of the dispersion suppressor sections is 566\,m following the requirements of the FCC hadron collider, which also needs a different curvature radius of $R=15.06$\,km instead of the regular curvature radius of the arc sections which is $R=13.15$\,km.
In the booster lattice 10 FODO cells with 56.6\,m length and less bending strength are installed to follow the geometry.
A quadrupole based dispersion suppressor in the last five cells is used to match the optics to the straight FODO cells. 
The transition of the optics of the arcs to these long FODO cells is shown in Fig.~\ref{fig:fccboosterds} for the \ang{60} optics.

In order to avoid orbit amplitudes and optics distortions due to the energy sawtooth the strength of dipoles and quadrupoles in the collider rings is adjusted to the local beam energy to obtain the same bending angle all over the arcs \cite{fcceeDesignReport,oidelattice1,oidelattice2}. 
As the booster is supposed to be a rapid cycling synchrotron such scaling might be difficult to synchronise because of the changing beam energy. 
In the case of the booster we assume lower requirements on orbit and optics stability since there are no mini-beta insertions in the lattice.
Therefore, such a "tapering" of magnetic strengths is not foreseen for the booster synchrotron.

Tunes, chromaticities and momentum compactionfactor of each optics are compared in \tab{tab:comparisonfodos}.

 \begin{table}[tbp]
 \caption{Global machine parameters for the three investigated optics. The table summarises tunes $Q_u$, natural chromaticities $\xi_u$, and momentum compaction factor.}
 \begin{center}
 \begin{tabular}{lSSS}
 \toprule
 	&	\ang{90}/\ang{60} & \ang{90}/\ang{90} & \ang{60}/\ang{60}\\
 \midrule
 $Q_x$	&	444.225	 & 459.225	& 296.225\\
 $Q_y$	&	295.290	 & 457.290	& 295.290\\
 $\xi_x$ 	&	-496.864 & -584.643	& -327.141\\
 $\xi_y$	&	-409.400	 & -582.910	& -326.332 \\
 $\alpha_c$ /$10^{-6}$	&	6.5	 & 6.7	& 13.8 \\
 \bottomrule
   \end{tabular}
 \end{center}
 \label{tab:comparisonfodos}
 \end{table}

 \section{Sextupole schemes for the arc sections}
 Several sextupole arrangements have been investigated for the chromaticity correction scheme of the FCC-ee booster synchrotron in order to achieve maximum dynamic aperture.
 For the study, next to each quadrupole a sextupole magnet is installed on each side.
 This arrangement allows to investigate different sextupole configurations.
 For all three optics three types of chromaticity correction scheme were studied for the FCC-ee booster synchrotron:
 \begin{compactenum}
 \item an interleaved sextupole scheme
 \item a "partially" non-interleaved sextupole scheme and
 \item a "completely" non-interleaved sextupole scheme
 \end{compactenum}

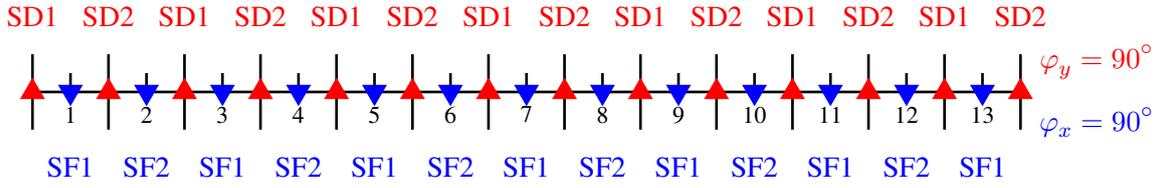
\begin{figure}[tbp]
   \centering
   \begin{tikzpicture}
     \def\scale{1.0}
     \draw[line width=1pt] (0*\scale,0)--(13*\scale,0);
     \draw[line width=1pt] (0*\scale,0.5*\scale)--(0*\scale,-0.5*\scale);
     \draw[line width=1pt] (1*\scale,0.5*\scale)--(1*\scale,-0.5*\scale);
     \draw[line width=1pt] (2*\scale,0.5*\scale)--(2*\scale,-0.5*\scale);
     \draw[line width=1pt] (3*\scale,0.5*\scale)--(3*\scale,-0.5*\scale);
     \draw[line width=1pt] (4*\scale,0.5*\scale)--(4*\scale,-0.5*\scale);
     \draw[line width=1pt] (5*\scale,0.5*\scale)--(5*\scale,-0.5*\scale);
     \draw[line width=1pt] (6*\scale,0.5*\scale)--(6*\scale,-0.5*\scale);
     \draw[line width=1pt] (7*\scale,0.5*\scale)--(7*\scale,-0.5*\scale);
     \draw[line width=1pt] (8*\scale,0.5*\scale)--(8*\scale,-0.5*\scale);
     \draw[line width=1pt] (9*\scale,0.5*\scale)--(9*\scale,-0.5*\scale);
     \draw[line width=1pt] (10*\scale,0.5*\scale)--(10*\scale,-0.5*\scale);
     \draw[line width=1pt] (11*\scale,0.5*\scale)--(11*\scale,-0.5*\scale);
     \draw[line width=1pt] (12*\scale,0.5*\scale)--(12*\scale,-0.5*\scale);
     \draw[line width=1pt] (13*\scale,0.5*\scale)--(13*\scale,-0.5*\scale);
     \draw[line width=1pt] (0.5*\scale,0.25*\scale)--(0.5*\scale,-0.*\scale);
     \draw[line width=1pt] (1.5*\scale,0.25*\scale)--(1.5*\scale,-0.*\scale);
     \draw[line width=1pt] (2.5*\scale,0.25*\scale)--(2.5*\scale,-0.*\scale);
     \draw[line width=1pt] (3.5*\scale,0.25*\scale)--(3.5*\scale,-0.*\scale);
     \draw[line width=1pt] (4.5*\scale,0.25*\scale)--(4.5*\scale,-0.*\scale);
     \draw[line width=1pt] (5.5*\scale,0.25*\scale)--(5.5*\scale,-0.*\scale);
     \draw[line width=1pt] (6.5*\scale,0.25*\scale)--(6.5*\scale,-0.*\scale);
     \draw[line width=1pt] (7.5*\scale,0.25*\scale)--(7.5*\scale,-0.*\scale);
     \draw[line width=1pt] (8.5*\scale,0.25*\scale)--(8.5*\scale,-0.*\scale);
     \draw[line width=1pt] (9.5*\scale,0.25*\scale)--(9.5*\scale,-0.*\scale);
     \draw[line width=1pt] (10.5*\scale,0.25*\scale)--(10.5*\scale,-0.*\scale);
     \draw[line width=1pt] (11.5*\scale,0.25*\scale)--(11.5*\scale,-0.*\scale);
     \draw[line width=1pt] (12.5*\scale,0.25*\scale)--(12.5*\scale,-0.*\scale);
     \node at (0.5*\scale,-0.3*\scale) {\footnotesize{1}};
     \node at (1.5*\scale,-0.3*\scale) {\footnotesize{2}};
     \node at (2.5*\scale,-0.3*\scale) {\footnotesize{3}};
     \node at (3.5*\scale,-0.3*\scale) {\footnotesize{4}};
     \node at (4.5*\scale,-0.3*\scale) {\footnotesize{5}};
     \node at (5.5*\scale,-0.3*\scale) {\footnotesize{6}};
     \node at (6.5*\scale,-0.3*\scale) {\footnotesize{7}};
     \node at (7.5*\scale,-0.3*\scale) {\footnotesize{8}};
     \node at (8.5*\scale,-0.3*\scale) {\footnotesize{9}};
     \node at (9.5*\scale,-0.3*\scale) {\footnotesize{10}};
     \node at (10.5*\scale,-0.3*\scale) {\footnotesize{11}};
     \node at (11.5*\scale,-0.3*\scale) {\footnotesize{12}};
     \node at (12.5*\scale,-0.3*\scale) {\footnotesize{13}};
     \node [red] at (0.0*\scale,1.0*\scale) {SD1};
     \node [red] at (1.0*\scale,1.0*\scale) {SD2};
     \node [red] at (2.0*\scale,1.0*\scale) {SD1};
     \node [red] at (3.0*\scale,1.0*\scale) {SD2};
     \node [red] at (4.0*\scale,1.0*\scale) {SD1};
     \node [red] at (5.0*\scale,1.0*\scale) {SD2};
     \node [red] at (6.0*\scale,1.0*\scale) {SD1};
     \node [red] at (7.0*\scale,1.0*\scale) {SD2};
     \node [red] at (8.0*\scale,1.0*\scale) {SD1};
     \node [red] at (9.0*\scale,1.0*\scale) {SD2};
     \node [red] at (10.0*\scale,1.0*\scale) {SD1};
     \node [red] at (11.0*\scale,1.0*\scale) {SD2};
     \node [red] at (12.0*\scale,1.0*\scale) {SD1};
     \node [red] at (13.0*\scale,1.0*\scale) {SD2};
     \node [blue] at (0.5*\scale,-1.0*\scale) {SF1};
     \node [blue] at (1.5*\scale,-1.0*\scale) {SF2};
     \node [blue] at (2.5*\scale,-1.0*\scale) {SF1};
     \node [blue] at (3.5*\scale,-1.0*\scale) {SF2};
     \node [blue] at (4.5*\scale,-1.0*\scale) {SF1};
     \node [blue] at (5.5*\scale,-1.0*\scale) {SF2};
     \node [blue] at (6.5*\scale,-1.0*\scale) {SF1};
     \node [blue] at (7.5*\scale,-1.0*\scale) {SF2};
     \node [blue] at (8.5*\scale,-1.0*\scale) {SF1};
     \node [blue] at (9.5*\scale,-1.0*\scale) {SF2};
     \node [blue] at (10.5*\scale,-1.0*\scale) {SF1};
     \node [blue] at (11.5*\scale,-1.0*\scale) {SF2};
     \node [blue] at (12.5*\scale,-1.0*\scale) {SF1};
     \node [red] at (14.0*\scale,0.4*\scale) {$\varphi_y =90^\degree$};
     \node [blue] at (14.0*\scale,-0.4*\scale) {$\varphi_x =90^\degree$};
     
     \node[mark size=5pt,color=red] at (0,0) {\pgfuseplotmark{triangle*}};
     \node[mark size=5pt,color=red] at (1,0) {\pgfuseplotmark{triangle*}};
     \node[mark size=5pt,color=red] at (2,0) {\pgfuseplotmark{triangle*}};
     \node[mark size=5pt,color=red] at (3,0) {\pgfuseplotmark{triangle*}};
     \node[mark size=5pt,color=red] at (4,0) {\pgfuseplotmark{triangle*}};
     \node[mark size=5pt,color=red] at (5,0) {\pgfuseplotmark{triangle*}};
     \node[mark size=5pt,color=red] at (6,0) {\pgfuseplotmark{triangle*}};
     \node[mark size=5pt,color=red] at (7,0) {\pgfuseplotmark{triangle*}};
     \node[mark size=5pt,color=red] at (8,0) {\pgfuseplotmark{triangle*}};
     \node[mark size=5pt,color=red] at (9,0) {\pgfuseplotmark{triangle*}};
     \node[mark size=5pt,color=red] at (10,0) {\pgfuseplotmark{triangle*}};
     \node[mark size=5pt,color=red] at (11,0) {\pgfuseplotmark{triangle*}};
     \node[mark size=5pt,color=red] at (12,0) {\pgfuseplotmark{triangle*}};
     \node[mark size=5pt,color=red] at (13,0) {\pgfuseplotmark{triangle*}};

     \node[mark size=5pt,color=blue,rotate=180] at (0.5,0) {\pgfuseplotmark{triangle*}};
     \node[mark size=5pt,color=blue,rotate=180] at (1.5,0) {\pgfuseplotmark{triangle*}};
     \node[mark size=5pt,color=blue,rotate=180] at (2.5,0) {\pgfuseplotmark{triangle*}};
     \node[mark size=5pt,color=blue,rotate=180] at (3.5,0) {\pgfuseplotmark{triangle*}};
     \node[mark size=5pt,color=blue,rotate=180] at (4.5,0) {\pgfuseplotmark{triangle*}};
     \node[mark size=5pt,color=blue,rotate=180] at (5.5,0) {\pgfuseplotmark{triangle*}};
     \node[mark size=5pt,color=blue,rotate=180] at (6.5,0) {\pgfuseplotmark{triangle*}};
     \node[mark size=5pt,color=blue,rotate=180] at (7.5,0) {\pgfuseplotmark{triangle*}};
     \node[mark size=5pt,color=blue,rotate=180] at (8.5,0) {\pgfuseplotmark{triangle*}};
     \node[mark size=5pt,color=blue,rotate=180] at (9.5,0) {\pgfuseplotmark{triangle*}};
     \node[mark size=5pt,color=blue,rotate=180] at (10.5,0) {\pgfuseplotmark{triangle*}};
     \node[mark size=5pt,color=blue,rotate=180] at (11.5,0) {\pgfuseplotmark{triangle*}};
     \node[mark size=5pt,color=blue,rotate=180] at (12.5,0) {\pgfuseplotmark{triangle*}};
   \end{tikzpicture}
   \caption{Schematic of an interleaved sextupole scheme for a FODO cell lattice with $\varphi=90^\degree$ phase advance in both planes. After every quadrupole a sextupole magnet is installed leading to a maximum number of sextupoles. The sextupoles that are separated by a phase advance of $\pi$ form a family. In this case there are two families per planes.}
   \label{fig:sextschemeinter}
 \end{figure}

 \noindent The interleaved sextupole scheme uses all available sextupoles. 
 Sextupoles located in positions with $\pi$ phase advance form families with the same strength in order to cancel geometric effects of the sextupoles.
 However, the members of one family are interlaced with the ones of the other families in both planes as illustrated for a 90\textsuperscript{\degree}/90\textsuperscript{\degree} FODO lattice in Fig.~\ref{fig:sextschemeinter}.
 While the interlaced sextupoles can disturb the cancellation of the geometric effects, 
 the big advantage of this scheme is the high number of sextupoles and consequently a low required strength leading to low non-linear fields in the lattice.

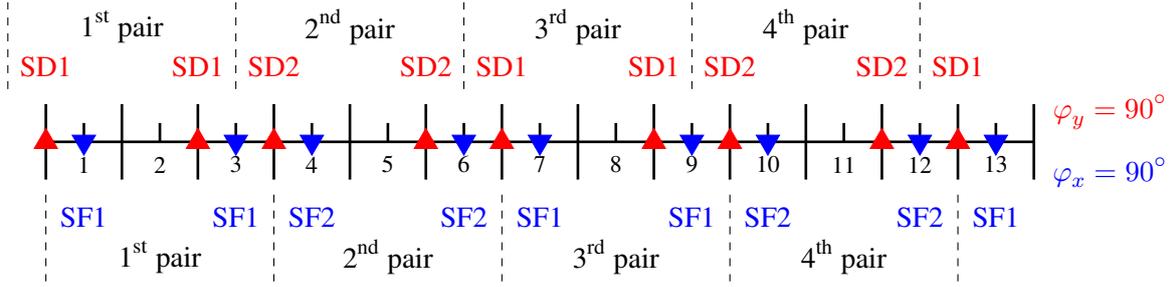
\begin{figure}[tbp]
   \centering
   \begin{tikzpicture}
     \def\scale{1.0}
     \draw[line width=1pt] (0*\scale,0)--(13*\scale,0);
     \draw[line width=1pt] (0*\scale,0.5*\scale)--(0*\scale,-0.5*\scale);
     \draw[line width=1pt] (1*\scale,0.5*\scale)--(1*\scale,-0.5*\scale);
     \draw[line width=1pt] (2*\scale,0.5*\scale)--(2*\scale,-0.5*\scale);
     \draw[line width=1pt] (3*\scale,0.5*\scale)--(3*\scale,-0.5*\scale);
     \draw[line width=1pt] (4*\scale,0.5*\scale)--(4*\scale,-0.5*\scale);
     \draw[line width=1pt] (5*\scale,0.5*\scale)--(5*\scale,-0.5*\scale);
     \draw[line width=1pt] (6*\scale,0.5*\scale)--(6*\scale,-0.5*\scale);
     \draw[line width=1pt] (7*\scale,0.5*\scale)--(7*\scale,-0.5*\scale);
     \draw[line width=1pt] (8*\scale,0.5*\scale)--(8*\scale,-0.5*\scale);
     \draw[line width=1pt] (9*\scale,0.5*\scale)--(9*\scale,-0.5*\scale);
     \draw[line width=1pt] (10*\scale,0.5*\scale)--(10*\scale,-0.5*\scale);
     \draw[line width=1pt] (11*\scale,0.5*\scale)--(11*\scale,-0.5*\scale);
     \draw[line width=1pt] (12*\scale,0.5*\scale)--(12*\scale,-0.5*\scale);
     \draw[line width=1pt] (13*\scale,0.5*\scale)--(13*\scale,-0.5*\scale);
     \draw[line width=1pt] (0.5*\scale,0.25*\scale)--(0.5*\scale,-0.*\scale);
     \draw[line width=1pt] (1.5*\scale,0.25*\scale)--(1.5*\scale,-0.*\scale);
     \draw[line width=1pt] (2.5*\scale,0.25*\scale)--(2.5*\scale,-0.*\scale);
     \draw[line width=1pt] (3.5*\scale,0.25*\scale)--(3.5*\scale,-0.*\scale);
     \draw[line width=1pt] (4.5*\scale,0.25*\scale)--(4.5*\scale,-0.*\scale);
     \draw[line width=1pt] (5.5*\scale,0.25*\scale)--(5.5*\scale,-0.*\scale);
     \draw[line width=1pt] (6.5*\scale,0.25*\scale)--(6.5*\scale,-0.*\scale);
     \draw[line width=1pt] (7.5*\scale,0.25*\scale)--(7.5*\scale,-0.*\scale);
     \draw[line width=1pt] (8.5*\scale,0.25*\scale)--(8.5*\scale,-0.*\scale);
     \draw[line width=1pt] (9.5*\scale,0.25*\scale)--(9.5*\scale,-0.*\scale);
     \draw[line width=1pt] (10.5*\scale,0.25*\scale)--(10.5*\scale,-0.*\scale);
     \draw[line width=1pt] (11.5*\scale,0.25*\scale)--(11.5*\scale,-0.*\scale);
     \draw[line width=1pt] (12.5*\scale,0.25*\scale)--(12.5*\scale,-0.*\scale);
     \node at (0.5*\scale,-0.3*\scale) {\footnotesize{1}};
     \node at (1.5*\scale,-0.3*\scale) {\footnotesize{2}};
     \node at (2.5*\scale,-0.3*\scale) {\footnotesize{3}};
     \node at (3.5*\scale,-0.3*\scale) {\footnotesize{4}};
     \node at (4.5*\scale,-0.3*\scale) {\footnotesize{5}};
     \node at (5.5*\scale,-0.3*\scale) {\footnotesize{6}};
     \node at (6.5*\scale,-0.3*\scale) {\footnotesize{7}};
     \node at (7.5*\scale,-0.3*\scale) {\footnotesize{8}};
     \node at (8.5*\scale,-0.3*\scale) {\footnotesize{9}};
     \node at (9.5*\scale,-0.3*\scale) {\footnotesize{10}};
     \node at (10.5*\scale,-0.3*\scale) {\footnotesize{11}};
     \node at (11.5*\scale,-0.3*\scale) {\footnotesize{12}};
     \node at (12.5*\scale,-0.3*\scale) {\footnotesize{13}};
     %
     \node [red] at (0.0*\scale,1.0*\scale) {SD1};
     \node [red] at (2.0*\scale,1.0*\scale) {SD1};
     \node [red] at (3.0*\scale,1.0*\scale) {SD2};
     \node [red] at (5.0*\scale,1.0*\scale) {SD2};
     \node [red] at (6.0*\scale,1.0*\scale) {SD1};
     \node [red] at (8.0*\scale,1.0*\scale) {SD1};
     \node [red] at (9.0*\scale,1.0*\scale) {SD2};
     \node [red] at (11.0*\scale,1.0*\scale) {SD2};
     \node [red] at (12.0*\scale,1.0*\scale) {SD1};
     \node [blue] at (0.5*\scale,-1.0*\scale) {SF1};
     \node [blue] at (2.5*\scale,-1.0*\scale) {SF1};
     \node [blue] at (3.5*\scale,-1.0*\scale) {SF2};
     \node [blue] at (5.5*\scale,-1.0*\scale) {SF2};
     \node [blue] at (6.5*\scale,-1.0*\scale) {SF1};
     \node [blue] at (8.5*\scale,-1.0*\scale) {SF1};
     \node [blue] at (9.5*\scale,-1.0*\scale) {SF2};
     \node [blue] at (11.5*\scale,-1.0*\scale) {SF2};
     \node [blue] at (12.5*\scale,-1.0*\scale) {SF1};
     \node [red] at (14.0*\scale,0.4*\scale) {$\varphi_y =90^\degree$};
     \node [blue] at (14.0*\scale,-0.4*\scale) {$\varphi_x =90^\degree$};
     \draw[dashed] (-0.5*\scale, 0.7*\scale)--(-0.5*\scale, 1.85*\scale);
     \draw[dashed] (2.5*\scale, 0.7*\scale)--(2.5*\scale, 1.85*\scale);
     \draw[dashed] (5.5*\scale, 0.7*\scale)--(5.5*\scale, 1.85*\scale);
     \draw[dashed] (8.5*\scale, 0.7*\scale)--(8.5*\scale, 1.85*\scale);
     \draw[dashed] (11.5*\scale, 0.7*\scale)--(11.5*\scale, 1.85*\scale);
     \draw[dashed] (0.0*\scale,-0.7*\scale)--(0.0*\scale,-1.85*\scale);
     \draw[dashed] (3.0*\scale,-0.7*\scale)--(3.0*\scale,-1.85*\scale);
     \draw[dashed] (6.0*\scale,-0.7*\scale)--(6.0*\scale,-1.85*\scale);
     \draw[dashed] (9.0*\scale,-0.7*\scale)--(9.0*\scale,-1.85*\scale);
     \draw[dashed] (12.0*\scale,-0.7*\scale)--(12.0*\scale,-1.85*\scale);
     \node [] at (1.0*\scale,1.525*\scale) {1\textsuperscript{st} pair};
     \node [] at (4*\scale,1.525*\scale) {2\textsuperscript{nd} pair};
     \node [] at (7*\scale,1.525*\scale) {3\textsuperscript{rd} pair};
     \node [] at (10*\scale,1.525*\scale) {4\textsuperscript{th} pair};
     \node [] at (1.5*\scale,-1.525*\scale) {1\textsuperscript{st} pair};
     \node [] at (4.5*\scale,-1.525*\scale) {2\textsuperscript{nd} pair};
     \node [] at (7.5*\scale,-1.525*\scale) {3\textsuperscript{rd} pair};
     \node [] at (10.5*\scale,-1.525*\scale) {4\textsuperscript{th} pair};
     \node[mark size=5pt,color=red] at (0,0) {\pgfuseplotmark{triangle*}};
     \node[mark size=5pt,color=red] at (2,0) {\pgfuseplotmark{triangle*}};
     \node[mark size=5pt,color=red] at (3,0) {\pgfuseplotmark{triangle*}};
     \node[mark size=5pt,color=red] at (5,0) {\pgfuseplotmark{triangle*}};
     \node[mark size=5pt,color=red] at (6,0) {\pgfuseplotmark{triangle*}};
     \node[mark size=5pt,color=red] at (8,0) {\pgfuseplotmark{triangle*}};
     \node[mark size=5pt,color=red] at (9,0) {\pgfuseplotmark{triangle*}};
     \node[mark size=5pt,color=red] at (11,0) {\pgfuseplotmark{triangle*}};
     \node[mark size=5pt,color=red] at (12,0) {\pgfuseplotmark{triangle*}};

     \node[mark size=5pt,color=blue,rotate=180] at (0.5,0) {\pgfuseplotmark{triangle*}};
     \node[mark size=5pt,color=blue,rotate=180] at (2.5,0) {\pgfuseplotmark{triangle*}};
     \node[mark size=5pt,color=blue,rotate=180] at (3.5,0) {\pgfuseplotmark{triangle*}};
     \node[mark size=5pt,color=blue,rotate=180] at (5.5,0) {\pgfuseplotmark{triangle*}};
     \node[mark size=5pt,color=blue,rotate=180] at (6.5,0) {\pgfuseplotmark{triangle*}};
     \node[mark size=5pt,color=blue,rotate=180] at (8.5,0) {\pgfuseplotmark{triangle*}};
     \node[mark size=5pt,color=blue,rotate=180] at (9.5,0) {\pgfuseplotmark{triangle*}};
     \node[mark size=5pt,color=blue,rotate=180] at (11.5,0) {\pgfuseplotmark{triangle*}};
     \node[mark size=5pt,color=blue,rotate=180] at (12.5,0) {\pgfuseplotmark{triangle*}};
   \end{tikzpicture}
   \caption{Schematic of a non-interleaved sextupole scheme for a FODO cell lattice with $\varphi=90^\degree$ phase advance in both planes. In each plane sextupole pairs are installed with a distance of $\pi$ phase advance. The sextupoles are considered to only act in one plane and are interlaced with the ones of the other plane. }
   \label{fig:sextschemenoninter}
 \end{figure}

 In case of the "partially" non-interleaved sextupole scheme the sextupole pairs are non-interleaved within each plane, but interleaved with the sextupole pairs of the other respective plane as shown in Fig.~\ref{fig:sextschemenoninter}.
 The chromatic effect of the sextupoles is scaled with the value of the beta function. 
 Therefore, the sextupoles are usually installed at positions where the beta functions are large in the plane where they are focusing.
 As the beta function in the defocussing plane is small at this positions their effect in this plane can be assumed to be small compared to the effect in the focusing plane, which is why the sextupole pairs of the different planes can be interleaved.

 The "completely" non-interleaved sextupole scheme features single sextupole pairs in a distance of $\mu_x = \mu_y=\pi$ phase advance. 
 Obviously this requirement is only possible to meet, if the phase advance is the same in both planes.
 The sextupole pairs alternately act on the two planes. Within two sextupoles of one pair no other sextupoles are installed.
 This scheme features the best cancellation of the geometric aberrations.
 When the number of sextupoles is large enough to distribute the non-linear fields this scheme should provide the larges dynamic aperture.
 In the case of the 90\textdegree/60\textdegree~optics a similar arrangement can be found that features $\mu_x = \pi$ and $\mu_y=2\pi$, which also allows to cancel the geometric aberrations.
In addition, with the smallest number of sextupoles this sextupole scheme is the most cost-efficient one because the least hardware is needed.

 \begin{figure}[tbp]
   \centering
   \begin{tikzpicture}
     \def\scale{1.0}
     \draw[line width=1pt] (0*\scale,0)--(13*\scale,0);
     \draw[line width=1pt] (0*\scale,0.5*\scale)--(0*\scale,-0.5*\scale);
     \draw[line width=1pt] (1*\scale,0.5*\scale)--(1*\scale,-0.5*\scale);
     \draw[line width=1pt] (2*\scale,0.5*\scale)--(2*\scale,-0.5*\scale);
     \draw[line width=1pt] (3*\scale,0.5*\scale)--(3*\scale,-0.5*\scale);
     \draw[line width=1pt] (4*\scale,0.5*\scale)--(4*\scale,-0.5*\scale);
     \draw[line width=1pt] (5*\scale,0.5*\scale)--(5*\scale,-0.5*\scale);
     \draw[line width=1pt] (6*\scale,0.5*\scale)--(6*\scale,-0.5*\scale);
     \draw[line width=1pt] (7*\scale,0.5*\scale)--(7*\scale,-0.5*\scale);
     \draw[line width=1pt] (8*\scale,0.5*\scale)--(8*\scale,-0.5*\scale);
     \draw[line width=1pt] (9*\scale,0.5*\scale)--(9*\scale,-0.5*\scale);
     \draw[line width=1pt] (10*\scale,0.5*\scale)--(10*\scale,-0.5*\scale);
     \draw[line width=1pt] (11*\scale,0.5*\scale)--(11*\scale,-0.5*\scale);
     \draw[line width=1pt] (12*\scale,0.5*\scale)--(12*\scale,-0.5*\scale);
     \draw[line width=1pt] (13*\scale,0.5*\scale)--(13*\scale,-0.5*\scale);
     \draw[line width=1pt] (0.5*\scale,0.25*\scale)--(0.5*\scale,-0.*\scale);
     \draw[line width=1pt] (1.5*\scale,0.25*\scale)--(1.5*\scale,-0.*\scale);
     \draw[line width=1pt] (2.5*\scale,0.25*\scale)--(2.5*\scale,-0.*\scale);
     \draw[line width=1pt] (3.5*\scale,0.25*\scale)--(3.5*\scale,-0.*\scale);
     \draw[line width=1pt] (4.5*\scale,0.25*\scale)--(4.5*\scale,-0.*\scale);
     \draw[line width=1pt] (5.5*\scale,0.25*\scale)--(5.5*\scale,-0.*\scale);
     \draw[line width=1pt] (6.5*\scale,0.25*\scale)--(6.5*\scale,-0.*\scale);
     \draw[line width=1pt] (7.5*\scale,0.25*\scale)--(7.5*\scale,-0.*\scale);
     \draw[line width=1pt] (8.5*\scale,0.25*\scale)--(8.5*\scale,-0.*\scale);
     \draw[line width=1pt] (9.5*\scale,0.25*\scale)--(9.5*\scale,-0.*\scale);
     \draw[line width=1pt] (10.5*\scale,0.25*\scale)--(10.5*\scale,-0.*\scale);
     \draw[line width=1pt] (11.5*\scale,0.25*\scale)--(11.5*\scale,-0.*\scale);
     \draw[line width=1pt] (12.5*\scale,0.25*\scale)--(12.5*\scale,-0.*\scale);
     \node at (0.5*\scale,-0.3*\scale) {\footnotesize{1}};
     \node at (1.5*\scale,-0.3*\scale) {\footnotesize{2}};
     \node at (2.5*\scale,-0.3*\scale) {\footnotesize{3}};
     \node at (3.5*\scale,-0.3*\scale) {\footnotesize{4}};
     \node at (4.5*\scale,-0.3*\scale) {\footnotesize{5}};
     \node at (5.5*\scale,-0.3*\scale) {\footnotesize{6}};
     \node at (6.5*\scale,-0.3*\scale) {\footnotesize{7}};
     \node at (7.5*\scale,-0.3*\scale) {\footnotesize{8}};
     \node at (8.5*\scale,-0.3*\scale) {\footnotesize{9}};
     \node at (9.5*\scale,-0.3*\scale) {\footnotesize{10}};
     \node at (10.5*\scale,-0.3*\scale) {\footnotesize{11}};
     \node at (11.5*\scale,-0.3*\scale) {\footnotesize{12}};
     \node at (12.5*\scale,-0.3*\scale) {\footnotesize{13}};
     \node at (0.5*\scale,-1*\scale) {FODO cells};
     \node [red] at (0.0*\scale,1.0*\scale) {SD1};
     \node [red] at (2.0*\scale,1.0*\scale) {SD1};
     \node [red] at (5.0*\scale,1.0*\scale) {SD2};
     \node [red] at (7.0*\scale,1.0*\scale) {SD2};
     \node [red] at (10.0*\scale,1.0*\scale) {SD1};
     \node [red] at (12.0*\scale,1.0*\scale) {SD1};
     \node [blue] at (2.5*\scale,-1.0*\scale) {SF1};
     \node [blue] at (4.5*\scale,-1.0*\scale) {SF1};
     \node [blue] at (7.5*\scale,-1.0*\scale) {SF2};
     \node [blue] at (9.5*\scale,-1.0*\scale) {SF2};
     \node [blue] at (12.5*\scale,-1.0*\scale) {SF1};
     \node [red] at (14.0*\scale,0.4*\scale) {$\varphi_y =90^\degree$};
     \node [blue] at (14.0*\scale,-0.4*\scale) {$\varphi_x =90^\degree$};
     \draw[dashed] (-0.5*\scale, 0.7*\scale)--(-0.5*\scale, 1.85*\scale);
     \draw[dashed] (2.5*\scale, 0.7*\scale)--(2.5*\scale, 1.85*\scale);
     \draw[dashed] (4.5*\scale, 0.7*\scale)--(4.5*\scale, 1.85*\scale);
     \draw[dashed] (7.5*\scale, 0.7*\scale)--(7.5*\scale, 1.85*\scale);
     \draw[dashed] (9.5*\scale, 0.7*\scale)--(9.5*\scale, 1.85*\scale);
     \draw[dashed] (12.5*\scale, 0.7*\scale)--(12.5*\scale, 1.85*\scale);
     \draw[dashed] (2.0*\scale,-0.7*\scale)--(2.0*\scale,-1.85*\scale);
     \draw[dashed] (5.0*\scale,-0.7*\scale)--(5.0*\scale,-1.85*\scale);
     \draw[dashed] (7.0*\scale,-0.7*\scale)--(7.0*\scale,-1.85*\scale);
     \draw[dashed] (10.0*\scale,-0.7*\scale)--(10.0*\scale,-1.85*\scale);
     \draw[dashed] (12.0*\scale,-0.7*\scale)--(12.0*\scale,-1.85*\scale);
     \node [] at (1.0*\scale,1.525*\scale) {1\textsuperscript{st} pair};
     \node [] at (6*\scale,1.525*\scale) {3\textsuperscript{rd} pair};
     \node [] at (11*\scale,1.525*\scale) {5\textsuperscript{th} pair};
     \node [] at (3.5*\scale,-1.525*\scale) {2\textsuperscript{nd} pair};
     \node [] at (8.5*\scale,-1.525*\scale) {4\textsuperscript{th} pair};
     \node[mark size=5pt,color=red] at (0,0) {\pgfuseplotmark{triangle*}};
     \node[mark size=5pt,color=red] at (2,0) {\pgfuseplotmark{triangle*}};
     \node[mark size=5pt,color=red] at (5,0) {\pgfuseplotmark{triangle*}};
     \node[mark size=5pt,color=red] at (7,0) {\pgfuseplotmark{triangle*}};
     \node[mark size=5pt,color=red] at (10,0) {\pgfuseplotmark{triangle*}};
     \node[mark size=5pt,color=red] at (12,0) {\pgfuseplotmark{triangle*}};

     \node[mark size=5pt,color=blue,rotate=180] at (2.5,0) {\pgfuseplotmark{triangle*}};
     \node[mark size=5pt,color=blue,rotate=180] at (4.5,0) {\pgfuseplotmark{triangle*}};
     \node[mark size=5pt,color=blue,rotate=180] at (7.5,0) {\pgfuseplotmark{triangle*}};
     \node[mark size=5pt,color=blue,rotate=180] at (9.5,0) {\pgfuseplotmark{triangle*}};
     \node[mark size=5pt,color=blue,rotate=180] at (12.5,0) {\pgfuseplotmark{triangle*}};
   \end{tikzpicture}
   \caption{Completely non-interleaved sextupole scheme for a FODO cell lattice with $\varphi=90^\degree$ phase advance in both planes. In order to optimise the cancellation of the sextupole's geometric effect, only linear elements are installed between two sextupoles forming a pair.}
   \label{fig:sextschemenoninterfull}
 \end{figure}
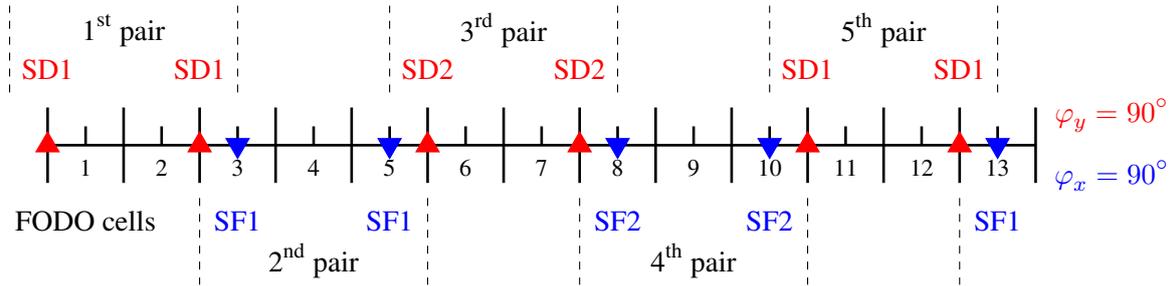

 \section{Dynamic, Momentum Aperture, and Frequency Map Analysis}
 \label{sec:da}
 Dynamic and momentum aperture are determined by performing single particle tracking for a certain number of turns. 
 If a particle survives the set number of turns without exceeding a set aperture (in our simulations at amplitudes of \SI{10}{\centi\meter} horizontally or vertically) the initial amplitudes of the individual particle are deemed stable initial conditions. 
 Thus, a border can be obtained for which particles with higher initial amplitudes are lost, and particles with smaller amplitudes will be stably stored in the machine. 
 Tracking was performed using PTC from within MAD-X for \num{2000} turns per particle \cite{madwebsite,madxusersguide}. 
 For the highest energy of \SI{182.5}{\giga\electronvolt}, this corresponds to \num{98} longitudinal damping times; for the lowest collider energy of \SI{45.6}{\giga\electronvolt} it corresponds to \num{0.64} longitudinal damping times.
 
For the study of frequency map analysis, we did not include radiation effects into the determination of the dynamic aperture. However, in the following section~\ref{sec:machimp}, where machine imperfections are studied, radiation effects are included showing that including radiation effects only increases the aperture. 
Frequency map analysis allows for identification of harmful resonances within the dynamic aperture. For each tracked particle, the diffusion rate 
 \begin{equation}
 d = \log_{10}\sqrt{\left(\nu_1^{(1)}-\nu_1^{(2)}\right)^2+\left(\nu_2^{(1)}-\nu_2^{(2)}\right)^2},
 \end{equation}
is calculated \cite{yannisNAFF}.
Here, $\nu_{1/2}^{(1 / 2)}$ are the tunes determined with high precision using numerical analysis of fundamental frequencies (NAFF) \cite{yannisNAFF} from the first / second half of the tracked turns. The fundamental frequency of motion is determined from the particles turn by turn position and then the diffusion is calculated based on the difference between the fundamental frequencies of first / second half of the tracked turns.
Diffusion rate is an indicator for the severity of a given resonance a particle is close to in tune space.

\subsection{Optimization of working point based on Diffusion Rate}
\begin{figure}[tbp]
   \subfloat[diffusion inside dynamic aperture]{\includegraphics[width=0.485\textwidth]{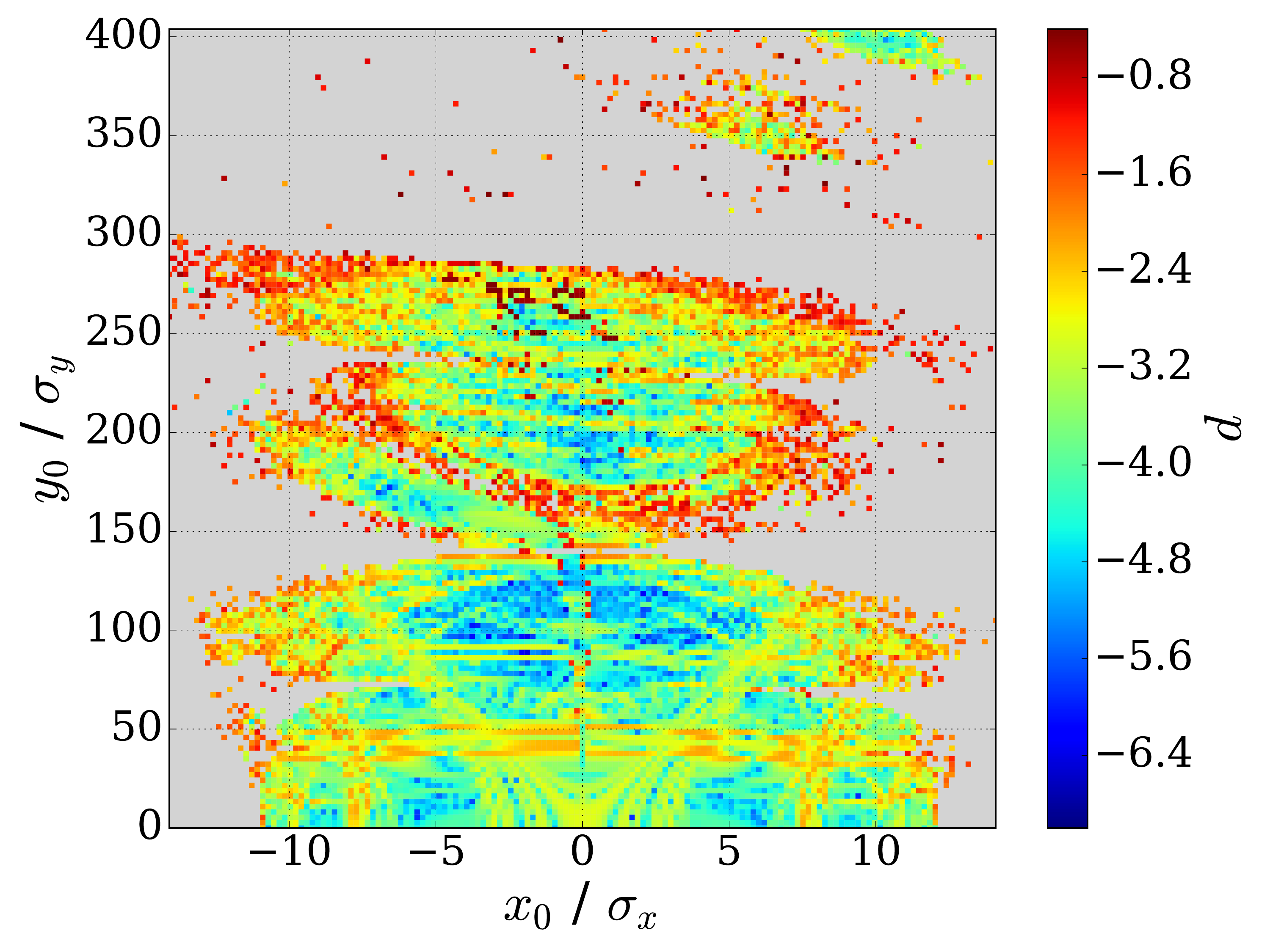}}
   \subfloat[diffusion as a function of working point]{\includegraphics[width=0.515\textwidth]{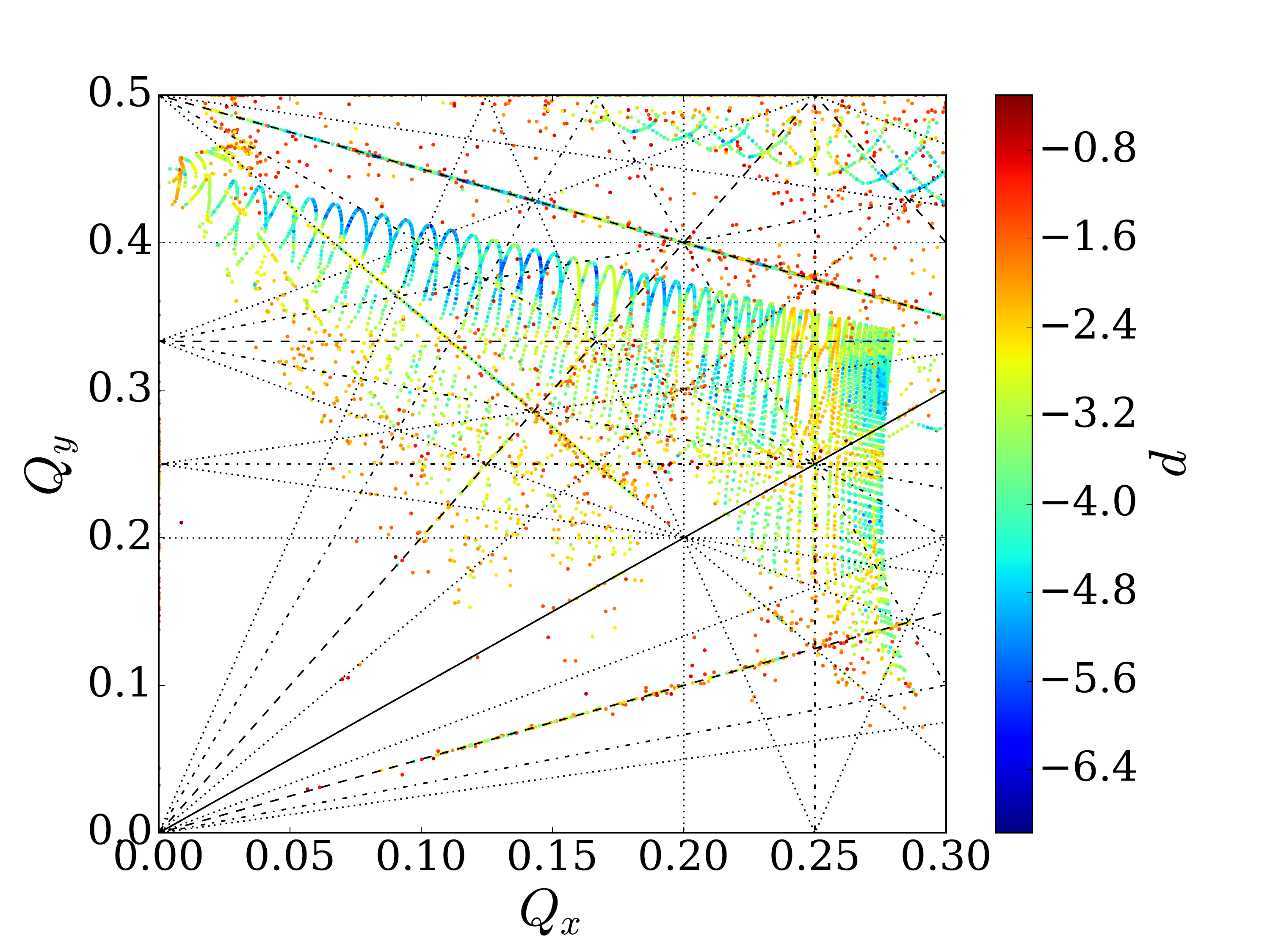}}
   \caption{Diffusion inside dynamic aperture (a) and as a function of working point (b) for interleaved sextupole scheme and \ang{90} / \ang{60} degrees phase advance.}
   \label{fig:DA9090Interleaved}
\end{figure}

As a first candidate, an interleaved sextupole scheme with \ang{90} phase advance in the horizontal plane and \ang{60} phase advance in the vertical plane was investigated towards its performance regarding dynamic aperture and frequency map.
As can be seen in Fig.~\ref{fig:DA9090Interleaved} (a), the dynamic aperture for this interleaved sextupole scheme is as large as $\sim\num{10}\,\sigma_x$ in the horizontal plane and $\sim\num{140}\,\sigma_y$ in the vertical plane (assuming \SI{1}{\percent} emittance coupling). 

If we compare the diffusion rate for on-axis particles in Fig.~ \ref{fig:DA9090Interleaved} (a) with the corresponding tune diagram in Fig.~ \ref{fig:DA9090Interleaved} (b), we see that the working point is not optimized since diffusion is high as indicated by colour. In operation this would lead to large vertical emittance for on-axis particles because the corresponding vertical tune is at $Q_y = \frac1{3}$ resonance.

\begin{figure}[tbp]
   \subfloat[diffusion as a function of working point]{\includegraphics[width=0.515\textwidth, height=6.4cm]{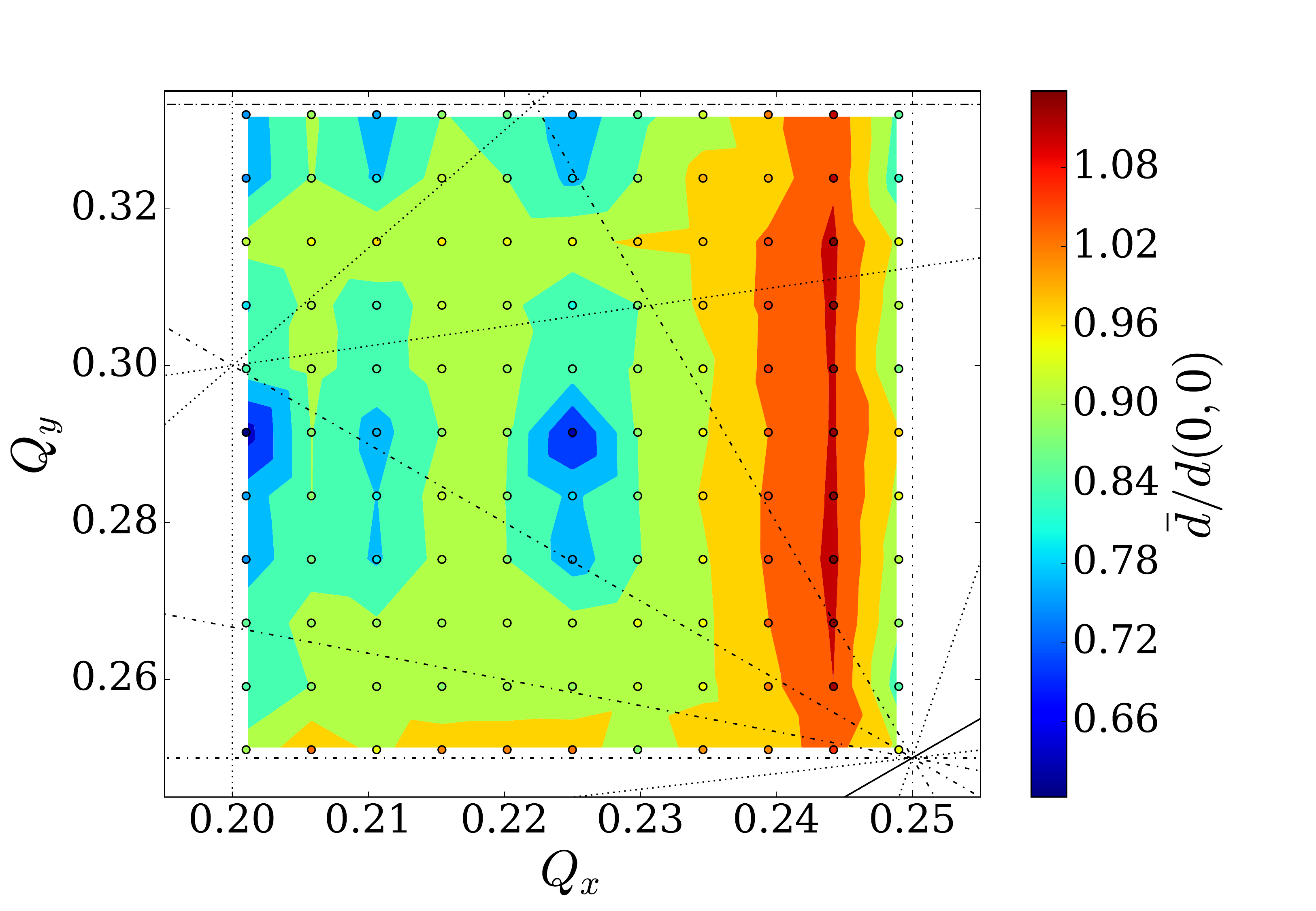}}
   \subfloat[diffusion inside dynamic aperture]{\includegraphics[width=0.485\textwidth]{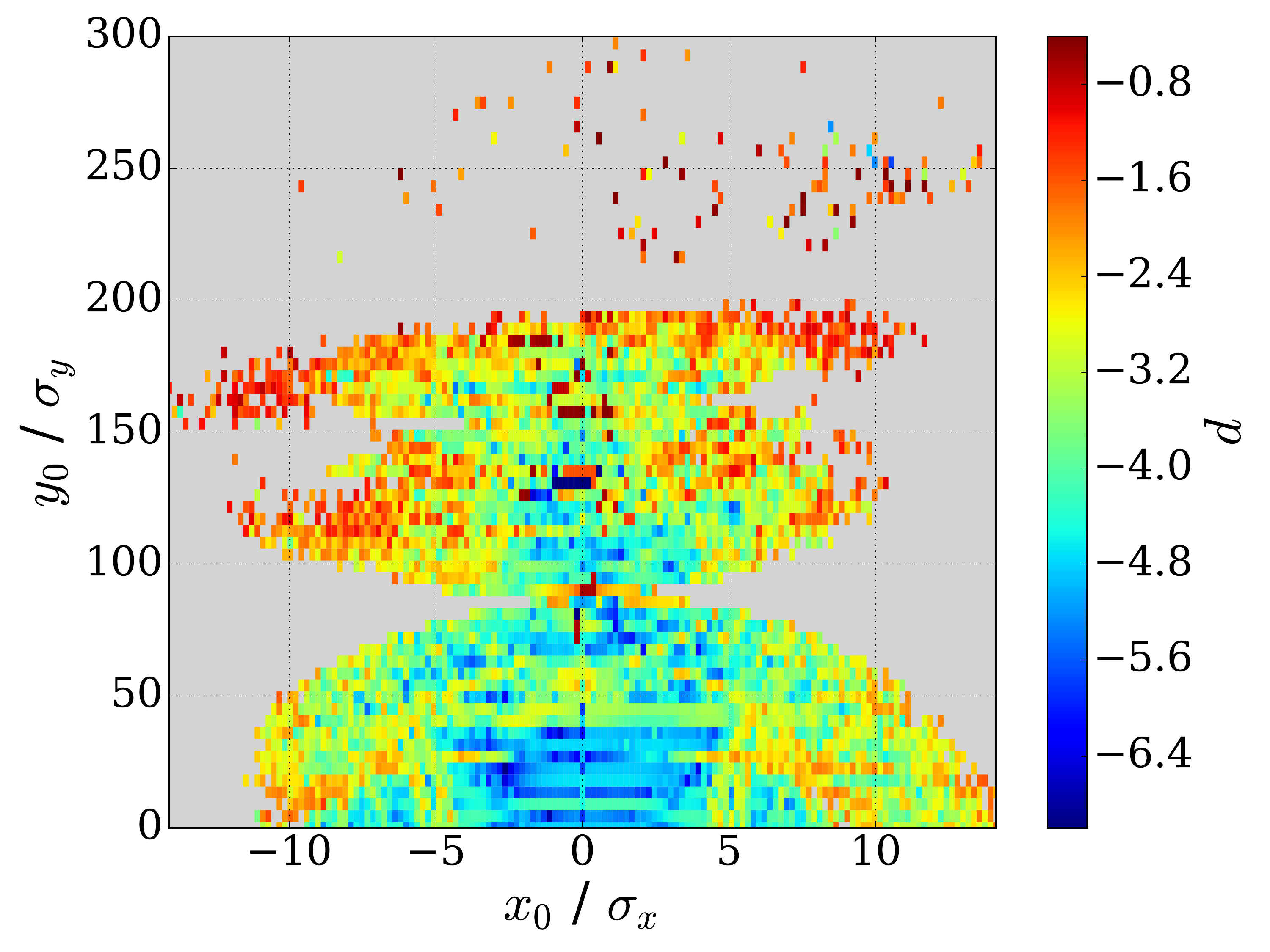}}
   \caption{Average diffusion as function of the initial working point for interleaved sextupole schema and \ang{90}/\ang{60} degrees phase advance (a) and diffusion inside dynamic aperture  for optimized working point (b).}
   \label{fig:matchTune}
\end{figure}

In this first version of the booster synchrotron, the working point was chosen to be $Q_x = 487.28$ / $Q_y = 327.34$ according to the tunes of the collider rings. 
For the booster this initial working point was not ideal and a scan of average diffusion rate as a function of working point was performed (comp. Fig.~\ref{fig:matchTune}~(a)).
As a result, the working point was chosen to be $Q_x = 487.225$ / $Q_y = 327.29$. This new working point yielded much less resonance interaction close to the reference orbit (comp. Fig.~\ref{fig:matchTune}~(b)). 
For all consecutive versions of the booster, the non integer part of the working point was kept at this optimized setting.

\subsection{Dynamic aperture for different optics and completely non-interleaved sextupole schemes}
The dynamic aperture has been investigated for all sextupole schemes and optics.
As expected the "completely" non-interleaved sextupole scheme provides the larges dynamic aperture, which will be presented in the following section.
 The phase advance of the arc FODO cells was chosen to be exactly 90\textsuperscript{\degree} or 60\textsuperscript{\degree}.
 Considering the arcs reach a length of more than \SI{16}{km} resonances might occur even within the arcs. If a different sextupole scheme might be wanted, a slight change of the phase advance per cell might help to avoid resonances.

In the following paragraphs, different versions regarding phase advance of a completely non-interleaved sextupole scheme are compared in terms of their dynamic aperture and resonance interaction. 

\begin{figure}[tbp]
   \subfloat[diffusion inside dynamic aperture]{\includegraphics[width=0.5\textwidth]{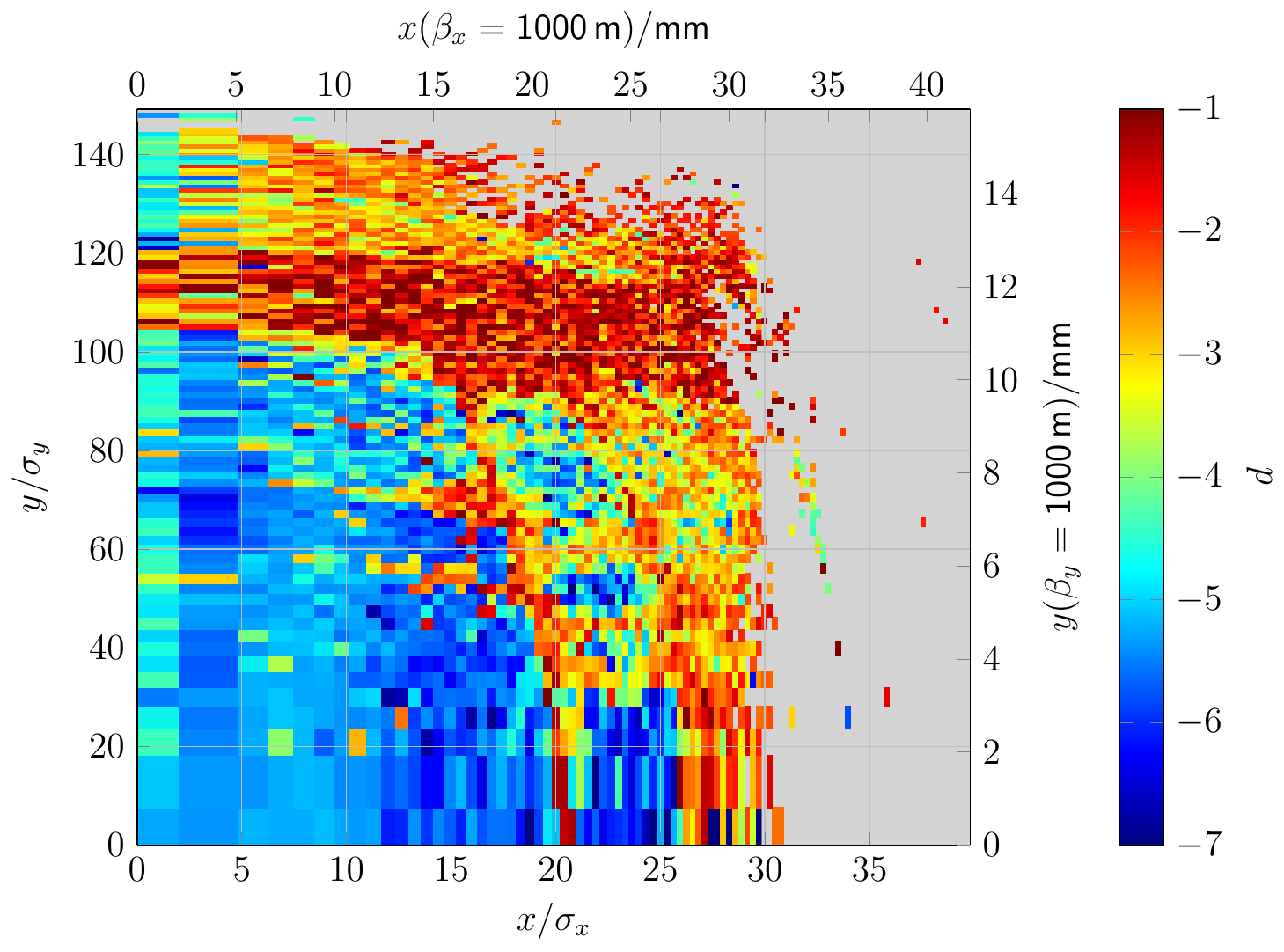}}
   \subfloat[diffusion as a function of working point]{\includegraphics[width=0.5\textwidth]{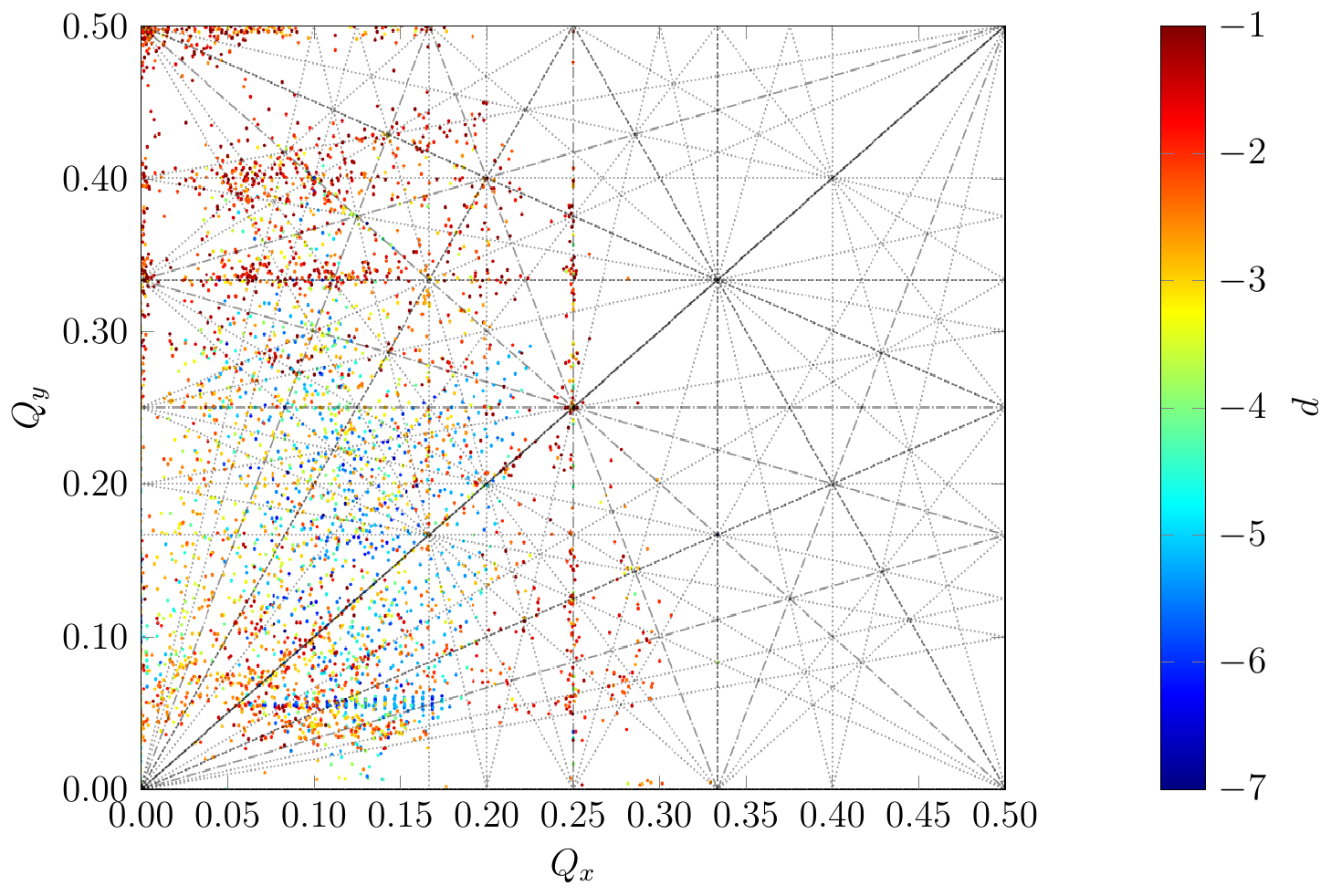}}
   \caption{Diffusion inside dynamic aperture (a) and as a function of working point (b) for completely non-interleaved sextupole schema and \ang{90} / \ang{60} degrees phase advance.}
   \label{fig:DA9060cni}
\end{figure}

The next candidate is for the same phase advances per plane as the previous versions, but the sextupole scheme has been altered from an interleaved to a non-interleaved scheme. 
The comparison of the resulting apertures in Fig.~\ref{fig:DA9060cni} with the previous one shows that dynamic aperture could be increased from \num{10}$\,\sigma_x$ to \num{30}$\,\sigma_x$ by introducing the non-interleaved sextupole scheme. 
Regarding tune footprint, this solution shows heavy resonance interaction for large apertures. Introducing machine imperfections will likely degrade aperture for this level of diffusion. 

\begin{figure}[tbp]
   \subfloat[diffusion inside dynamic aperture]{\includegraphics[width=0.5\textwidth]{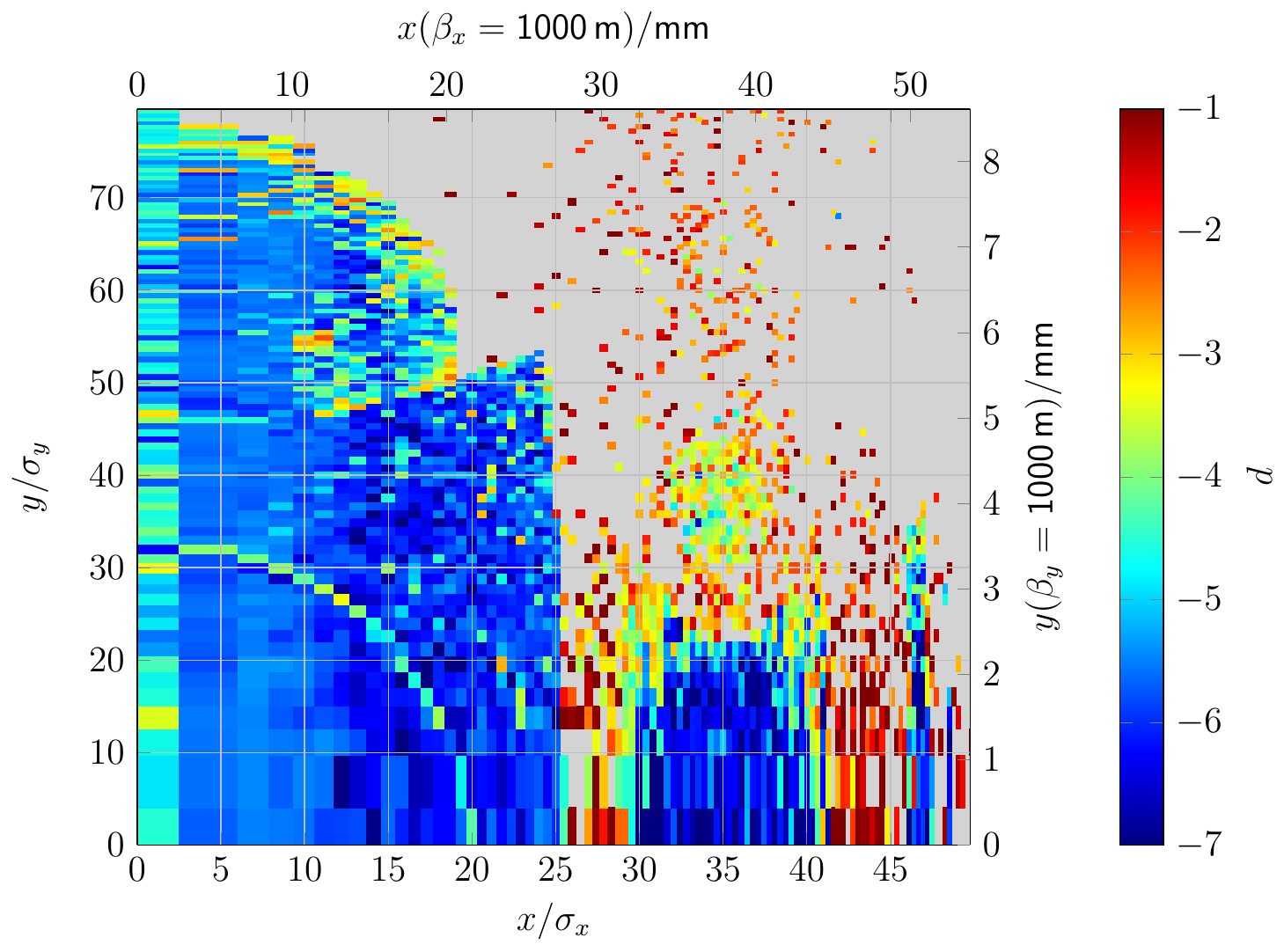}}
   \subfloat[diffusion as a function of working point]{\includegraphics[width=0.5\textwidth]{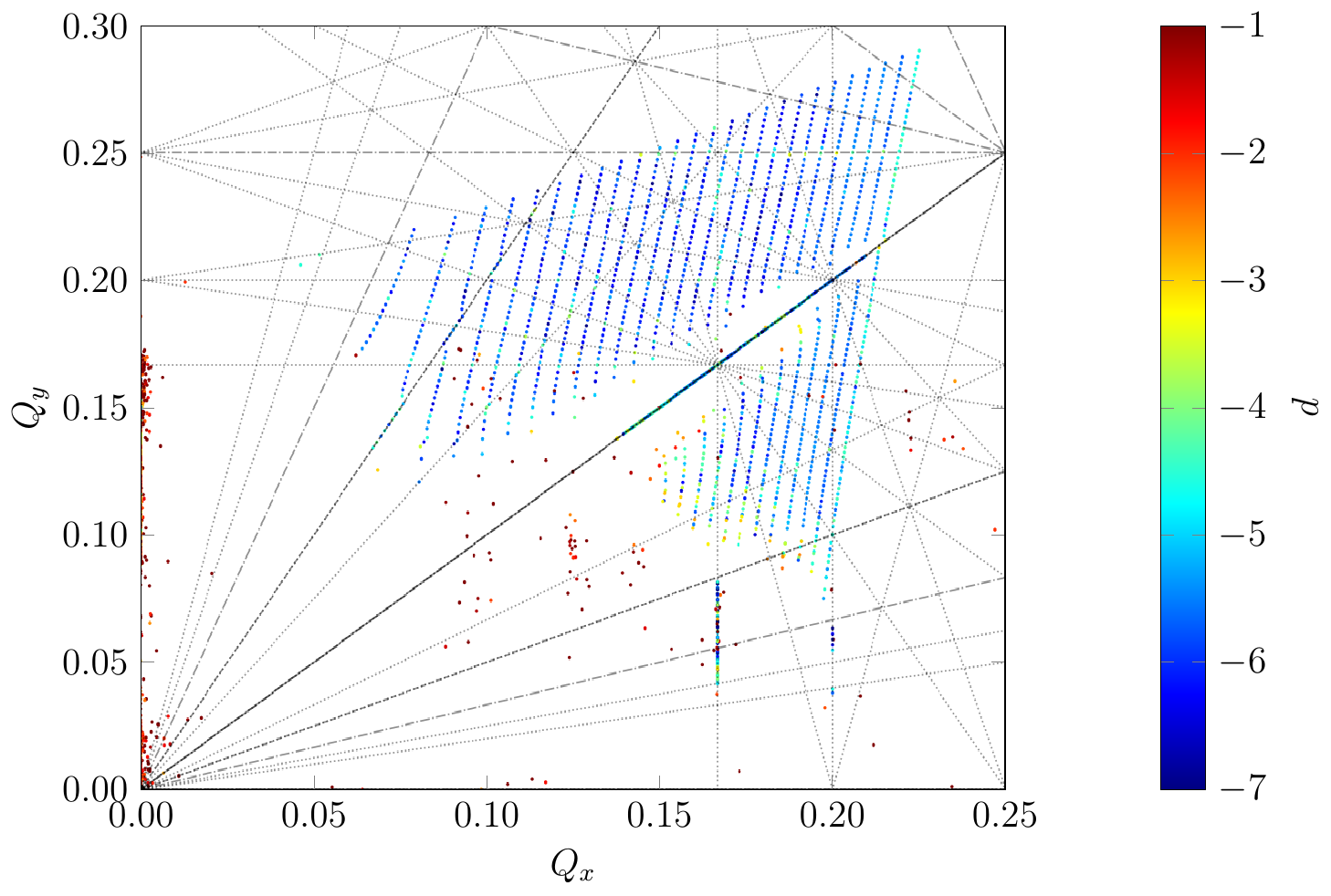}}
   \caption{Diffusion inside dynamic aperture (a) and as a function of working point (b) for completely non-interleaved sextupole schema and \ang{90} / \ang{90} degrees phase advance.}
   \label{fig:DA9090cni}
\end{figure}

Another candidate features a completely non-interleaved sextupole scheme with \ang{90} phase advance in each plane. 
The resulting aperture and tune footprint are presented in Fig.~\ref{fig:DA9090cni}. 
As the sextupole scheme is quite similar to the one of the collider, it also shows quite similar dynamic aperture and shape (comp. Fig. 1 in \cite{fcceeDAVancouver2018}).
It shows less diffusion rate within the aperture compared to previous candidates and is one of the candidates which are investigated for their performance including machine imperfections in Sec.~\ref{sec:machimp}.

\begin{figure}[tbp]
   \subfloat[diffusion inside dynamic aperture]{\includegraphics[width=0.5\textwidth]{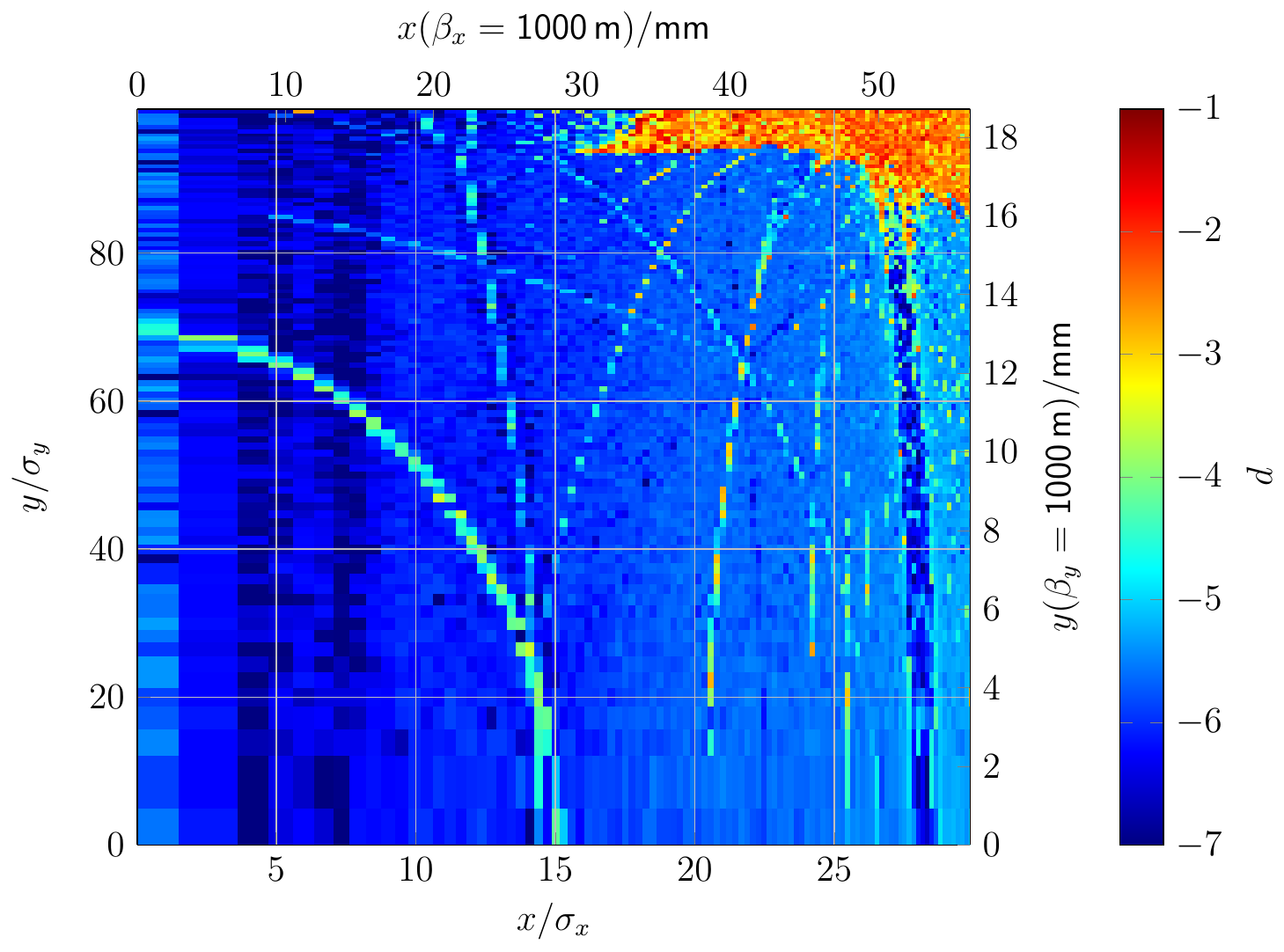}}
   \subfloat[diffusion as a function of working point]{\includegraphics[width=0.5\textwidth]{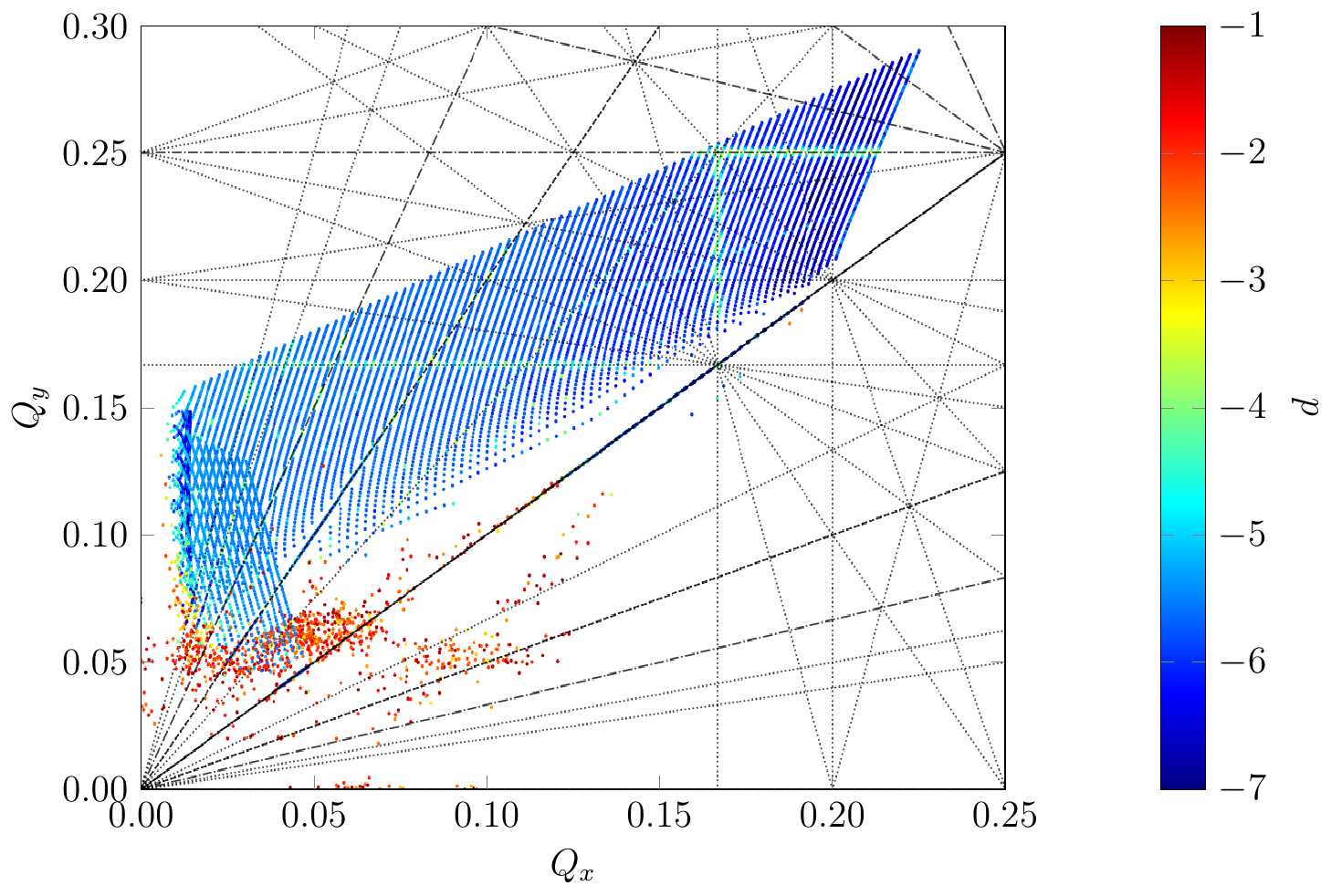}}
   \caption{Diffusion inside dynamic aperture (a) and as a function of working point (b) for completely non-interleaved sextupole schema and \ang{60} / \ang{60} degrees phase advance.}
   \label{fig:DA6060cni}
\end{figure}

Finally, another completely non-interleaved sextupole scheme is presented featuring \ang{60} phase advance in each plane.
The resulting aperture is comfortably large and the tune footprint shall be investigated further in the following:
Comparing Fig.~\ref{fig:DA6060cni} (a) with Fig.~\ref{fig:DA6060cni} (b), where the diffusion is presented as a function of working point, it is possible to identify resonance crossing due to tune shift with amplitude. 
The working point is properly chosen as the beam is free of resonances up to comparably high amplitudes of \num{15} horizontal and \num{70} vertical beam sizes. The first resonance showing an effect on the diffusion rate is the vertical quadrupole resonance at $Q_y=0.25$, represented as a circular line around the origin in Fig.~\ref{fig:DA6060cni} (a). 
Next but fainter is the horizontal sextupole resonance at $Q_x=0.\bar{16}$ which joins the vertical quadrupole resonance for $y=0$ and $x=15\,\sigma_x$ in Fig.~\ref{fig:DA6060cni} (a).
The tune footprint avoids the coupling resonance $Q_y=Q_x$ for vertical amplitudes below \num{100} vertical beam sizes and hits the integer resonance $Q_x=0.0$ for reasonably high horizontal amplitudes of \num{28} horizontal beam sizes.

\begin{figure}[tbp]
 \centering
 \includegraphics[width=0.6\textwidth]{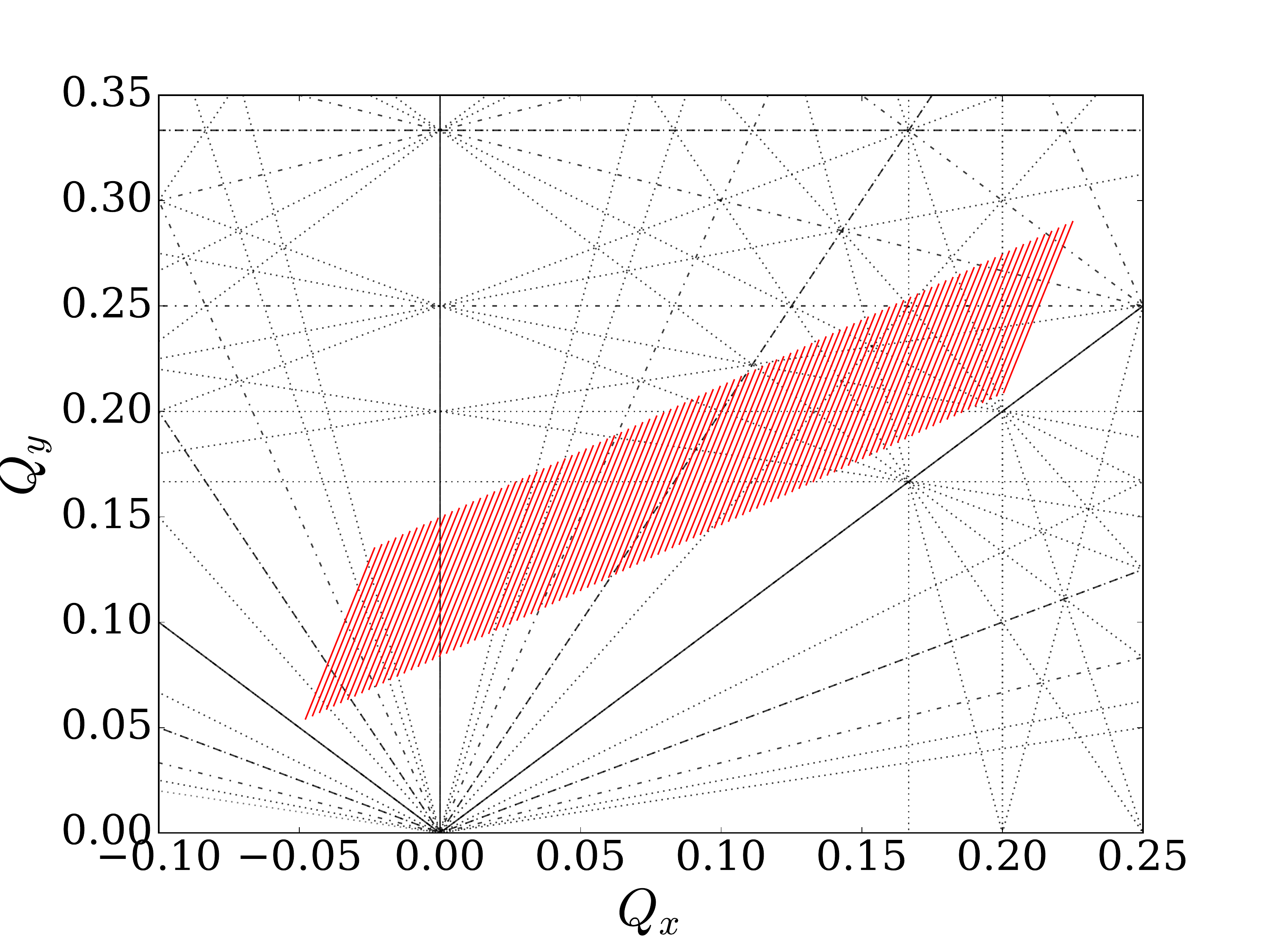}
 \caption{Calculated tuneshift with amplitude inside dynamic aperture for \ang{60} / \ang{60} phase advance and completely non interleaved sextupole scheme.}
 \label{fig:tune_cni33_6060_DA}
\end{figure}

\subsubsection{Tuneshift with amplitude}
The tracking has been performed inside an PTC environment within MAD-X.
The \texttt{ptc\_normal} command allows to estimate the tuneshift with amplitude. 
Since the resulting tune footprint of the completely non-interleaved sextupole scheme with \ang{60} phase advance in each plane is very clean, it is worthwhile to investigate the theoretical predictions made by \texttt{ptc\_normal}. 
 
First, one needs to calculate the action corresponding to the amplitude of concern. In our case, these are the amplitudes in the dynamic aperture. Action is defined as
\begin{equation}
2\cdot I_{x} = \gamma_{x}\cdot x^2+\beta_{x}\cdot xx'+\alpha_x x'^2.
\end{equation}
For tracking the dynamic aperture, we start particles with initial horizontal and vertical amplitude, but leave the angles equal to zero. Therefore, the calculation of the initial action becomes fairly simple:
\begin{equation}
2\cdot I_{x,\text{track}} = \gamma_x\cdot x^2,
\end{equation}
with $\gamma_x=\frac{1+\alpha_x^2}{\beta_x}$. The single particle emittance is now easily derived by multiplying action $I$ by \num{2}:
\begin{equation}
\varepsilon_{x,\texttt{track}}=2\cdot I_{x,\texttt{track}}=\gamma_x\cdot x_0^2.
\end{equation}

The results from \texttt{ptc\_normal} must now be translated to tune using a Taylor expansion:
\begin{align}
Q_x(\varepsilon_x,\varepsilon_y) =\, & Q_{x,0}+A_{100}\cdot \varepsilon_x+A_{010}\cdot \varepsilon_y \nonumber\\
&+\frac1{2}\cdot\Big[A_{200}\cdot \varepsilon_x^2+2\cdot A_{110}\cdot \varepsilon_x\varepsilon_y+A_{020}\cdot \varepsilon_y^2\Big]+\mathcal{O}(3).
\end{align}

The resulting tune footprint of this theoretical calculation is presented in Fig.~\ref{fig:tune_cni33_6060_DA} and can be compared with the tracking result in Fig.~\ref{fig:DA6060cni} (b). 
The results show good agreement apart from tunes below the integer resonance, which are displayed above integer resonance in Fig.~\ref{fig:DA6060cni} (this is an artefact of the frequency map analysis). 
This result shows how \texttt{ptc\_normal} can be used to determine tune footprint in order to optimize it in an iterative and more time effective manner  compared to time consuming tracking.

\section{Transverse misalignments of quadrupoles}\label{sec:machimp}
Simulations of the collider rings including random transverse and angular misalignments of the lattice have shown, that the beam is very sensitive to these misalignments and stable simulations need a sophisticated correction scheme \cite{Aumon:IPAC2016-THPOR001,Charles:IPAC2019-MOPRB001}.
The correction scheme developed for the collider rings requires "ramping up" the misalignments in 20 steps, coupling correction of the linear lattice using resonance driving terms, re-matching of the optics, dispersion free steering, ramping up of the sextupole strength, beta beat correction orbit correction, optics and tune re-matching before the emittance even can be calculated \cite{Sandra:LER}.
In addition, in order to obtain the very small vertical emittances a small coupling ratio of one per mille has to be achieved.
With this conditions in mind, the top-up booster synchrotron has been investigated towards its sensitivity to misaligned elements starting with the quadrupole magnets.

Transverse misaligments have been assigned to all quadrupole magnets of the lattice.
The alignment errors are randomly selected following a Gaussian distribution that was truncated at three sigma.
Different magnitudes of misalignments have been investigated: $\sigma=\SI{10}{\micro\metre}$, \SI{50}{\micro\metre}, \SI{100}{\micro\metre}, and \SI{150}{\micro\metre}.
For each case 100 seeds have been setup for each operation mode and both optics and completely non-interleaved sextupole scheme.
The orbit has been corrected with the MAD-X build-in optimizers MICADO and SVD in three consecutive runs.
Afterwards radiation effects have been activated and the tunes and chromaticity have been re-matched before the emittances have been calculated using the EMIT commend.

The values of the horizontal equilibrium emittance of the ideal lattice are compared to the average values of the 100 error seeds for each operation mode and misalignment magnitude in \tab{tab:emittancesbooster-misalign}. 
The emittances grow with beam energy as expected.
For beam energies up to \SI{80}{GeV} the misaligned quadrupoles do not affect the equilibrium emittance at all.
At higher energies it is difficult to account for a tendency: In case of the \ang{60}/\ang{60} optics the emittances seem to slightly decrease with larger misalignments, while for \SI{175}{GeV} beams and the \ang{90}/\ang{90} optics the emittance increases.

Tab.\,\ref{tab:emittancesbooster-xy} summarizes horizontal and vertical emittances and the emittance ratio for the $\sigma=\SI{100}{\micro\metre}$ case as an example.
The vertical emittances are about five orders of magnitude smaller than the horizontal emittances.
This means, the coupling between the transverse planes is two orders of magnitude below the goal of one per mille.
However, no angular misalignments and misalignments of sextupoles have been included yet, which also might add a significant contribution to the budget.

It is remarkable, that the emittance ratio is about a factor of 5-7 larger for the \ang{90}/\ang{90} optics.
Furthermore, there is an unexpected step for the \SI{120}{GeV} operation mode and the \ang{60}/\ang{60} optics, where the emittance ration reaches its larges value for all beam energies.
This step has also been observed for the $\sigma=\SI{10}{\micro\metre}$, \SI{50}{\micro\metre}, and \SI{150}{\micro\metre} cases.
As this step occurs only for \SI{120}{GeV} beam energy, this suggests an impact of the RF system. In order to explain this behaviour in detail, the RF settings should be carefully checked.

 \begin{table}[tbp]
 \caption{Equilibrium emittances of the top-up booster synchrotron for both the 60\textdegree/60\textdegree~optics and 90\textdegree/90\textdegree~optics in all operation modes and for different magnitudes of transverse misalignment of quadrupoles. The emittances of the ideal lattice are compared to the cases with randomly Gaussian distributed misalignments truncated at three $\sigma$ with $\sigma=\SI{10}{\micro\metre}$,\SI{50}{\micro\metre},\SI{100}{\micro\metre}, and \SI{150}{\micro\metre}.}
 \begin{center}
 \begin{tabular}{lS|SSSSS@{}}
     \toprule
         Optics & {Beam energy} & {ideal} 
         & \SI{10}{\micro\metre}   & \SI{50}{\micro\metre} & \SI{100}{\micro\metre}& \SI{150}{\micro\metre} \\
         & {(GeV)} & {(nm\,rad)}   & {(nm\,rad)} & {(nm\,rad)} & {(nm\,rad)} & {(nm\,rad)} \\
     \midrule
        {\ang{60}/\ang{60}}  & 20.0	 	& 0.045 &  0.045 &  0.045 & 0.045 & 0.045  \\
                             & 45.6 	& 0.230 &  0.230 &  0.230 & 0.230 & 0.230  \\
                             & 80.0	 	& 0.728 &  0.728 &  0.728 & 0.728 & 0.728  \\
                             & 120.0 	& 1.959 &  1.959 &  1.956 & 1.955 & 1.940  \\
                             & 175.0 	& 3.642 &  3.642 &  3.642 & 3.641 & 3.640 \\
     \midrule
        {\ang{90}/\ang{90}}  & 20.0	 	& 0.015 &  0.015 &  0.015 & 0.015 & 0.015  \\
                             & 45.6 	& 0.078 &  0.077 &  0.077 & 0.077 & 0.077  \\
                             & 80.0	 	& 0.242 &  0.242 &  0.242 & 0.242 & 0.242  \\
                             & 120.0 	& 0.547 &  0.547 &  0.546 & 0.546 & 0.546  \\
                             & 175.0 	& 1.188 &  1.193 &  1.190 & 1.195 & 1.201  \\
     \bottomrule
 \end{tabular}
 \end{center}
 \label{tab:emittancesbooster-misalign}
 \end{table}
 \begin{table}[tbp]
 \caption{Horizontal and vertical emittances and emittance ratio for all operation modes and both the 60\textdegree/60\textdegree~optics and 90\textdegree/90\textdegree~optics including transverse quadrupole misalignments of $\sigma=\SI{100}{\micro\metre}$.}
 \begin{center}
 \begin{tabular}{lS|SSSSS@{}}
     \toprule
         Optics & {Beam energy} & $\epsilon_x$ & $\epsilon_y$   & $\epsilon_y/\epsilon_x$ \\
         & {(GeV)} & {(nm\,rad)}   & {($10^{-6}$\,nm\,rad)} & \num{1e-6} \\
     \midrule
        {\ang{60}/\ang{60}}  & 20.0	 	& 0.045 &  0.2 &  \num{4.49} \\
                             & 45.6 	& 0.230 &  1.0 &  \num{4.54} \\
                             & 80.0	 	& 0.728 &  3.5 &  \num{4.82} \\
                             & 120.0 	& 1.955 &  12.2 &  \num{6.25} \\
                             & 175.0 	& 3.641 &  18.8 &  \num{5.16} \\
     \midrule
        {\ang{90}/\ang{90}}  & 20.0	 	& 0.015 &  0.5 &  \num{31.44} \\
                             & 45.6 	& 0.077 &  2.4 &  \num{31.56} \\
                             & 80.0	 	& 0.242 &  7.8 &  \num{32.09} \\
                             & 120.0 	& 0.546 &  18.2 &  \num{33.26} \\
                             & 175.0 	& 1.195 &  43.0 &  \num{35.97} \\
     \bottomrule
 \end{tabular}
 \end{center}
 \label{tab:emittancesbooster-xy}
 \end{table}

\begin{figure}[p]
     \centering
    \includegraphics[width=0.75\textwidth]{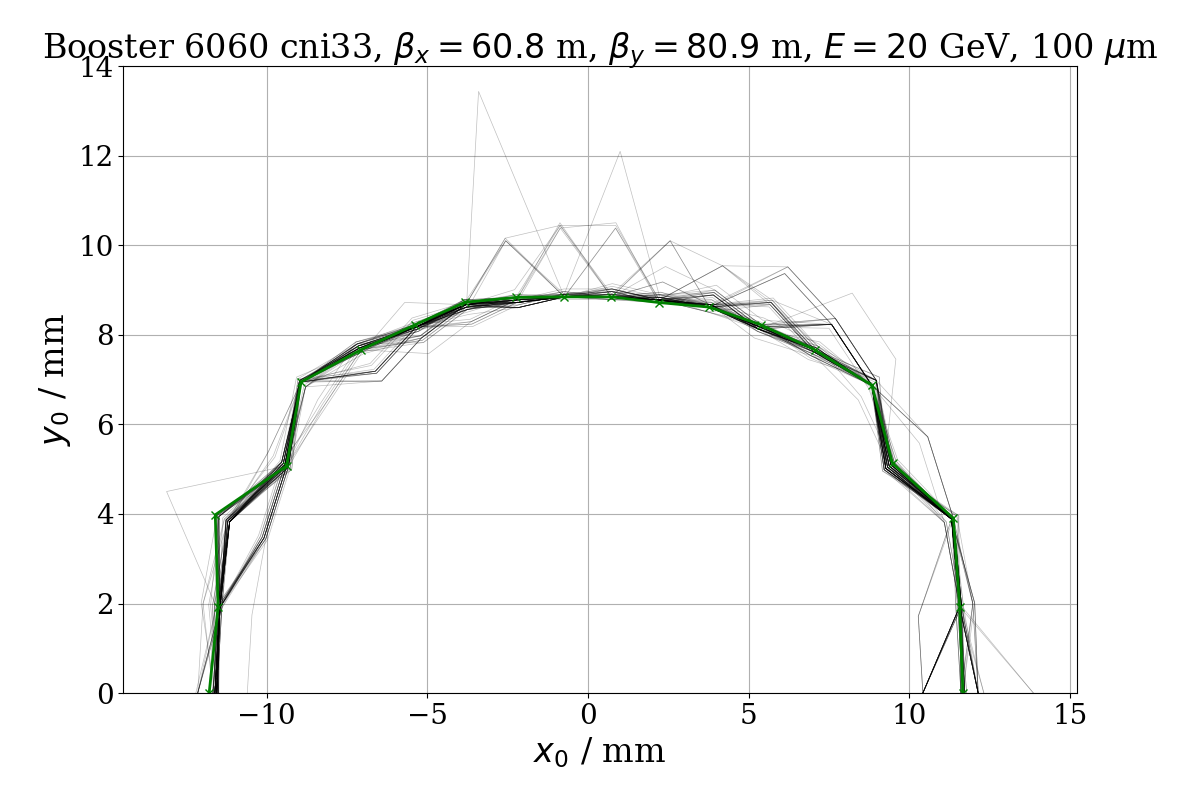}
    \caption{Dynamic aperture of the top-up booster synchrotron with \ang{60}/\ang{60} optics and completely non-interleaved sextupole scheme at \SI{20}{GeV} injection energy. The green line represents the ideal lattice, the grey lines the lattices with misaligned quadrupoles for 100 error seeds and $\sigma=$\SI{100}{\micro\meter}.}
     \label{fig:quadMisalignment100DA6060_mm}
 \end{figure}

\begin{figure}[p]
     \centering
    \includegraphics[width=0.75\textwidth]{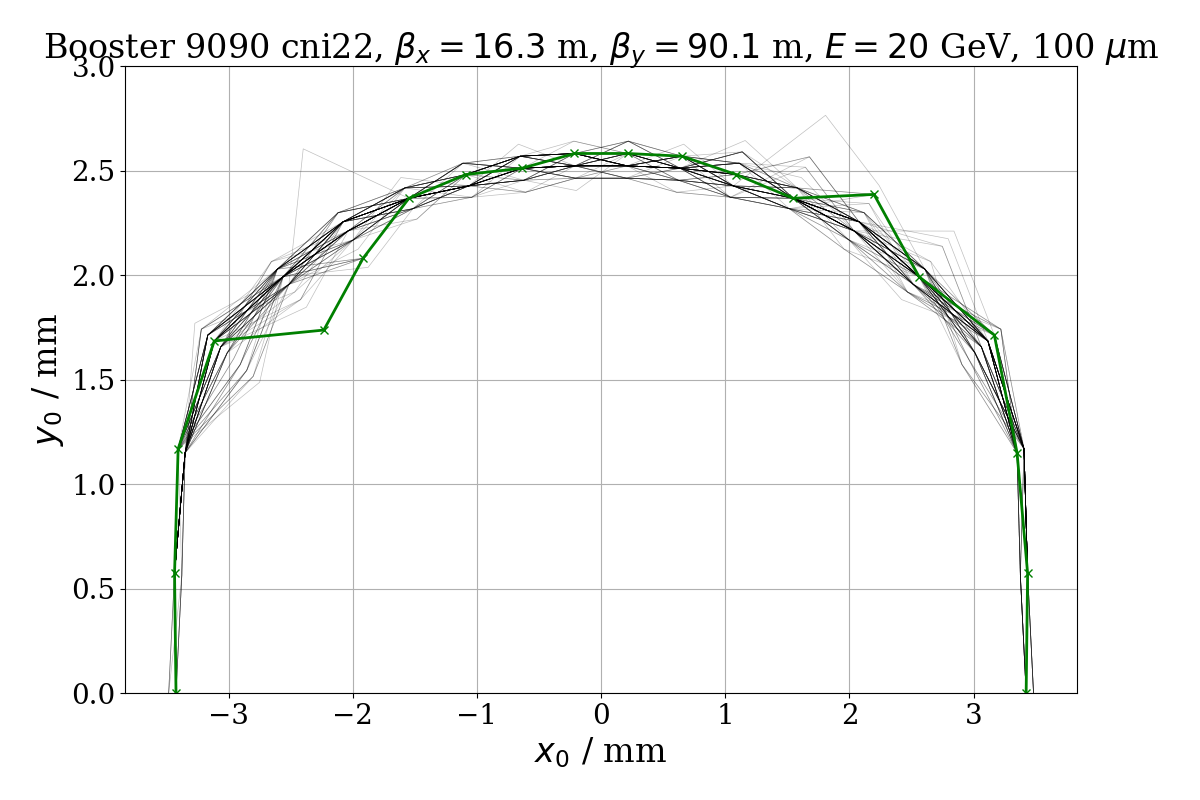}
    \caption{Dynamic aperture of the top-up booster synchrotron with \ang{90}/\ang{90} optics and completely non-interleaved sextupole scheme at \SI{20}{GeV} injection energy. The green line represents the ideal lattice, the grey lines the lattices with misaligned quadrupoles for 100 error seeds and $\sigma=$\SI{100}{\micro\meter}.}
     \label{fig:quadMisalignment100DA9090_mm}
 \end{figure}

As a next step, the effect of alignment errors on the dynamic aperture at injection energy has been investigated with 6D particle tracking including the effects of radiation damping and quantum excitation.
The particles were tracked using the MAD-X PTC module and the dynamic aperture was obtained by survival of the particles after 500 turns.
For both the \ang{60}/\ang{60} and the \ang{90}/\ang{90} optics the dynamic aperture has been measured for the ideal case and the same 100 lattices with misaligned quadrupoles as have been used for the emittance study.
The dynamic apertures of the $\sigma=\SI{100}{\micro\metre}$ case are shown in \fig{fig:quadMisalignment100DA6060_mm} for the \ang{60}/\ang{60} optics and in \fig{fig:quadMisalignment100DA9090_mm} for the \ang{90}/\ang{90} optics.
The green line represents the dynamic aperture of the ideal lattice, the grey lines belong to the misaligned lattices. 
The dynamic aperture of the ideal \ang{60}/\ang{60} lattice is about \SI{\pm12}{mm} in the horizontal plane and about \SI{9}{mm} in the vertical plane.
For the ideal \ang{90}/\ang{90} lattice the dynamic aperture is smaller than in the \ang{60}/\ang{60} lattice as expected from previous tracking campaigns and is about \SI{\pm4}{mm} in the horizontal plane and \SI{2.5}{mm} in the vertical plane.
For both optics the misalignments have no significant effect on the size of the dynamic aperture. 
The misalignments create some fluctuations at the borderline. Especially for the \ang{90}/\ang{90} optics they even seem to correct for the little dent at $x=\SI{-2.5}{mm}$.

\begin{figure}
     \centering
     \subfloat{\includegraphics[width=0.5\textwidth]{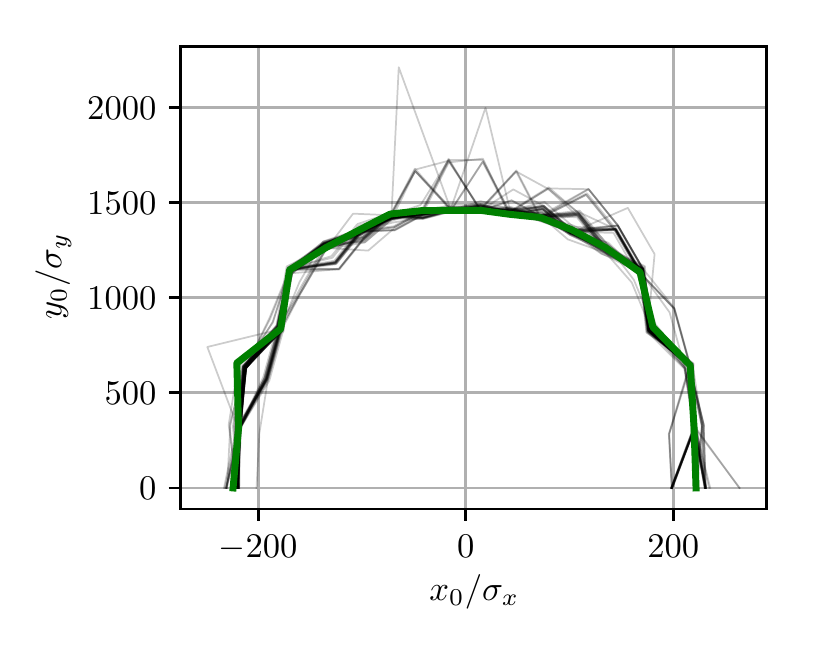}}
     \subfloat{\includegraphics[width=0.5\textwidth]{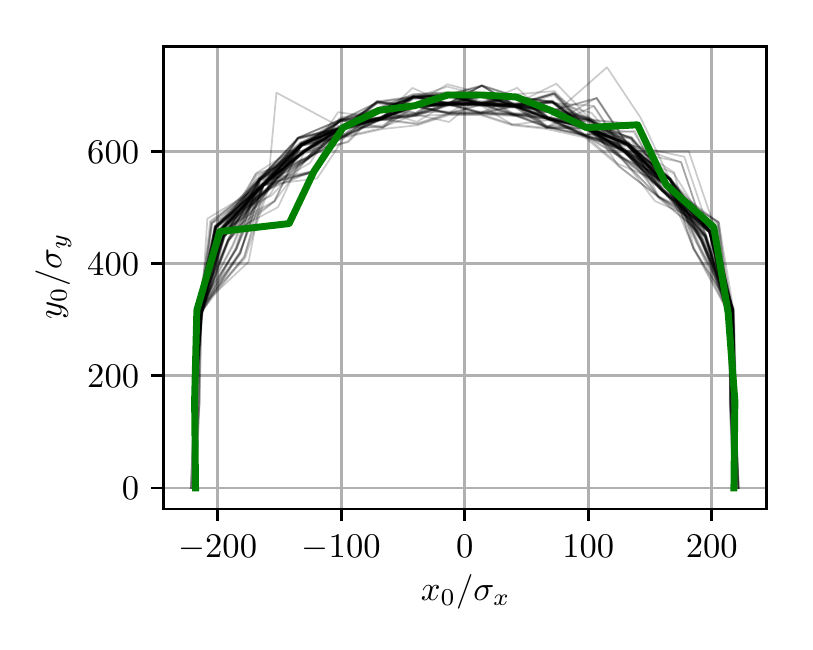}}
     \caption{Dynamic aperture in units of the beam size $\sigma$ for comparison to \fig{fig:quadMisalignment100DA6060_mm} and \fig{fig:quadMisalignment100DA9090_mm}. (a) shows the lattics with \ang{60}/\ang{60} optics, (b) the lattice with \ang{90}/\ang{90} optics. In both cases the misalignments were of $\sigma=$\SI{100}{\micro\meter}. The green line represents the ideal lattice, the grey lines the lattices with misaligned quadrupoles.}
     \label{fig:quadMisalignment100DA}
 \end{figure}

The reference point of the measurements were the begin of the straight section around point A.
At this reference point the betafunctions were $\beta_x=\SI{60.8}{m}$ and $\beta_y=\SI{80.9}{m}$ with the \ang{60}/\ang{60} optics and $\beta_x=\SI{16.3}{m}$ and $\beta_y=\SI{90.1}{m}$ with the \ang{90}/\ang{90} optics.
As dispersion is Zero in the straight sections, betafunctions and emittance can be used to convert the values of the dynamic aperture into units of the beam size $\sigma$.
The resulting plots with converted axes are given in \fig{fig:quadMisalignment100DA} (a) an (b) for comparison to \fig{fig:quadMisalignment100DA6060_mm} and \fig{fig:quadMisalignment100DA9090_mm}. 
Even though the absolute value of the horizontal DA was three times smaller for the \ang{90}/\ang{90} optics, its size is very similar in units of the beam size $\sigma$: 
about 200\,$\sigma$ in each case.
This is a consequence of the smaller values of emittance and betafunction.
In the vertical plane, the dynamic aperture is smaller in case of the \ang{90}/\ang{90} optics, but still spans comfortable 700\,$\sigma$.

\begin{figure}
    \centering
    \subfloat[\ang{60} / \ang{60} \SI{100}{\micro\meter} quadrupole misalignment]{\includegraphics[width=0.5\textwidth]{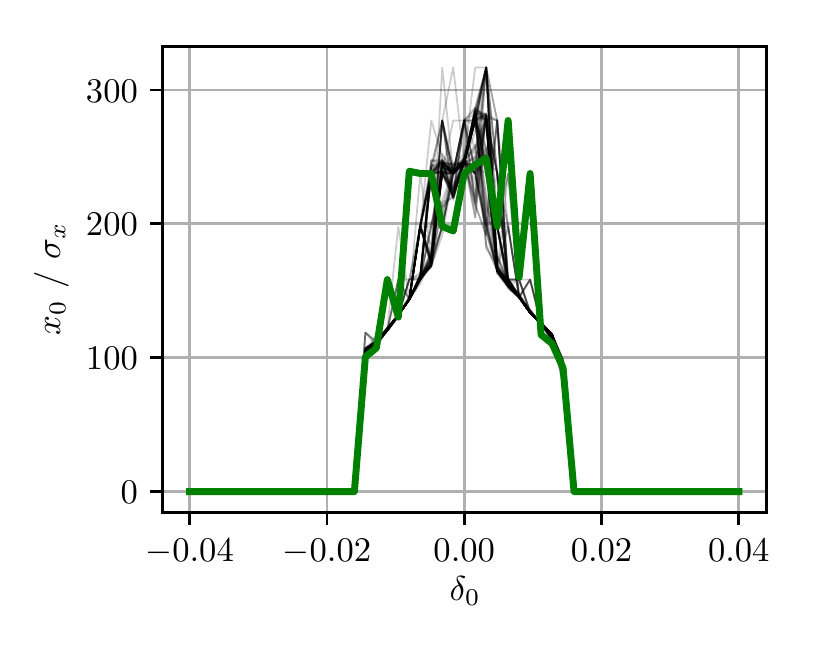}}
    \subfloat[\ang{90} / \ang{90} \SI{100}{\micro\meter} quadrupole misalignment]{\includegraphics[width=0.5\textwidth]{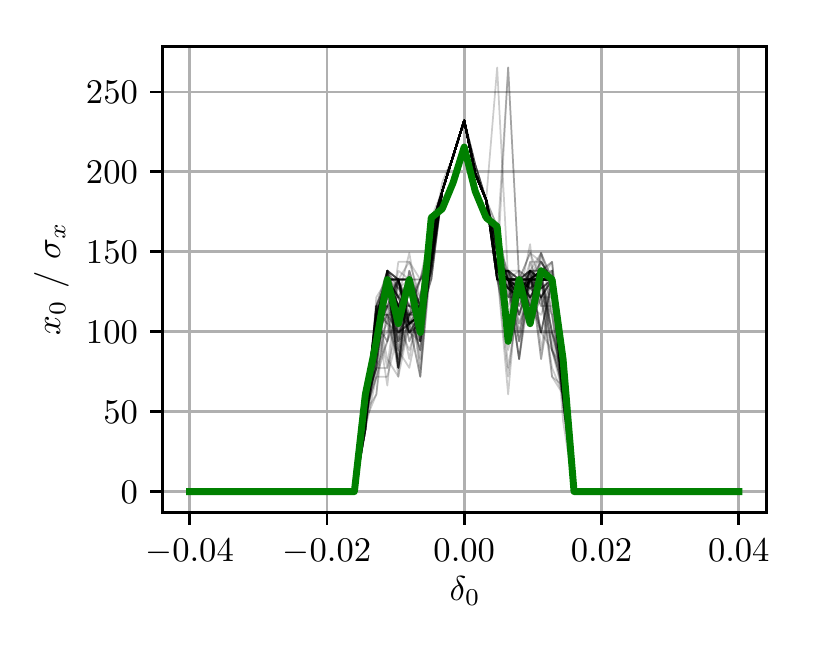}}
    \caption{Momentum aperture for 100 error seeds with quadrupole misalignments of amplitude \SI{100}{\micro\meter} for (a) \ang{60}/\ang{60} degrees phase advance and (b) \ang{90}/\ang{90} degrees phase advance and completely non-interleaved sextupole scheme. The green line represents the ideal lattice.}
    \label{fig:quadMisalignment100MA}
\end{figure}

\begin{figure}
    \centering
    \includegraphics[width=0.65\textwidth]{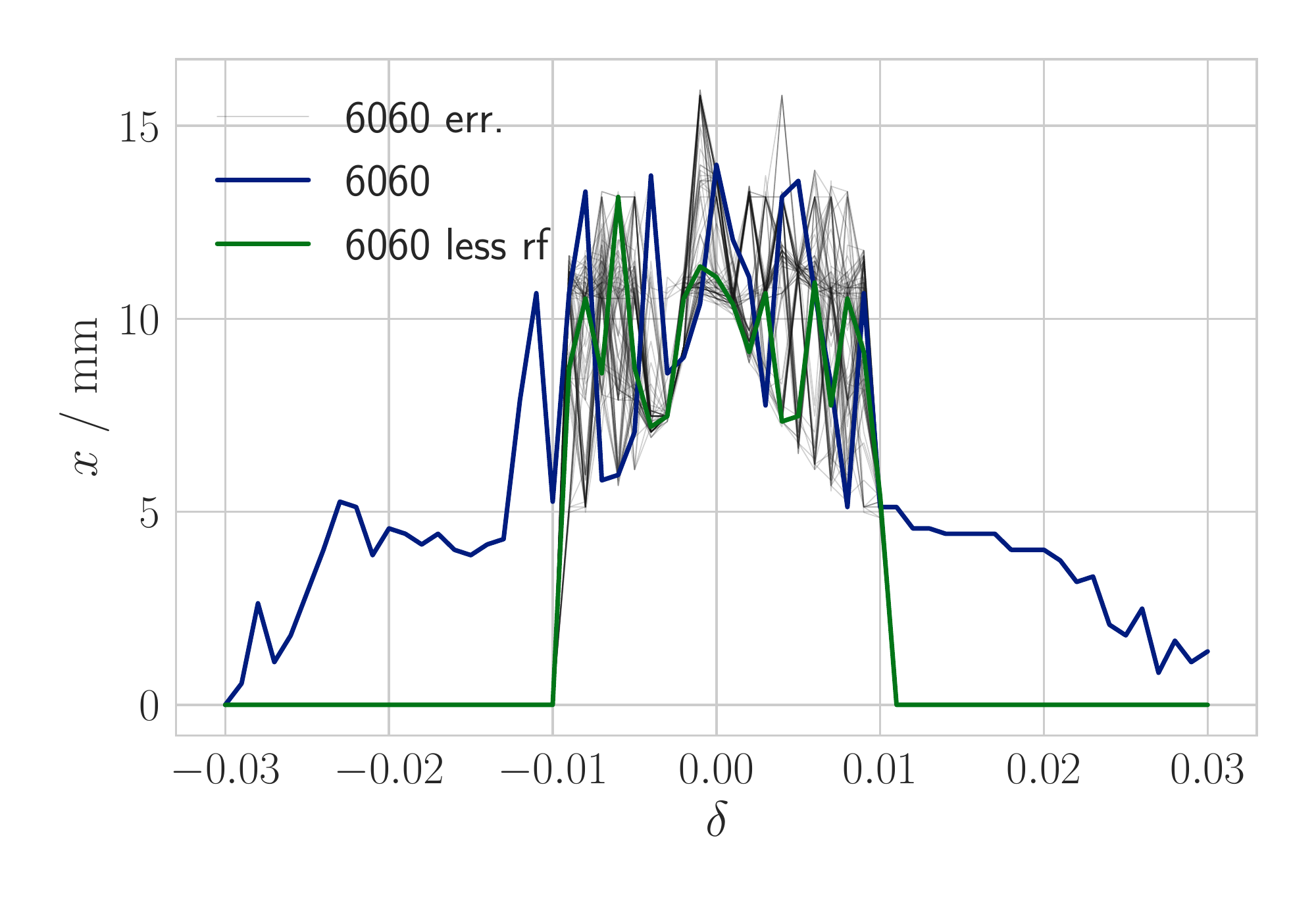}
    \caption{Momentum aperture for 100 error seeds with quadrupole misalignments of amplitude \SI{150}{\micro\meter} for \ang{60}/\ang{60} optics and completely non-interleaved sextupole scheme. The green line represents the ideal lattice with nominal RF voltage, the blue line the ideal lattice with increased voltage.}
    \label{fig:quadMisalignment100MA_moreRF}
\end{figure}

Apart from the dynamic aperture, the momentum acceptance has been measured as well.
Both optics allow particles with relative energy offsets up to $\delta=\pm\SI{1.5}{\percent}$ as shown in \fig{fig:quadMisalignment100MA} (a) and (b).
Again, the green line represents the ideal lattice and the grey lines the lattices with misaligned quadrupoles.
The misalignments do not decrease the momentum acceptance.
The steep edges however indicate, that the momentum acceptance during this tracking campaign is not limited by the lattice and the misalignments but rather by insufficiently strong RF voltage.
A preliminary run with increased RF voltage has been undertaken for the ideal lattice with the \ang{60}/\ang{60} optics. 
The result is presented in \fig{fig:quadMisalignment100MA_moreRF}, where the blue line 
represents the momentum acceptance with increased RF voltage, which could be increased up to $\delta=\pm\SI{3}{\percent}$.
This tracking result clearly indicates that if required a larger momentum acceptance could be obtained with more RF voltage.
The studies were not continued at this point since the RF voltage had to be increased anyway to allow for the installation of wigglers as will be discussed in the following section.

\clearpage
 
In this very first studies simple orbit correction was sufficient to close the orbit and calculate the emittances including the effects of synchrotron radiation.
It could be shown, that randomly distributed alignment errors with $\sigma$ up to \SI{150}{\micro\metre} only had tiny effects on the equilibrium values of the emittance.
Also dynamic aperture and momentum acceptance have not been affected by the quadrupole misalignments.
Of course, angular misalignments and  misalignments of the sextupole magnets should be studied for a comprehensive evaluation, but this very first set of quadrupole misalignments showed that the situation is much less critical for the top-up booster synchrotron compared to the collider rings presumably because the booster does not include the very ambitious mini-beta insertions.

 \section{Damping time at low energy and mitigation of intra-beam scattering with wigglers}
 Equilibrium emittance and the transverse damping time in synchrotron radiation dominated storage rings depend on the beam energy. 
 The transverse emittance is proportional to the fourth power of the beam energy, the damping time to one over the energy to the power of three:
 \begin{equation}
    U_0 \propto E^4    \qquad \text{and} \qquad \tau_{x,y} \propto \frac{1}{E^3}
 \end{equation}
 The damping times and the equilibrium emittances for both the 60\textdegree/60\textdegree~optics and the  90\textdegree/90\textdegree~optics for all intended operation modes are summarised in \tab{tab:emittancesbooster-initial}.
 %
 \begin{table}[tbp]
 \caption{Equilibrium emittances for both the 60\textdegree/60\textdegree~optics and 90\textdegree/90\textdegree~optics and the damping times for all operational scenarios and injection energy. The effect of intra-beam scattering is not included.}
 \begin{center}
 \begin{tabular}{lccS@{}}
     \toprule
         Beam energy & Eq.~emittance   & Eq.~emittance   & {Transv. damping time}  \\
         (GeV) & (nm\,rad)   & (nm\,rad)   &  {(s)} \\
         & 60\textdegree/60\textdegree~optics & 90\textdegree/90\textdegree~optics \\
     \midrule
         20.0	 	& 0.045 &  0.015 &  10.054  \\
         45.5 	    & 0.230 &  0.077 &  0.854  \\
         80.0	 	& 0.728 &  0.242 &  0.157  \\
         120.0 	    & 1.955 &  0.546 &  0.047  \\
         175.0 	    & 3.641 &  1.195 &  0.015  \\
     \bottomrule
 \end{tabular}
 \end{center}
 \label{tab:emittancesbooster-initial}
 \end{table}
 %
 For the injection beam energy of \SI{20}{GeV} the damping time reaches $\tau = \SI{10}{\second}$. 
 If the acceleration of the particle beams in the booster is intended to start from the equilibrium state to prevent unwanted bunch oscillations during the acceleration process, the damping time needs to be reduced in order to fulfill the required cycle time.
 A value of $\tau = \SI{0.1}{\second}$ has been proposed as an adequate target \cite{yannisprivate}.
 In order to achieve that damping time wigglers have to be installed. 

\begin{figure}[tbp]
   \begin{center}
   \includegraphics[width=0.6\textwidth]{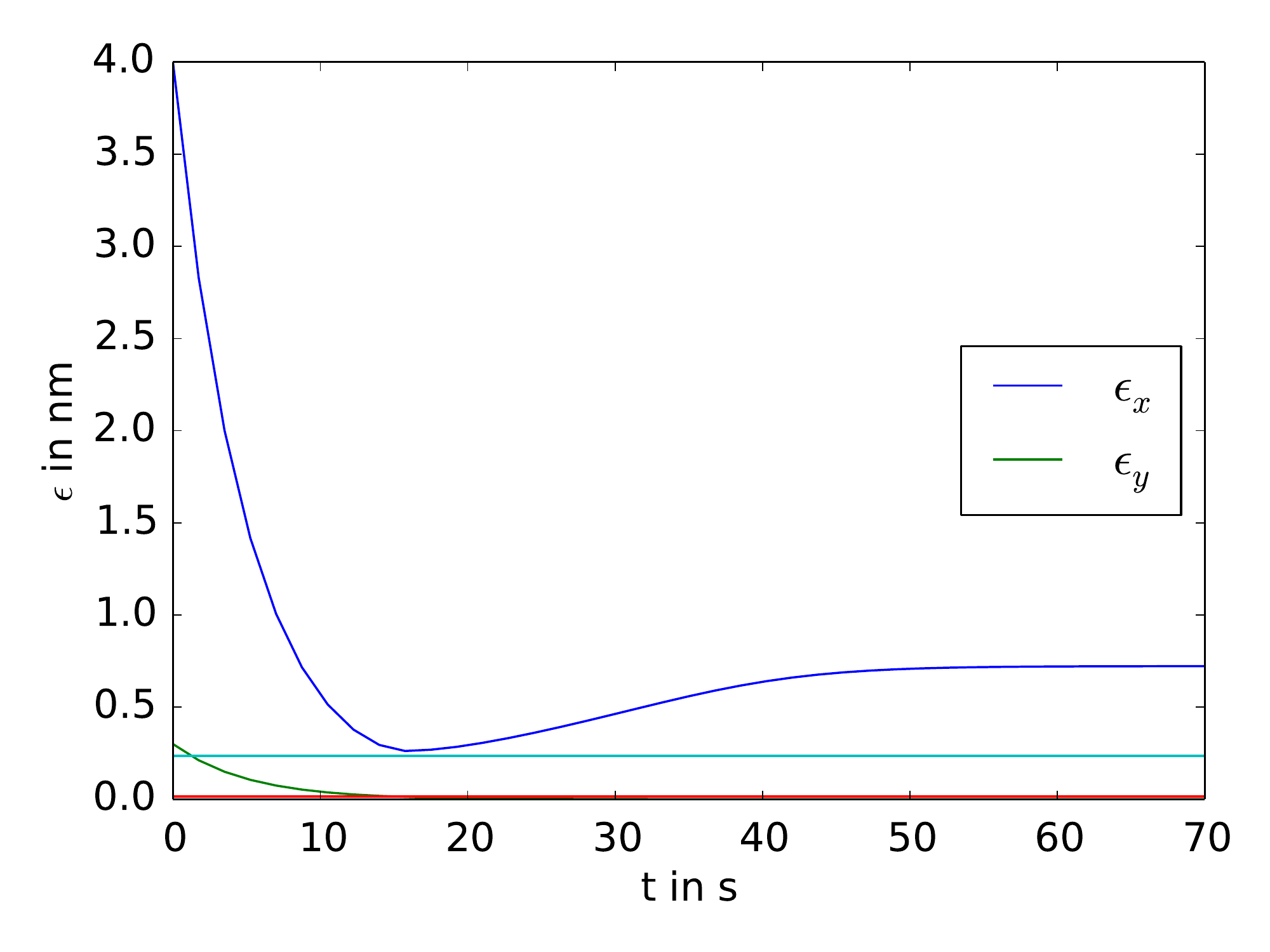}
   \end{center}
   \caption{Temporal evolution of the transverse beam emittances after injection into the FCC-ee top-up booster taking in account intra-beam scattering. While the vertical emittance is nicely damped, the horizontal emittance reaches a minimum after \SI{15}{s} and increases again up to \SI{722}{\pico\metre\radian} due to intra-beam-scattering.}
   \label{fig:IBS-withoutWigglers}
\end{figure}

 As a second consequence of the low synchrotron radiation power, the equilibrium beam emittance becomes very small. 
 For the injection beam energy of \SI{20}{GeV} the horizontal emittance shrinks down to $\epsilon_x = \SI{45}{\pico\metre\radian}$ for the 60\textdegree/60\textdegree~optics and \SI{45}{\pico\metre\radian} for the  90\textdegree/90\textdegree~optics
 because of the weak quantum excitation. 
 Such small values are the target values of ambitious next generation storage ring light sources \cite{Agapov:IPAC2019-TUPGW011} and have not been demonstrated yet. 
 One particular challenge for small beam sizes is emittance blow-up due to intra-beam scattering.
 This blow-up has been investigated\footnote{The MAD-X scripts used for this study were adapted LHC scripts that have been provided by Fanouria Antoniou. In addition, the authors would like to acknowledge the fruitful discussions with her.} for an electron beam injected from the pre-booster synchrotron. 
 The emittances were $\epsilon_x = \SI{4.0}{\nano\metre\radian}$ in the horizontal plane and $\epsilon_y = \SI{0.3}{\nano\metre\radian}$ in the vertical plane.
 As shown in Fig.~\ref{fig:IBS-withoutWigglers} the vertical emittance is nicely damped, but after reaching a minimum 15\,s after injection the horizontal emittance is blown-up because of intra-beam scattering.
 The new equilibrium is obtained at $\epsilon_x = \SI{722}{\pico\metre\radian}$.
 Keeping in mind that the equilibrium emittance for 45.6\,GeV beam energy is $\epsilon_x = \SI{235}{\pico\metre\radian}$ this is about three times the target value, which is indicated by the cyan horizontal line in Fig.~\ref{fig:IBS-withoutWigglers}.
 The wigglers used to decrease the damping time therefore need to also excite the beam to mitigate this emittance blow-up and establish a new equilibrium including intra-beam scattering below $\epsilon_x = \SI{235}{\pico\metre\radian}$.

 Several wiggler layouts with different pole length have been investigated.
 Following considerations drove the development of the current design:
 The damping time is given by
 \begin{equation}
 \tau_x=\frac{2}{J_x}\frac{E_0}{U_0}T_0,
 \end{equation}
 with $J_x=1$, the beam energy $E_0=\SI{20}{\giga\electronvolt}$, and the revolution time  $T_0=
 \SI{326}{\micro\second}$.
 For a damping time of $\tau_x=\SI{100}{\micro\second}$ the energy loss per turn must reach
 \begin{equation}
    U_0
    =\SI{130}{\mega\electronvolt},
 \end{equation}
 which is \num{0.65}\% of $E_0$ and \num{100} times more than the energy loss per turn without wigglers. 
 From this we can calculate the necessary change in $I_2$. 
 \begin{equation}
 U_0  = 
 \frac{C_\gamma}{2\pi}E_0^4\cdot I_2 \qquad
 \Rightarrow I_2  = 
 \frac{2\pi U_0}{C_\gamma E_0^4}
  = \SI{5.7323e-2}{\per\metre}
 \end{equation}
 with $C_\gamma=\SI{8.846e-5}{\metre\per\giga\electronvolt\cubed}$. 
 Without wigglers $I_2 = \SI{5.76e-4}{\per\metre}$, two orders of magnitude smaller as is the energy loss per turn. 
The energy loss per turn is defined by the magnetic strength and the overall length of the wiggler.
The equilibrium emittance is determined by the number of poles for constant magnetic length.
 It turned out, depending on the length of the wiggler poles sufficient intrinsic dispersion can be created to increase $\mathcal{I}_5$ and thus the emittance. 
 
 For the current design a normal conducting wiggler layout has been chosen, because it requires less infrastructure and the radiation losses are less critical.
 A pole tip field of \SI{1.8}{T} was assumed to minimise the number of wigglers.
 In addition, the wigglers have to be switched off during the energy ramp, which is easier to realise with normal conducting technology.
 The pole length is \SI{9.5}{\centi\metre} and the separation between two poles is $L_{\text{g}}=\SI{2}{\centi\metre}$.
 The gap $g$ was chosen to be \SI{5}{cm}, a rather small but critical value for the effective $B$ field on the beam axis:
 \begin{equation}
     B_{\text{eff}} = B_{\text{pole}}/ \cosh \left(\frac{\pi g}{2 L_{\text{pole}}+2 L_{\text{g}}}\right)
 \end{equation}
The parameters above lead to an effective $B$ field of \SI{1.45}{\tesla}.

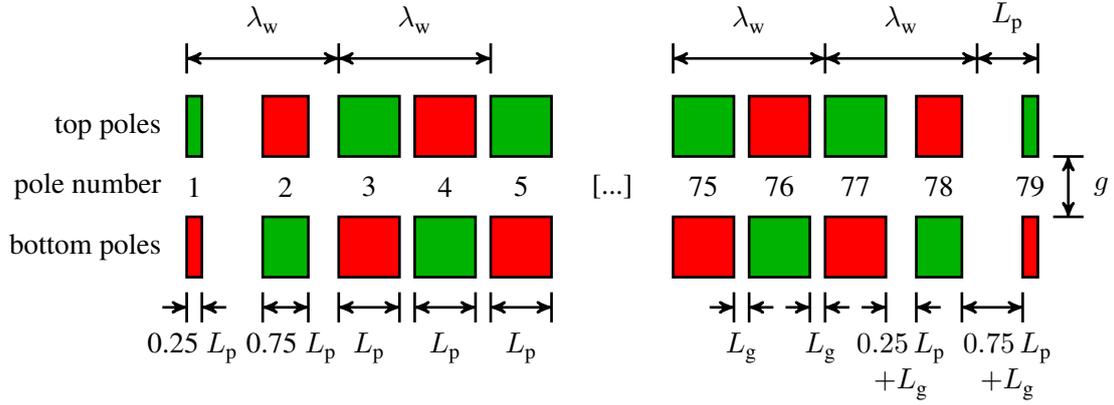
\begin{figure}[tbp]
   \begin{center}
   \begin{tikzpicture}[scale=0.8,yscale=0.5,every text node part/.style={align=center}]
     \draw [fill=green!70!black,line width=1pt] (0,-1) rectangle (0.25,1);
     \draw [fill=red,line width=1pt] (1.25,-1) rectangle (2,1);
     \draw [fill=green!70!black,line width=1pt] (2.5,-1) rectangle (3.5,1);
     \draw [fill=red,line width=1pt] (3.75,-1) rectangle (4.75,1);
     \draw [fill=green!70!black,line width=1pt] (5,-1) rectangle (6,1);
     \draw [fill=green!70!black,line width=1pt] (8,-1) rectangle (9,1);
     \draw [fill=red,line width=1pt] (9.25,-1) rectangle (10.25,1);
     \draw [fill=green!70!black,line width=1pt] (10.5,-1) rectangle (11.5,1);
     \draw [fill=red,line width=1pt] (12,-1) rectangle (12.75,1);
     \draw [fill=green!70!black,line width=1pt] (13.75,-1) rectangle (14,1);
    
     \node [left] at (-0.25,0) {top poles};
   \begin{scope}[yshift=-2.0cm]
     \node at (0.125,0) {1};
     \node at (1.625,0) {2};
     \node at (3.0,0) {3};
     \node at (4.25,0) {4};
     \node at (5.5,0) {5};
     \node at (7,0) {[...]}; 
     \node at (8.5,0) {75};
     \node at (9.75,0) {76};
     \node at (11,0) {77};
     \node at (12.375,0) {78};
     \node at (13.875,0) {79};

     \draw [line width=1pt] (14.25,1) -- (14.75,1);
     \draw [line width=1pt] (14.25,-1) -- (14.75,-1);
     \draw [stealth'-stealth',line width=1pt] (14.5,-1) -- (14.5,1);
     \node [right,xshift=0.2cm] at (14.5,0) {$g$};
     \node [left] at (-0.25,0) {pole number};
   \end{scope}
   \begin{scope}[yshift=-4.0cm]
     \draw [fill=red,line width=1pt] (0,-1) rectangle (0.25,1);
     \draw [fill=green!70!black,line width=1pt] (1.25,-1) rectangle (2,1);
     \draw [fill=red,line width=1pt] (2.5,-1) rectangle (3.5,1);
     \draw [fill=green!70!black,line width=1pt] (3.75,-1) rectangle (4.75,1);
     \draw [fill=red,line width=1pt] (5,-1) rectangle (6,1);
     \draw [fill=red,line width=1pt] (8,-1) rectangle (9,1);
     \draw [fill=green!70!black,line width=1pt] (9.25,-1) rectangle (10.25,1);
     \draw [fill=red,line width=1pt] (10.5,-1) rectangle (11.5,1);
     \draw [fill=green!70!black,line width=1pt] (12,-1) rectangle (12.75,1);
     \draw [fill=red,line width=1pt] (13.75,-1) rectangle (14,1);

     \node [left] at (-0.25,0) {bottom poles};
   \end{scope}
   \begin{scope}[yshift=-6cm]
     \draw [line width=1pt] (0,-0.5) -- (0,0.5);
     \draw [line width=1pt] (0.25,-0.5) -- (0.25,0.5);
     \draw [stealth'-,line width=1pt] (0,0) -- (-0.4,0);
     \draw [stealth'-,line width=1pt] (0.25,0) -- (0.65,0);
     \node [below,yshift=-0.2cm] at (0.1,0) {0.25 $L_{\text{p}}$};
     \draw [line width=1pt] (1.25,-0.5) -- (1.25,0.5);
     \draw [line width=1pt] (2,-0.5) -- (2,0.5);
     \draw [stealth'-stealth',line width=1pt] (1.25,0) -- (2,0);
     \node [below,yshift=-0.2cm] at (1.725,0) {0.75 $L_{\text{p}}$};
     \draw [line width=1pt] (2.5,-0.5) -- (2.5,0.5);
     \draw [line width=1pt] (3.5,-0.5) -- (3.5,0.5);
     \draw [stealth'-stealth',line width=1pt] (2.5,0) -- (3.5,0);
     \node [below,yshift=-0.2cm] at (3,0) {$L_{\text{p}}$};
     \draw [line width=1pt] (3.75,-0.5) -- (3.75,0.5);
     \draw [line width=1pt] (4.75,-0.5) -- (4.75,0.5);
     \draw [stealth'-stealth',line width=1pt] (4.75,0) -- (3.75,0);
     \node [below,yshift=-0.2cm] at (4.25,0) {$L_{\text{p}}$};
     \draw [line width=1pt] (5,-0.5) -- (5,0.5);
     \draw [line width=1pt] (6,-0.5) -- (6,0.5);
     \draw [stealth'-stealth',line width=1pt] (5,0) -- (6,0);
     \node [below,yshift=-0.2cm] at (5.5,0) {$L_{\text{p}}$};

     \draw [line width=1pt] (9,-0.5) -- (9,0.5);
     \draw [line width=1pt] (9.25,-0.5) -- (9.25,0.5);
     \draw [stealth'-,line width=1pt] (9,0) -- (8.6,0);
     \draw [stealth'-,line width=1pt] (9.25,0) -- (9.65,0);
     \node [below,yshift=-0.2cm] at (9.125,0) {$L_{\text{g}}$};
     \draw [line width=1pt] (10.25,-0.5) -- (10.25,0.5);
     \draw [line width=1pt] (10.5,-0.5) -- (10.5,0.5);
     \draw [stealth'-,line width=1pt] (10.25,0) -- (9.85,0);
     \draw [stealth'-,line width=1pt] (10.5,0) -- (10.9,0);
     \node [below,yshift=-0.2cm] at (10.425,0) {$L_{\text{g}}$};
     \draw [line width=1pt] (11.5,-0.5) -- (11.5,0.5);
     \draw [line width=1pt] (12,-0.5) -- (12,0.5);
     \draw [stealth'-,line width=1pt] (11.5,0) -- (11.1,0);
     \draw [stealth'-,line width=1pt] (12,0) -- (12.4,0);
     \node [below,yshift=-0.2cm] at (11.75,0) {$0.25\,L_{\text{p}}$\\$+L_{\text{g}}$};
     \draw [line width=1pt] (12.75,-0.5) -- (12.75,0.5);
     \draw [line width=1pt] (13.75,-0.5) -- (13.75,0.5);
     \draw [stealth'-stealth',line width=1pt] (12.75,0) -- (13.75,0);
     \node [below,yshift=-0.2cm] at (13.5,0) {$0.75\,L_{\text{p}}$\\$+L_{\text{g}}$};
   \end{scope}

   \begin{scope}[yshift=2.25cm]
     \draw [line width=1pt] (0,-0.5) -- (0,0.5);
     \draw [line width=1pt] (2.5,-0.5) -- (2.5,0.5);
     \draw [stealth'-stealth',line width=1pt] (0,0) -- (2.5,0);
     \node [above,yshift=0.2cm] at (1.25,0) {$\lambda_{\text{w}}$};
     \draw [line width=1pt] (5,-0.5) -- (5,0.5);
     \draw [stealth'-stealth',line width=1pt] (2.5,0) -- (5,0);
     \node [above,yshift=0.2cm] at (3.75,0) {$\lambda_{\text{w}}$};

     \draw [line width=1pt] (8,-0.5) -- (8,0.5);
     \draw [line width=1pt] (10.5,-0.5) -- (10.5,0.5);
     \draw [stealth'-stealth',line width=1pt] (8,0) -- (10.5,0);
     \node [above,yshift=0.2cm] at (9.25,0) {$\lambda_{\text{w}}$};
     \draw [line width=1pt] (13,-0.5) -- (13,0.5);
     \draw [stealth'-stealth',line width=1pt] (10.5,0) -- (13,0);
     \node [above,yshift=0.2cm] at (11.75,0) {$\lambda_{\text{w}}$};
     \draw [line width=1pt] (14,-0.5) -- (14,0.5);
     \draw [stealth'-stealth',line width=1pt] (13,0) -- (14,0);
     \node [above,yshift=0.2cm] at (13.5,0) {$L_{\text{p}}$};
   \end{scope}
   \end{tikzpicture}
   \end{center}
     \caption{Schematic of the wiggler design for the FCC-ee booster synchrotron. Each wiggler has 79 poles. The first and last poles have one quarter and the second and second last poles three quarter of the regular length for on-axis oscillation of the beam.}
     \label{fig:schematicwiggler}
 \end{figure}

\begin{figure}[htb!]
 \centering
 \includegraphics[width=0.6\textwidth]{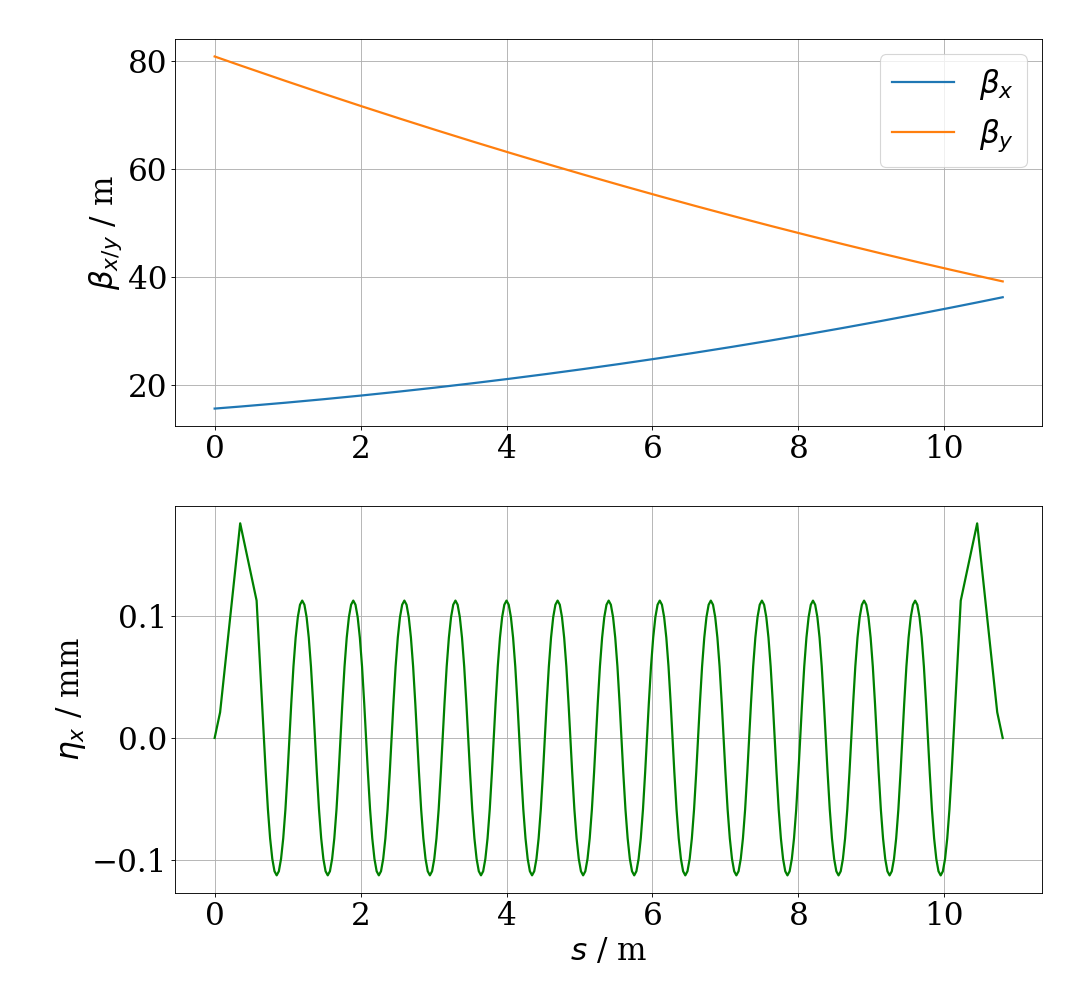}
 \caption{Beta functions and horizontal dispersion function of an on-axis wiggler in the first half of a straight FODO cell.}
 \label{fig:betadisponaxis}
 \end{figure}

The wigglers consist of 79 poles as illustrated in Fig.~\ref{fig:schematicwiggler} leading to an overall length of \SI{9.065}{m}.
The in-coupling poles have $\frac1{4}$ / $\frac{3}{4}$ pole length to obtain on axis oscillation of both trajectory and dispersion function (comp. Fig.~\ref{fig:betadisponaxis}). 
This method will give less emittance heating due to smaller dispersion. In addition, it will leave more margin to aperture, be it dynamic or geometric.


 The FCC-ee booster lattice includes $n_{\text{w}}=16$ of these wigglers, which are installed in the same straight sections as the RF cavities, D and J.
 Since the energy loss due to the wigglers is 0.65\% of the beam energy it was beneficial to install them close to the cavities, where the energy is re-fed to the beam.
 Otherwise the beam would have a large energy offset in the arc sections, that either leads to optics distortions or has to be compensated for.
 The arrangement of cavities and wigglers including absorber scheme to protect the superconducting cavities from the wiggler's radiation power still has to be investigated.

This wiggler layout with $n_{\text{p}}=79$ poles described above gives a $\Delta \mathcal{I}_2$ of
\begin{equation}
    \Delta \mathcal{I}_2 = \oint\frac1{\rho_{\text{w}}} ~\text{d}s \approx \frac{n_\text{w}(n_{\text{p}}-2)L_{\text{p}}}{\rho_{\text{w}}^2} = n_\text{w}(n_{\text{p}}-2)L_{\text{p}}\left(\frac{eB}{ p}\right)^2 = \SI{0.0554}{\per\metre}.
\end{equation}
Recalling $\mathcal{I}_{2,0} = \SI{0.00058}{\per\metre}$ the new value of $\mathcal{I}_{2}$ is given by 
\begin{equation}
    \mathcal{I}_2 = \mathcal{I}_{2,0} + \Delta \mathcal{I}_2 = \SI{0.0560}{\per\metre},
\end{equation}
which nicely fits to the simulation result of MAD-X of $\mathcal{I}_2 = \SI{0.0558}{\per\metre}$. 
With the second synchrotron radiation integral the additional energy loss per turn caused by the wigglers is 
\begin{equation}
    \Delta U_0 = \frac{C_{\gamma}}{2\pi}E^4 \Delta\mathcal{I}_2 = \SI{124.7}{MeV}.
\end{equation}
Adding the energy loss in in the  arcs the total energy loss per turn is 
\begin{equation}
    U_0= U_{0,0} + \Delta U_0= \SI{124.7}{MeV} + \SI{1.3}{MeV} = \SI{126.0}{MeV},
\end{equation}
which again nicely fits to the MAD-X simulation giving $U_0 = \SI{125.6}{MeV}$.
The transverse damping time is then given by 
\begin{equation}
    \tau_x = \frac{2}{J_x}\frac{E_0}{U_0}T_0 = \SI{103.5}{ms},
\end{equation}
and MAD-X gives $\tau_x =\SI{103.8}{ms}$.

The calculation of the equilibrium emittance 
\begin{equation}
    \epsilon_x = C_{\text{q}}\frac{\gamma^2}{J_x}\frac{\mathcal{I}_{5}}{\mathcal{I}_2}
\end{equation}
is a little bit more complicated because it involves the 5\textsuperscript{th} synchrotron radiation integral
\begin{equation}
    \mathcal{I}_5 = \oint \frac1{|\rho|^3}\mathcal{H}\,\text{d}s,
\end{equation}
which not only depends on the local bending radius but also the local optics represented by the chromatic invariant, the 'curly-H' function 
\begin{equation}
    \mathcal{H} = \beta D'^2 + 2 \alpha DD'+ \gamma D^2.
\end{equation}

\begin{figure}[tbp]
   \begin{center}
   \includegraphics[width=0.6\textwidth]{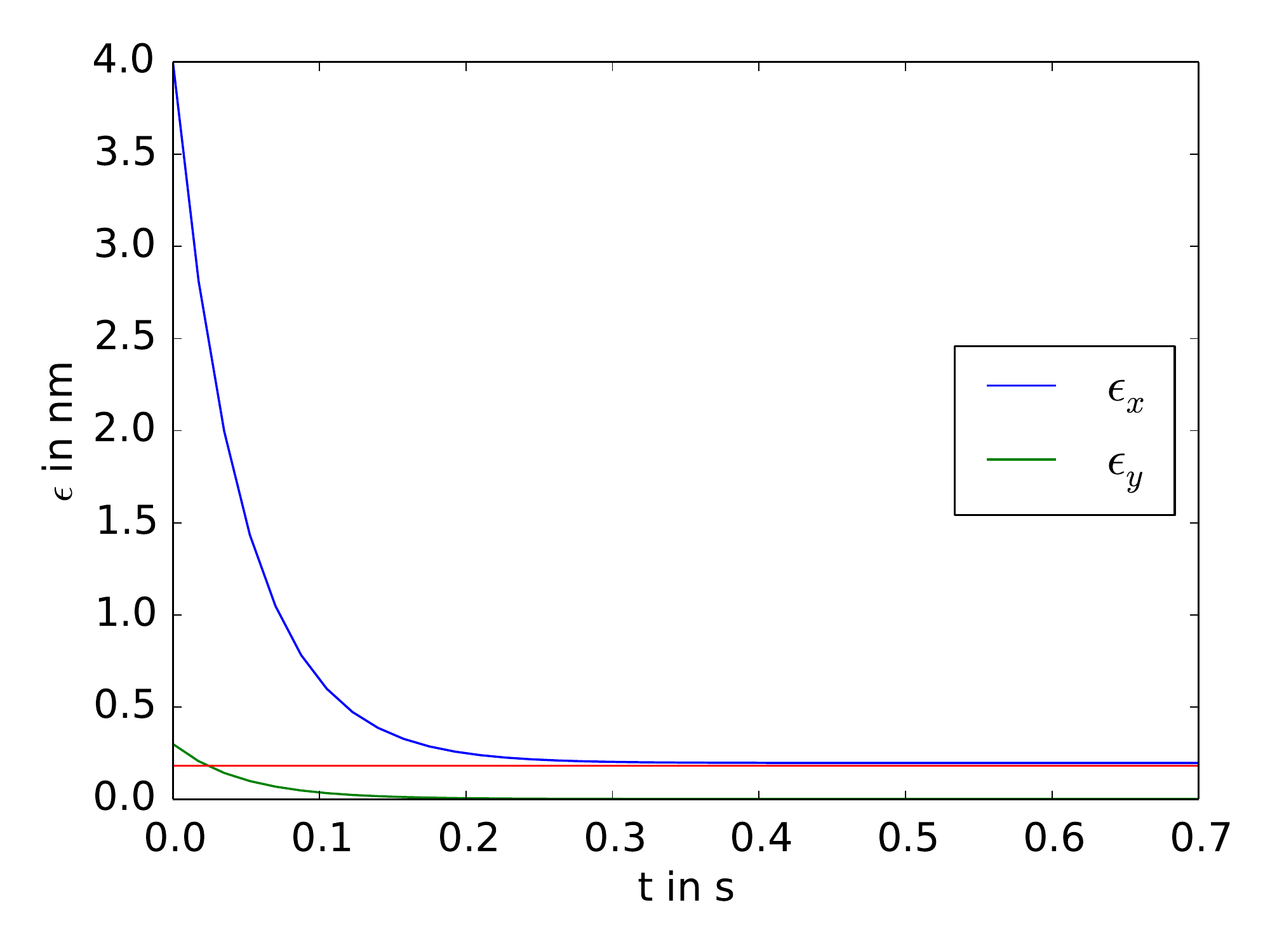}
   \end{center}
   \caption{Evolution of the beam emittance with time. In this case wigglers are included to the lattice (\SI{60}{\degree} optics) to increase the equilibrium emittance and mitigate the effects of intra-beam-scattering. As seen in the plot, the emittance reaches a value of \SI{242}{\pico\metre\radian} which is slightly more than the goal value of \SI{235}{\pico\metre\radian} indicated by the red line.}
  \label{fig:IBS-withWigglers}
\end{figure}

The current layout was optimised for a previous version of the booster lattice.
The goal value for the emittance were \SI{235}{\pico\metre\radian}, which is the equilibrium value at \SI{45.6}{GeV} beam energy with $60^\degree$ optics.
A value of \SI{242}{pm} has been reached with the wigglers with $60^\degree$ optics and \SI{193}{pm} with the $90^\degree$ optics.
As illustrated in \fig{fig:IBS-withWigglers}, the horizontal emittance is now nicely damped until it reaches its equilibrium value at about \SI{0.3}{s}, slightly above the goal value indicated by the red line.
%

In the new layout the optics at the position of the wigglers is a little bit different, which then results in emittance values of \SI{300}{\pico\metre\radian} and \SI{290}{\pico\metre\radian} for the $60^\degree$ and the $90^\degree$ optics, respectively.
The emittance can be decreased again by adapting the pole length and the number of poles as discussed earlier.
For example, a shorter pole length of $L_{\text{p}}=\SI{0.9}{m}$ and 89 poles lead to $\epsilon_x = \SI{255}{pm}$ and a damping time of $\tau_x = \SI{105}{ms}$.
However, such an updated design of the wigglers is not yet included the current lattice files.
The basic parameters of the wiggler design as well as energy loss, damping time and equilibrium emittance  for 16 wigglers are summarised in \tab{tab:wigglerParameter}.
In general, the emittance values for the $90^\degree$ optics are less important, as this optics will only be used at the high energy operation modes where the emittance will be larger than \SI{500}{\nano\metre\radian} anyway.

\begin{table}
\caption{Design parameters of the FCC-ee damping wiggler as well as energy loss, damping time and equilibrium emittance  for 16 wigglers.}
\centering
 \begin{tabular}{lc}
 \toprule
 Pole length  &  0.095\,m \\
 Pole separation   &  0.020\,m \\
 Gap    & 0.050\,m\\
 Number of poles  & 79  \\
 Wiggler length    & \SI{9.065}{m} \\
 Magnetic field (beam axis) &  1.45\,T\\
 \midrule
 Energy loss per turn  & 126\, MeV \\
 Hor. damping time  &  104\,ms \\
 Hor. emittance (60\textdegree optics)     & 300\,pm\,rad \\
 \bottomrule
 \end{tabular}
\label{tab:wigglerParameter}
\end{table}

 \section{RF infrastructure}
 The current layout of the RF infrastructure \cite{taviancryolayout,rfarrangementtunnel} has been included to the MAD-X model of the FCC-ee booster synchrotron.
 In order to reduce the costs for superconducting infrastructure only two RF sections are foreseen instead of equally distributed RF stations.
 These two RF sections are located in the \SI{2.4}{km} long straight sections around points D and J.
The energy loss due to synchrotron radiation creates a local energy deviation that leads to a sawtooth orbit, which in case of the collider rings is corrected by tapering \cite{fcceeDesignReport,oidelattice1,oidelattice2}. 
 The local magnetic field of the dipole and quadrupole magnets are adjusted to the local beam energy in order to obtain the same bending radius and focusing strength in all magnets.
 For the FCC-ee booster synchrotron this is not foreseen so far, as it is operated as a rapid cycling synchrotron and operation is considered easier without tapering.

 The energy loss of the particles that has to be compensated by the RF system depends on the fourth power of the particle energy. 
 As a result a total power in the range from \SI{140}{MV} up to \SI{10.9}{GV} is necesssary for stable operation at the four different operation modes.
 In order to mitigate wakefield effects introduced by the cavities, a staging scenario has been defined to upgrade the RF system in-between the different operation modes \cite{taviancryolayout}.
 This in particular is necessary for operation at the $Z$ pole where the beams will have the highest intensity with 16640 bunches and a bunch population of $2.13\times\SI{e10}{}$.

 Most important in the scope of this paper is the number of cavities, the number of cryo modules and their arrangement. 
 This information was provided by the RF group and is summarised in Tab.~\ref{tab:rfparameters} for each operation mode for the booster synchrotron \cite{taviancryolayout}.
 It might be noted that for operation at \SI{45.6}{GeV} beam energy only an RF voltage of \SI{100}{MV} is foreseen for the collider.
 The booster needs more RF voltage, \SI{140}{MV}, because of the additional synchrotron radiation losses created by the damping wigglers during accumulation.

 \begin{table}
     \centering
     \caption{Parameters of the RF system for the FCC-ee booster synchrotron. A more comprehensive table can be found in \cite{taviancryolayout}.}
     \begin{tabular}{lccccc}
          \toprule 
          & Z & W & H & ttbar\textsubscript{1} & ttbar\textsubscript{2} \\
          \midrule
          Total RF voltage (MV) & 140 & 750 & 2000 & 9500 & 10930 \\
          \textbf{frequency (MHz)} & \multicolumn{5}{c}{\textbf{400}} \\
          RF voltage (MV) & 140 & 750 & 2000 & 2000 & 2000 \\
          $E_{\text{acc}}$ (MV/m) & 8.0 & 9.6 & 9.8 & 10.0 & 10.0 \\
          \# CM & 3 & 13 & 34 & 34 & 34 \\
          \# cavities & 12 & 52 & 136 & 136 & 136 \\
          \# cells/cav. & 4 & 4 & 4 & 4 & 4 \\
          \textbf{frequency (MHz)} & \multicolumn{5}{c}{\textbf{800}} \\
          RF voltage (MV) &  &  &  & 7500 & 8930 \\
          $E_{\text{acc}}$ (MV/m) & & & & 20 & 19.8 \\
          \# CM &  &  &  & 100 & 120 \\
          \# cavities &  &  &  & 400 & 480 \\
          \# cells/cav. &  &  &  & 5 & 5 \\
          \bottomrule
     \end{tabular}
     \label{tab:rfparameters}
 \end{table}

 Table \ref{tab:rfparameters} shows that two types of RF systems are foreseen for FCC-ee:
 On one hand a \SI{400}{MHz} RF system and on the other hand a \SI{800}{MHz} RF system, which will exclusively be used for operation at the $t\bar{t}$ threshold at maximum beam energy.
 The \SI{400}{MHz} cavities consist of four cells, the \SI{800}{MHz} of five cells. 
 As a result of the different frequencies the length of one cell of each cavity is different: for the \SI{400}{MHz} RF system one cell has a length of \SI{0.375}{m}, in case of the \SI{800}{MHz} RF system the cell length is \SI{0.1875}{m}.
 Therefore, the length of the two cavities is different and since in both cases four cavities are housed in one cryo module the length of one cryo module is \SI{12}{m} in case of the \SI{400}{MHz} RF system and \SI{9}{m} in case of the \SI{800}{MHz} RF system \cite{taviancryolayout}.

 \begin{table}
     \centering
     \caption{Details of the RF system that are important for the lattice design of the FCC-ee booster synchrotron.}
     \begin{tabular}{lcc}
         \toprule
         frequency (MHz) & 400 & 800\\
         \midrule
         length of cryo module (m) & 12 & 9 \\ 
         no. of cavities per module & 4 & 4 \\
         no. of cells per cavity & 4 &5 \\
         cell length (m) & 0.375 & 0.1875 \\
          \bottomrule
     \end{tabular}
     \label{tab:cavitydetails}
 \end{table}

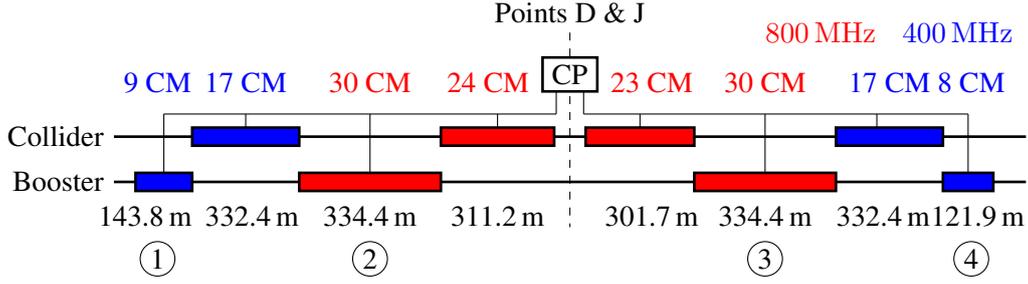
\begin{figure}[tbp]
   \begin{center}
   \begin{tikzpicture}[scale=0.6]
     \def\cma{12}
     \def\cmb{9}
     \def\shrink{0.0115}
     \def\centre{30}
     \draw [line width=1pt] (-10.0,0.0) -- (10.0,0.0);
     \draw [line width=1pt] (-10.0,-1.0) -- (10.0,-1.0);
     \node [left] at (-10,0) {Collider};
     \node [left] at (-10,-1) {Booster};
     \draw [line width=1pt,fill=red] (-\centre*\shrink,0.2) rectangle (-\centre*\shrink-\shrink*\cmb*24,-0.2);
     \draw [line width=1pt,fill=red] (-\centre*\shrink-\shrink*54*\cmb,-0.8) rectangle (-\centre*\shrink-\shrink*\cmb*24,-1.2);
     \node [red] at (-\centre*\shrink-\shrink*\cmb*14,1.2) {24 CM};
     \node [red] at (-\centre*\shrink-\shrink*\cmb*39,1.2) {30 CM};
     \draw [line width=1pt,fill=blue] (-\centre*\shrink-\shrink*54*\cmb,-0.2) rectangle (-\centre*\shrink-\shrink*\cmb*54-\shrink*\cma*17,0.2);
     \draw [line width=1pt,fill=blue] (-\centre*\shrink-\shrink*54*\cmb-\shrink*\cma*26,-1.2) rectangle (-\centre*\shrink-\shrink*\cmb*54-\shrink*\cma*17,-0.8);
     \node [blue] at (-\centre*\shrink-\shrink*\cmb*54-\shrink*\cma*8.5,1.2) {17 CM};
     \node [blue] at (-\centre*\shrink-\shrink*\cmb*54-\shrink*\cma*22.5,1.2) {9 CM};
     \draw [dashed] (0,2.3) -- (0,-2);
     \node [above] at (0,2.3) {Points D \& J};
     \draw [line width=1pt, fill=white] (0.6,1.0) rectangle (-0.6,1.75);
     \draw (-0.3,0.5) -- (-0.3,1.0);
     \draw (-0.3,0.5) -- (-\centre*\shrink-\shrink*\cmb*54-\shrink*\cma*21.5,0.5);
     \draw (-\centre*\shrink-\shrink*\cmb*54-\shrink*\cma*21.5,-0.8) -- (-\centre*\shrink-\shrink*\cmb*54-\shrink*\cma*21.5,0.5);
     \draw (-\centre*\shrink-\shrink*\cmb*54-\shrink*\cma*8.5,0.2) -- (-\centre*\shrink-\shrink*\cmb*54-\shrink*\cma*8.5,0.5);
     \draw (-\centre*\shrink-\shrink*\cmb*39,-0.8) -- (-\centre*\shrink-\shrink*\cmb*39,0.5);
     \draw (-\centre*\shrink-\shrink*\cmb*12,0.2) -- (-\centre*\shrink-\shrink*\cmb*12,0.5);
     \draw (0.3,0.5) -- (\centre*\shrink+\shrink*\cmb*53+\shrink*\cma*21,0.5);
     \draw (0.3,0.5) -- (0.3,1.0);
     \draw (\centre*\shrink+\shrink*\cmb*53+\shrink*\cma*21,-0.8) -- (\centre*\shrink+\shrink*\cmb*53+\shrink*\cma*21,0.5);
     \draw (\centre*\shrink+\shrink*\cmb*53+\shrink*\cma*6.5,0.2) -- (\centre*\shrink+\shrink*\cmb*53+\shrink*\cma*6.5,0.5);
     \draw (\centre*\shrink+\shrink*\cmb*38,-0.8) -- (\centre*\shrink+\shrink*\cmb*38,0.5);
     \draw (\centre*\shrink+\shrink*\cmb*11.5,0.2) -- (\centre*\shrink+\shrink*\cmb*11.5,0.5);
     \node at (0,1.375) {CP};
     \node [below] at (-\centre*\shrink-\shrink*\cmb*12,-1.3) {311.2\,m};
     \node [below] at (-\centre*\shrink-\shrink*\cmb*39,-1.3) {334.4\,m};
     \node [below] at (-\centre*\shrink-\shrink*\cmb*54-\shrink*\cma*7.5,-1.3) {332.4\,m};
     \node [below] at (-\centre*\shrink-\shrink*\cmb*54-\shrink*\cma*24.5,-1.3) {143.8\,m};
     \node [below] at (\centre*\shrink+\shrink*\cmb*14,-1.3) {301.7\,m};
     \node [below] at (\centre*\shrink+\shrink*\cmb*38,-1.3) {334.4\,m};
     \node [below] at (\centre*\shrink+\shrink*\cmb*53+\shrink*\cma*7.5,-1.3) {332.4\,m};
     \node [below] at (\centre*\shrink+\shrink*\cmb*53+\shrink*\cma*22.5,-1.3) {121.9\,m};
     \node [blue] at (8.5,2.3) {\SI{400}{MHz}};
     \node [red] at (5.5,2.3) {\SI{800}{MHz}};
     \draw [line width=1pt,fill=red] (\centre*\shrink,0.2) rectangle (\centre*\shrink+\shrink*\cmb*23,-0.2);
     \draw [line width=1pt,fill=red] (\centre*\shrink+\shrink*53*\cmb,-0.8) rectangle (+\centre*\shrink+\shrink*\cmb*23,-1.2);
     \node [red] at (\centre*\shrink+\shrink*\cmb*14,1.2) {23 CM};
     \node [red] at (\centre*\shrink+\shrink*\cmb*38,1.2) {30 CM};
     \draw [line width=1pt,fill=blue] (\centre*\shrink+\shrink*53*\cmb,-0.2) rectangle (\centre*\shrink+\shrink*\cmb*53+\shrink*\cma*17,0.2);
     \draw [line width=1pt,fill=blue] (\centre*\shrink+\shrink*53*\cmb+\shrink*\cma*25,-1.2) rectangle (\centre*\shrink+\shrink*\cmb*53+\shrink*\cma*17,-0.8);
     \node [blue] at (\centre*\shrink+\shrink*\cmb*53+\shrink*\cma*8.5,1.2) {17 CM};
     \node [blue] at (\centre*\shrink+\shrink*\cmb*53+\shrink*\cma*21.5,1.2) {8 CM};
  
     \node [circle, draw, black, fill=white, minimum width=13pt, inner sep=0] at (-\centre*\shrink-\shrink*\cmb*54-\shrink*\cma*22.5,-2.7) {1};
     \node [circle, draw, black, fill=white, minimum width=13pt, inner sep=0] at (-\centre*\shrink-\shrink*\cmb*39,-2.7) {2};
     \node [circle, draw, black, fill=white, minimum width=13pt, inner sep=0] at (\centre*\shrink+\shrink*\cmb*38,-2.7) {3};
     \node [circle, draw, black, fill=white, minimum width=13pt, inner sep=0] at (\centre*\shrink+\shrink*\cmb*53+\shrink*\cma*21.5,-2.7) {4};
   \end{tikzpicture}
   \end{center}
     \caption{Cryogenic layout for FCC-ee at the points D and J for maximum RF power and 182.5\,GeV beam energy. The cryo plant (CP) is located at the centre of the straight section and feeds the cryo modules at both sides \cite{taviancryolayout}. Four sections have been defined to allow the discussion of the RF layout for the booster synchrotron in more detail.}
     \label{fig:cryogeniclayout}
 \end{figure}

The diameters of the cryo modules are \SI{1.14}{\metre} and \SI{1.09}{\metre} for the \SI{400}{MHz} and the \SI{800}{MHz} cavities respectively.
The cryo modules of booster and collider therefore cannot be installed in parallel, they have to be staggered \cite{rfarrangementtunnel}. 
In this case a minimum distance of \SI{1.26}{m} between booster and collider is required. [(180216\_layout\_in\_the\_RF\_straights.pdf)]
The RF sections are organised as illustrated in Fig.~\ref{fig:cryogeniclayout}. 
  The more detailed plan of the arrangement of the RF installations is presented in \cite{taviancryolayout}.
 The cryo plant is located at the centre of the straight section. 
  Directly next to it the \SI{800}{MHz} cavities of the collider rings are installed on both sides in 24 and 23 cryo modules respectively.
 Then the \SI{800}{MHz} cavities of the booster follow in the outwards direction in 30 cryo modules each. 
 In the later discussion these sections will be referred to as sections 2 and 3.
 For the \SI{400}{MHz} RF infrastructure again the cavities of the collider are first installed and the cavities of the booster form the last layer.
 These sections comprise 9 and 8 cryo modules and will later referred to as sections 1 and 4.  
 
\begin{figure}[tbp]
   \begin{center}
   \begin{subfloat}[Arrangement of the cryo modules for \SI{400}{MHz} cavities in sections 2 and 3]{
   \begin{tikzpicture}[scale=0.6]
     \draw [line width=1pt] (0,0.0) -- (20.0,0.0);
     \draw [line width=1pt,fill=white] (0.0,0.4) rectangle (1.0,-0.4);
     \draw [line width=1pt,fill=white] (19.0,0.4) rectangle (20.0,-0.4);
     \node at (0.5,0) {Q};
     \node at (19.5,0) {Q};
     \draw [line width=1pt,fill=blue] (1.5,0.2) rectangle (6.5,-0.2);
     \draw [line width=1pt,fill=blue] (7.0,0.2) rectangle (12.0,-0.2);
     \draw [line width=1pt,fill=blue] (12.5,0.2) rectangle (17.5,-0.2);
     \draw [dashed] (0.5,2.3) -- (0.5,-2.5);
     \draw [stealth'-] (0.5,-2.0) -- (9.0,-2.0);
     \node at (10.0,-2.0) {50\,m};
     \draw [-stealth'] (11,-2.0) -- (19.5,-2.0);
     \draw [dashed] (19.5,2.3) -- (19.5,-2.5);
     \draw (1.0,0.0) -- (1.0,1.25);
     \draw [-stealth'] (0.5,0.75) -- (1.0,0.75);
     \node at (1.3,1.75) {0.3\,m};
     \draw [stealth'-] (1.5,0.75) -- (2.0,0.75);
     \draw (1.5,0) -- (1.5,1.25);
     \begin{scope}[xshift=5.5cm]
     \draw (1.0,0.0) -- (1.0,1.25);
     \draw [-stealth'] (0.5,0.75) -- (1.0,0.75);
     \node at (1.3,1.75) {0.6\,m};
     \draw [stealth'-] (1.5,0.75) -- (2.0,0.75);
     \draw (1.5,0) -- (1.5,1.25);
     \end{scope}
     \begin{scope}[xshift=11cm]
     \draw (1.0,0.0) -- (1.0,1.25);
     \draw [-stealth'] (0.5,0.75) -- (1.0,0.75);
     \node at (1.3,1.75) {0.6\,m};
     \draw [stealth'-] (1.5,0.75) -- (2.0,0.75);
     \draw (1.5,0) -- (1.5,1.25);
     \end{scope}
     \draw (17.5,0.0) -- (17.5,1.25);
     \draw [stealth'-stealth'] (17.5,0.75) -- (19.0,0.75);
     \node at (18.25,1.75) {11\,m};
     \draw (19.0,0) -- (19.0,1.25);
     \draw (1.5,0) -- (1.5,-1.5);
     \draw [stealth'-] (1.5,-1) -- (3,-1.0);
     \node at (4.0,-1.0) {12\,m};
     \draw [-stealth'] (5.0,-1) -- (6.5,-1.0);
     \draw (6.5,0) -- (6.5,-1.5);
     \begin{scope}[xshift=5.5cm]
     \draw (1.5,0) -- (1.5,-1.5);
     \draw [stealth'-] (1.5,-1) -- (3,-1.0);
     \node at (4.0,-1.0) {12\,m};
     \draw [-stealth'] (5.0,-1) -- (6.5,-1.0);
     \draw (6.5,0) -- (6.5,-1.5);
     \end{scope}
     \begin{scope}[xshift=11cm]
     \draw (1.5,0) -- (1.5,-1.5);
     \draw [stealth'-] (1.5,-1) -- (3,-1.0);
     \node at (4.0,-1.0) {12\,m};
     \draw [-stealth'] (5.0,-1) -- (6.5,-1.0);
     \draw (6.5,0) -- (6.5,-1.5);
     \end{scope}
   \end{tikzpicture}}
   \end{subfloat}
   \begin{subfloat}[Arrangement of the cryo modules for \SI{800}{MHz} cavities in sections 2 and 3]{
   \begin{tikzpicture}[scale=0.6]
     \draw [line width=1pt] (0,0.0) -- (20.0,0.0);
     \draw [line width=1pt,fill=white] (0.0,0.4) rectangle (1.0,-0.4);
     \draw [line width=1pt,fill=white] (19.0,0.4) rectangle (20.0,-0.4);
     \node at (0.5,0) {Q};
     \node at (19.5,0) {Q};
     \draw [line width=1pt,fill=red] (1.5,0.2) rectangle (4.5,-0.2);
     \draw [line width=1pt,fill=red] (5.0,0.2) rectangle (8.0,-0.2);
     \draw [line width=1pt,fill=red] (8.5,0.2) rectangle (11.5,-0.2);
     \draw [line width=1pt,fill=red] (12.0,0.2) rectangle (15.0,-0.2);
     \draw [line width=1pt,fill=red] (15.5,0.2) rectangle (18.5,-0.2);
     \draw [dashed] (0.5,2.3) -- (0.5,-2.5);
     \draw [stealth'-] (0.5,-2.0) -- (9.0,-2.0);
     \node at (10.0,-2.0) {50\,m};
     \draw [-stealth'] (11,-2.0) -- (19.5,-2.0);
     \draw [dashed] (19.5,2.3) -- (19.5,-2.5);
     \draw (1.0,0.0) -- (1.0,1.25);
     \draw [-stealth'] (0.5,0.75) -- (1.0,0.75);
     \node at (1.3,1.75) {0.3\,m};
     \draw [stealth'-] (1.5,0.75) -- (2.0,0.75);
     \draw (1.5,0) -- (1.5,1.25);
     \begin{scope}[xshift=3.5cm]
     \draw (1.0,0.0) -- (1.0,1.25);
     \draw [-stealth'] (0.5,0.75) -- (1.0,0.75);
     \node at (1.3,1.75) {0.6\,m};
     \draw [stealth'-] (1.5,0.75) -- (2.0,0.75);
     \draw (1.5,0) -- (1.5,1.25);
     \end{scope}
     \begin{scope}[xshift=7cm]
     \draw (1.0,0.0) -- (1.0,1.25);
     \draw [-stealth'] (0.5,0.75) -- (1.0,0.75);
     \node at (1.3,1.75) {0.6\,m};
     \draw [stealth'-] (1.5,0.75) -- (2.0,0.75);
     \draw (1.5,0) -- (1.5,1.25);
     \end{scope}
     \begin{scope}[xshift=10.5cm]
     \draw (1.0,0.0) -- (1.0,1.25);
     \draw [-stealth'] (0.5,0.75) -- (1.0,0.75);
     \node at (1.3,1.75) {0.6\,m};
     \draw [stealth'-] (1.5,0.75) -- (2.0,0.75);
     \draw (1.5,0) -- (1.5,1.25);
     \end{scope}
     \begin{scope}[xshift=14cm]
     \draw (1.0,0.0) -- (1.0,1.25);
     \draw [-stealth'] (0.5,0.75) -- (1.0,0.75);
     \node at (1.3,1.75) {0.6\,m};
     \draw [stealth'-] (1.5,0.75) -- (2.0,0.75);
     \draw (1.5,0) -- (1.5,1.25);
     \end{scope}
     \draw (18.5,0.0) -- (18.5,1.25);
     \draw [-stealth'] (18.0,0.75) -- (18.5,0.75);
     \node at (18.7,1.75) {0.8\,m};
     \draw [stealth'-] (19,0.75) -- (19.5,0.75);
     \draw (19.0,0) -- (19.0,1.25);
     \draw (1.5,0) -- (1.5,-1.5);
     \draw [stealth'-] (1.5,-1) -- (2.25,-1.0);
     \node at (3.0,-1.0) {9\,m};
     \draw [-stealth'] (3.75,-1) -- (4.5,-1.0);
     \draw (4.5,0) -- (4.5,-1.5);
     \begin{scope}[xshift=3.5cm]
     \draw (1.5,0) -- (1.5,-1.5);
     \draw [stealth'-] (1.5,-1) -- (2.25,-1.0);
     \node at (3.0,-1.0) {9\,m};
     \draw [-stealth'] (3.75,-1) -- (4.5,-1.0);
     \draw (4.5,0) -- (4.5,-1.5);
     \end{scope}
     \begin{scope}[xshift=7cm]
     \draw (1.5,0) -- (1.5,-1.5);
     \draw [stealth'-] (1.5,-1) -- (2.25,-1.0);
     \node at (3.0,-1.0) {9\,m};
     \draw [-stealth'] (3.75,-1) -- (4.5,-1.0);
     \draw (4.5,0) -- (4.5,-1.5);
     \end{scope}
     \begin{scope}[xshift=10.5cm]
     \draw (1.5,0) -- (1.5,-1.5);
     \draw [stealth'-] (1.5,-1) -- (2.25,-1.0);
     \node at (3.0,-1.0) {9\,m};
     \draw [-stealth'] (3.75,-1) -- (4.5,-1.0);
     \draw (4.5,0) -- (4.5,-1.5);
     \end{scope}
     \begin{scope}[xshift=14cm]
     \draw (1.5,0) -- (1.5,-1.5);
     \draw [stealth'-] (1.5,-1) -- (2.25,-1.0);
     \node at (3.0,-1.0) {9\,m};
     \draw [-stealth'] (3.75,-1) -- (4.5,-1.0);
     \draw (4.5,0) -- (4.5,-1.5);
     \end{scope}
   \end{tikzpicture}}
   \end{subfloat}
   \end{center}
     \caption{Schematic arrangement of the cryo modules in a FODO half-cell in the RF straight section around points D and J of the FCC-ee booster synchrotron. The FODO cell lenght is \SI{100}{m}.}
     \label{fig:cryogeniclayoutcell}
 \end{figure}
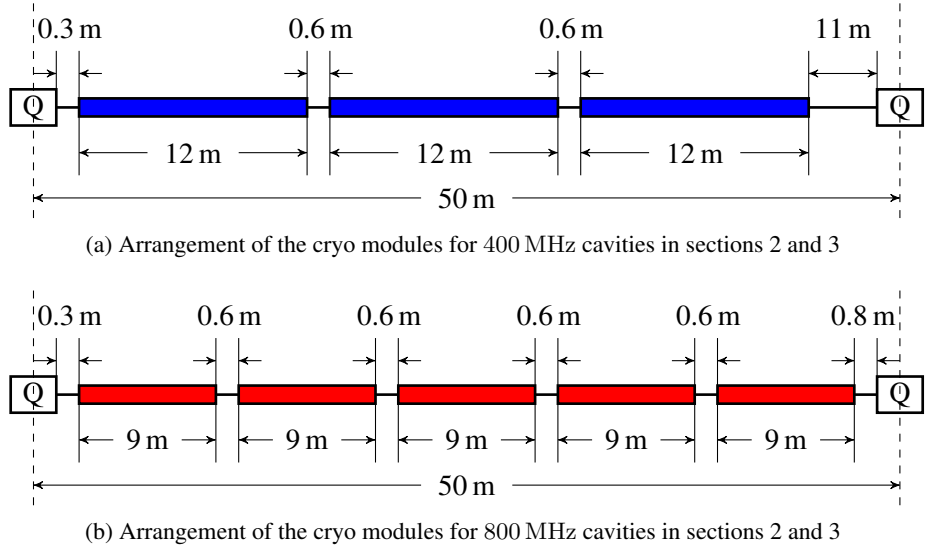

 In-between the cryo modules a minimum distance of \SI{0.6}{m} is required for interconnections.
 The arrangement of the cryo modules in the FODO cells are illustrated in the schematic in Fig.~\ref{fig:cryogeniclayoutcell}.
 Because of their different length only three cryo modules for the \SI{400}{MHz} cavities fit in one half-cell, while five cryo modules for the \SI{800}{MHz} cavities can be installed.
 
 According to the staging scenario an arrangement of the cryo modules for both the collider and the booster has been proposed in \cite{taviancryolayout} and then been adapted to the lattice of the booster for each operation mode:
 \begin{enumerate}
     \item \SI{45.6}{GeV}: 12 \SI{400}{MHz} cavities in three cryo modules deliver an RF voltage of \SI{140}{MV} in the straight section around point J. Two cryo modules are installed in section 1 and one is installed in section 4.
     \item \SI{80.0}{GeV}: 40 additional \SI{400}{MHz} cavities will be installed to obtain a total number of 52 cavities and an accelerating voltage of \SI{750}{MV}. The number of cryo modules in the straight section around point J will be increased to seven in section 1 and six in section 4.
     \item \SI{120.0}{GeV}: In this stage the same RF infrastructure will be installed in both straight sections around J and D: 136 \SI{400}{MHz} cavities in 9 and 8 cryo modules in sections 1 and 4 respectively. The total RF voltage is \SI{2}{GV}.
     \item \SI{182.5}{GeV}: To compensate the massive energy loss at top energy the \SI{800}{MHz} cavities added in both point D and J. A total number of 480 cavities is installed in 30 cryo modules in both section 2 and 3 obtaining an RF voltage of \SI{10.9}{GV} in combination with the \SI{400}{MHz} cavities.
 \end{enumerate}
The staging scenario also includes modifications to the collider RF system \cite{taviancryolayout}, which are not discussed in this context. 
Lattice files with each RF configuration have been set up for the FCC-ee booster synchrotron and are available.

\section{Conclusion}
This report presented the status of the top-up booster synchrotron that will serve as full-energy injector for the FCC-ee collider rings as presented in the FCC-ee Conceptual Design Report.
Based on the boundary conditions imposed by the compatibility with both the hadron and the lepton colliders of FCC a lattice for the booster has been developed and described in detail.
Three optics with \SI{90}{\degree}/\SI{60}{\degree}, \SI{60}{\degree}/\SI{60}{\degree}, and \SI{90}{\degree}/\SI{90}{\degree} phase advance per cell have been implemented and several sextupole schemes have been investigated towards their dynamic aperture.
A first misalignment study showed, that the dynamic aperture is not significantly reduced for Gaussian distributed transverse quadrupole misalignments of $\sigma =\SI{100}{\micro\metre}$.
Calculations of IBS at injection energy of \SI{20}{MeV} showed, that the equilibrium emittance has to be increased in order to avoid emittance blow-up beyond the equilibrium value at extraction. 
Therefore, excitation wigglers have been proposed and implemented, which also help to limit the damping time to am maximum of \SI{0.1}{s}, which is required to allow full filling of both rings in \SI{20}{min}.
The wigglers could be installed in the long straight sections which also contain the RF installations, which have been included to the model according to the design of the RF colleagues.

As next steps in the design process the injection and extraction sections have to be added to the lattice.
For machine protection the necessity of collimators should be investigated as well as an emergency beam dump option to safely extract the beam in case of technical problems.
At the moment the wigglers are installed in the dispersion suppressor sections to the straight sections which also contain the superconducting RF. 
The potential interaction of the wiggler's synchrotron radiation light fan and the cavities is a subject to be further studied.
As a last step, we would propose start-to-end simulations including injection, accumulation, energy ramp and extraction to have a full overview about the beam dynamics in all different use-cases of the top-up booster synchrotron for FCC-ee.

\section{Acknowledgements}
The authors would like to acknowledge the fruitful discussion with the colleagues working on FCC-ee.
In particular they would like to thank Yannis Papaphilippou, Panos Zisopoulos Fanouria Antoniou for the discussions and support for NAFF analysis and for the IBS calculations, respectively.


\end{document}